\documentclass{article}
\usepackage[margin = 1in]{geometry}
\usepackage{rotating}
\usepackage{graphicx,footmisc}
\usepackage{lineno}
\usepackage{pifont}
\usepackage[caption=false]{subfig}
\usepackage{color}
\usepackage{natbib,url}
\usepackage{amssymb}
\usepackage{amsmath}
\usepackage{float}
\usepackage[normalem]{ulem}
\usepackage{longtable}
\usepackage{placeins}
\usepackage{epstopdf}
\usepackage{tikz}
\usepackage{soul}
\usepackage{setspace}
\def\be{\begin{equation}}
\def\ee{\end{equation}}
\def\begineqn{\begin{equation*}}
\def\endeqn{\end{equation*}}

\def\beginar{\begin{eqnarray}}
\def\endar{\end{eqnarray}}
\def\beginarn{\begin{eqnarray*}}
\def\endarn{\end{eqnarray*}}

\newcommand{\TOr}[1]{\textcolor{black}{#1}}

\def\lb{\left ( }
\def\rb{\right ) }
\def\ls{\left[ \vphantom] }
\def\rs{\right] }

\def\ep{\epsilon}

\def\Rat{\widetilde{Ra}}

\def\Ra{Ra}

\def\Ek{Ek}

\def\ub{\mathbf{u}}

\def\hz{{\bf\widehat z}}

\def\Gstd{{ {\Gamma}^{\prime}_z }}

\def\rcmb{{ {r_{cmb}}}}

\def\gvec{{\hat{\mathbf{g}}}}
\def\avec{{\hat{\mathbf{a}}}}
\def\gchar{{g}}
\def\achar{{a}}

\def\hc{\mathcal{H}}
\newcommand\hlm[2]{h_{#1,#2}}

\newcommand\phatLM{\pi^{\prime}_{\ell,m}}
\newcommand\PhatLM{\Pi^{\prime}_{\ell}}
\newcommand\hLM{h_{\ell,m}}

\def\title#1{%
{\centering \Large\bf #1 \vskip14pt}
\def\thetitle{#1}}
\def\authors#1{{\centering \normalsize\bf #1\vskip12pt}}
\def\affil#1{$^{#1}$\ignorespaces}
\def\affiliation#1#2{\vskip-.5\parskip\relax{\centering{\footnotesize
$^{#1}$#2\relax}\vskip-\parskip}}

\def\@fnsymbol#1{\ensuremath{\ifcase#1\or \dagger\or \ddagger\or
   \mathsection\or \mathparagraph\or \|\or **\or \dagger\dagger
   \or \ddagger\ddagger \else\@ctrerr\fi}}

\begin{document}
\title{Effects of global core-mantle boundary topography on outer-core convection and topographic torques}
\authors{
Tobias G. Oliver\affil{1}\footnote{Corresponding Author: tobias.oliver@colorado.edu}, 
Eric G. Blackman\affil{2}, 
John A. Tarduno\affil{3}, 
and Michael A. Calkins\affil{1}
}
\affiliation{1}{Department of Physics, University of Colorado, Boulder, Colorado 80309}
\affiliation{2}{Department of Physics and Astronomy, University of Rochester, Rochester, NY 14627}
\affiliation{3}{Department of Earth and Environmental Sciences, University of Rochester, Rochester, NY 14627}


\begin{abstract}
\TOr{Topography at the core-mantle boundary (CMB) couples the outer core to the mantle and likely generates observable variations in the length of day ($\Delta$LOD) and the geomagnetic field, though these effects remain poorly understood.
We use direct numerical simulations of rotating shell convection with finite-amplitude CMB topography to investigate dynamical effects on the outer core.
A range of topographic shapes is used, including individual spherical harmonics and a model representing seismically inferred heterogeneities in the deep mantle.
As predicted by prior linear theory in the rotating annulus model, a new instability arises for Rayleigh numbers below the onset of convection; we confirm its existence in a global geometry, though the predicted scalings are quantitatively modified.
The shape of the geostrophic contours --- lines of constant axial height --- plays a central role: deformed contours allow buoyancy to do work on the time-averaged flow, driving increases in Reynolds and Nusselt numbers of up to $\sim$100\% relative to a spherical boundary.
Previous work showed that topographic torques scale linearly with topographic amplitude and quadratically with flow speeds; we confirm this scaling and extend it with new theory that estimates the torques for global, spectrally broad topography.
When extrapolated to core conditions, the predicted torques are consistent with the magnitude required to drive observed decadal and subdecadal  $\Delta$LOD variations.}
\end{abstract}
\section*{Plain Language Summary}

The boundary between the liquid core and the mantle of the Earth is neither perfectly spherical nor smooth.
We expect that structure in the mantle generates topography at this boundary, and that this topography generates new mechanisms by which the fluid motion in the core can interact with the mantle. In this study, we present numerical simulations of the liquid core in which we implement boundary topography. We find that topography can increase flow speeds and the amount of heat that is transported from the inner core to the mantle. In addition, we investigate the pressure torques exerted by the core on the topography. In agreement with previous studies, we determine that the pressure torques are large enough to explain variations in the Earth's length of day on timescales of roughly 6 to 60 years.

\section{Introduction}

Topography on the Earth's core-mantle boundary (CMB) likely plays an important role in the dynamics of the core, with possible observable effects in both the geomagnetic field \citep{aA77} and changes in the length of day ($\Delta$LOD) \citep{dJ89,HdV05,pR12,sR23,fM25}. 
\TOr{\cite{jT15} hypothesized that CMB topography can generate reversed magnetic flux patches and the decadal and subdecadal fluctuations in $\Delta$LOD have been attributed to core coupling \citep[e.g.][]{HdV05}. $\Delta$LOD is known to fluctuate on a broad range of timescales; however, different timescales are associated with different interactions \citep{rG15}. High frequency variations on the scale of days to months are attributed to interactions with the atmosphere \citep[e.g.][]{jD94} and, to a lesser extent, the oceans \citep{rG04}. Slow variations, on the scale of millennia and longer, occur due to the lunar forcing of the tides and glacial isostatic adjustment \citep[e.g.][]{rG15}, as well as the redistribution of terrestrial ice \citep{mKS26}.
	The decadal and subdecadal fluctuations, which are $1-2$ ms in amplitude, are attributed to interactions with the core \citep{HdV05, sR23}, although the particular coupling mechanism is unknown.
}

\TOr{\cite{rH69} proposed that mechanical coupling between the core and CMB topography could generate topographic torques, although other couplings exist. 
}
Gravitational coupling between the inner core and the mantle \citep{bB96,jM03} and electromagnetic coupling between core-generated magnetic field and electrically conducting regions at the base of the mantle \citep{bB07,sB70, dJ15,wK19} both likely play a role in generating $\Delta$LOD. 
\TOr{The core and mantle are coupled by viscous forces as well; however, the torques generated are likely too small to account for the observed signal in $\Delta$LOD \citep{bB07,tO25}.} 
Recent work has demonstrated that topographic torques exhibit scalings that indicate they are of sufficient magnitude to play a significant role in $\Delta$LOD \citep{tO25}, 
\TOr{ although this study only considered a fixed-width, isolated topography whereas CMB topography is likely globally distributed.}
\TOr{Other} studies investigating the ramifications of CMB topography have been limited to the linear regime \citep{pB96,wK01}, or simplified two-dimensional models that may not fully capture all dynamical effects \citep{wK93,jH98,mW03,mC12}.
The primary goal of the present work is to determine how globally distributed topography influences both the topographic torques and core dynamics using a suite of fully three-dimensional, turbulent numerical models. 

While the shape of the CMB is poorly constrained, seismic investigations and geodynamic modeling of the deep mantle point to possible structure on the CMB. 
Seismological studies show considerable variation in inferred CMB structure, with typical topographic amplitudes up to a few kilometers \citep{kK03,sT10}. 
A possible cause for CMB topography is convection in the mantle. 
It has been proposed that regions of slow seismic velocity deep in the mantle, the so-called Large Low Shear Velocity Provinces (LLSVPs) \citep{sC16,pK21}, contribute to the CMB deformation. 
The largest LLSVPs are located at the southern tip of Africa and in the southern Pacific. These findings suggest that CMB topography may consist of a few large structures. 
\TOr{
Geodynamic modeling of the deep mantle paints a more complicated picture; the LLSVPs may generate short wavelength topographic features depending on the thermal and compositional characteristics (\citealp{tL07}; \citealp{tL10}). }
Furthermore, the edges of the LLSVPs may be quite sharp, which may imply lateral variations in topography over tens to hundreds of kilometers \citep{sC16}. 

%


The canonical model of core dynamics consists of a rotating, self-gravitating fluid bounded by spherically symmetric surfaces that is heated from within. 
For sufficiently strong heating, the basic instability in this system consists of convective Rossby waves that are characterized by an approximately axially invariant structure with short longitudinal wavelength \citep{pR68,fB70}. 
These waves give rise to broadband turbulence as the system is forced more strongly \citep[e.g.][]{tG16,jN24}, and the resulting motions are known to readily generate magnetic fields that bear similarities with the geomagnetic field for certain parameter values (\citealp{nS17}; \citealp{cF23}; \citealp{juA23}; \citealp{nG22}; \citealp{nG24}; \citealp{yL25}). While this system continues to serve as the `zeroth' order model for the geodynamo, it is known that unique dynamical effects arise when the CMB deviates from axial symmetry \citep{dJ20,rM25}. 
A well-known result for rapidly rotating flows is that the streamlines follow a particular set of geometric curves of constant axial height known as geostrophic contours \citep{hG68}. 
In a perfect sphere these curves form circles about the origin, and a mechanism to break geostrophy is a necessary ingredient for the convective instability. 
In the hydrodynamic case, \TOr{the convective instability arises when} inertia, buoyancy, and viscosity perturb the geostrophic balance to yield unstable lengths scales of size $O\lb Ek^{1/3}\rb,$ where $Ek$ is the Ekman number -- a ratio of viscous to Coriolis forces \citep{sC61}. 
\TOr{As the Coriolis force increases in strength (faster rotation), the convective scales decrease; however, it is known that}
the scales tend to increase as the system becomes more turbulent \citep{tG16, cG19, tO23, jN24, cG25}. 

\cite{pB96} and \cite{aB96} showed that boundary topography can lead to the generation of instabilities that are distinct from the convective Rossby waves. These instabilities are more unstable in comparison to the Rossby waves. They are characterized by length scales that are comparable to the underlying topography and are not directly controlled by viscosity.
Topography breaks the axisymmetry of the geostrophic contours and allows buoyancy to directly drive flow along the contours.
Deviations from rotational symmetry are likely present on the CMB, implying that the geostrophic contours may be characterized by complex shapes. The asymmetric flows inferred from geomagnetic field observations indicate core-mantle coupling is occurring (\citealp{rH11}; \citealp{pL17}; \citealp{cF12}; \citealp{cF23}), with CMB topography representing one of the primary coupling mechanisms. 
A known mechanism is that resonant interactions between the flow and topography can generate time-independent flows which likely have consequences for the structure and dynamics of the geomagnetic field \citep{mC12}.

The present work studies the influence of CMB topography on core convection and the resulting torques through direct numerical simulations that allow for a non-spherical boundary. 
The paper is organized as follows. 
In Section \ref{s:methods}, we introduce the governing equations for rotating convection, discuss the implementation of CMB topography, and define relevant quantities for the investigation.
Results are given in Section \ref{s:results}, where we discuss heat and momentum transport, flow morphology, and topographic torques. To compare with the theoretical predictions of \cite{pB96}, we first present results from a set of simulations in which the thermal forcing is weak. 
The results from the turbulent simulations are then presented, which are all performed at the same rotation rate and thermal forcing.
Five different topographic shapes are considered, including four that are represented by single spherical harmonics, and a fifth topographic shape that is informed by studies of LLSVPs. 
For each topographic shape we vary the topographic amplitude.
\TOr{In total, we present results from 51 numerical simulations, which are summarized in Tables \ref{t:p_space}-\ref{t:p_space3}.}

We discuss the role of shape and amplitude on the global transport of momentum and heat, as well as the topographic torques exerted on the CMB. 
We demonstrate that the deformation of the geostrophic contours generally serves to increase the heat and momentum transport, but has a more complicated effect on the torque for certain parameters. 
We provide further evidence and justification for the dynamical scaling law which suggests a linear dependence on topographic amplitude and a quadratic dependence on flow speed \citep{tO25}.
We propose a procedure by which we can determine the magnitude of the torques given a particular topographic shape.
In Section \ref{s:disc} we discuss our results and provide geophysical context.

\section{Methods}
\label{s:methods}

\subsection{Governing equations}

We perform numerical simulations of convection in deformed spherical shell geometries rotating with constant rotation vector $\boldsymbol{\Omega} = \Omega \, \hz$. The non-dimensional inner and outer radii are denoted $r_i$ and $r_o$, respectively. In all cases presented the aspect ratio of the geometry is fixed to the core-like value of $\chi = r_i/r_o = 0.35$, such that $r_i = \chi/(1-\chi) = 7/13$ and $r_o = 1/(1-\chi) = 20/13$. The inner boundary is spherical in all cases, whereas the outer boundary is deformed according to the relationship
\be
r_{cmb} = r_o - \epsilon  h\lb \theta,\phi\rb,
\label{e:rcmb}
\ee
where $h\lb \theta,\phi\rb$ is the topographic shape function and $\epsilon$ is the dimensionless maximum topographic amplitude. 
\TOr{We note that a positive value of $h\lb\theta,\phi\rb$ corresponds to a topographic intrusion into the core, while a negative value generates positive \textit{mantle-side} topography; throughout this study, we use the terms \textit{CMB topography} and \textit{topography} to indiscriminately refer to nonzero values of $h$. }
The co-latitude and azimuthal angle are $\theta$ and $\phi,$ respectively. In a co-rotating frame the non-dimensional governing equations for a Boussinesq fluid are
given by
\begin{subequations}
\begin{equation}
	\frac{\partial \mathbf{u}}{\partial t} + \mathbf{u}\cdot \nabla \mathbf{u} + \frac{2}{Ek}\hat{\mathbf{z}} \times \mathbf{u} = -\nabla p + \frac{\Ra}{Pr}\frac{\mathbf{r}}{r_{o}} \vartheta + \nabla^{2}\mathbf{u} ,
	\label{e:mom}
\end{equation}
\begin{equation}
	\frac{\partial \vartheta}{\partial t} + \mathbf{u}\cdot \nabla \vartheta = \frac{1}{Pr }\nabla^{2}\vartheta ,
	\label{e:energy}
\end{equation}
\begin{equation}
	\nabla\cdot\mathbf{u} = 0,
	\label{e:incom}
\end{equation}
\end{subequations}
where the velocity, pressure and temperature are denoted by $\ub $, $p$, and $\vartheta$, respectively, and the position vector is $\mathbf{r}$. The equations have been non-dimensionalized using the shell depth $d$, viscous diffusion time $d^2/\nu$ (where $\nu$ is the kinematic viscosity), and temperature $\Delta T$, which is the temperature difference between the inner and outer boundaries.
We use $u,v,$ and $w$ to refer to the $\hat{\mathbf{x}},\hat{\mathbf{y}}$, and $\hat{\mathbf{z}}$ components of the velocity field respectively, and $u_{r}, u_{s},$ and $u_{\phi}$ to refer to the
$\hat{\mathbf{r}},\hat{\mathbf{s}}$, and $\hat{\boldsymbol{\phi}}$ components, where $s$ refers to the cylindrical radial coordinate. 

The non-dimensional control parameters appearing above are the Ekman, Rayleigh, and Prandtl numbers defined, respectively, as
\be
Ek = \frac{\nu}{\Omega d^{2}},\;\; Ra = \frac{g_o \alpha \Delta T d^{3}}{\nu\kappa}, \;\; Pr = \frac{\nu}{\kappa},
\ee
where $g_o$ is the gravitational acceleration at the CMB, $\alpha$ is the coefficient of thermal expansion, and $\kappa$ is the thermal diffusivity of the fluid. 
\TOr{ 
It is useful to define the conductive temperature profile 
\be
\vartheta_{c}\lb r\rb  = \frac{r_{i}r_{o}}{ r} - r_{i},
\label{e:tec}
\ee
which satisfies equations \eqref{e:mom}-\eqref{e:incom} in the absence of fluid motion and temporal variation ($\mathbf{u} = 0,$ $\frac{\partial}{\partial t} = 0$). Per our choice, $\vartheta_c$ is non-dimensionalized by $\Delta T$.
}

The primary goal of the present study is to determine the role played by the topographic parameters $\epsilon $ and $h\lb \theta,\phi\rb $. Thus, to narrow the parameter space, the Prandtl number is fixed at $Pr = 1$. 
With the exception of five simulations performed at $Ek = 10^{-4},$ the Ekman number is fixed at $Ek = 10^{-6}.$ 
No-slip, fixed temperature boundary conditions are employed. The temperature at $r_{cmb}$ is set to zero in all simulations except for a small set of laminar simulations which we discuss in Section \ref{s:subc}. 
\TOr{The inner temperature boundary condition is $\vartheta\lb r = r_{i}\rb  = 1$.} 
In the absence of topography, the onset of convection occurs at the critical Rayleigh number $Ra_c \approx 1.8\times 10^8$ and critical azimuthal wavenumber $m_c = 32$ \citep[e.g.][]{aB22}. 
Because the critical Rayleigh number increases with decreasing $Ek,$ it is useful to define the reduced Rayleigh number
$\Rat = Ra Ek^{4/3},$ 
which is an $O\lb 1\rb $ quantity near onset as $Ek\rightarrow0$ \citep[e.g.][]{eD04}. For convection in a perfect spherical shell of aspect ratio $\chi = 0.35$, the reduced critical Rayleigh number is $\Rat_{c} \approx 1.8$. 
However, as mentioned previously, unique instabilities arise in the deformed geometries which yield non-zero flows for Rayleigh numbers below this value. 

The results in this study are grouped into two categories. We first present results for small values of $\Rat,$ such that $\Rat<\Rat_{c}.$
As we show, some of these cases are unstable to fluid motion due to topographic instabilities. We will nevertheless refer to these simulations as ``subcritical'' because $\Rat<\Rat_{c}.$
The second group of results are turbulent, and for these cases we fix $\Rat = 40$ and $Ek = 10^{-6}$. Tables \ref{t:p_space} - \ref{t:p_space3} provide a comprehensive list of the run parameters.

\subsection{Numerical Methods}

The simulations are performed with the \texttt{Nek5000} code \citep{nek5000} that is based on the spectral element method \citep{pA84}. We use a cubed sphere to generate the mesh. In the radial direction we use $24$ elements distributed non-uniformly on a Chebyshev grid such that at least one element spans the Ekman boundary layer. In physical space this corresponds to a radial resolution of $192$ gridpoints.
The longitudinal and latitudinal dimensions are composed of $6$ blocks, each of which are comprised of $40\times 40$ elements. 
The total number of elements is $230\text{ }400$ and we use an order $7$ interpolating polynomial along each element so that the total resolution is $230\text{ }400\times8^3 \approx 1.2\times10^{8}$ points.

\subsection{Topography}

One of the primary goals in the present work is to better understand the impact of topography morphology on outer core convection.  
For most of the cases we therefore choose topographic shapes characterized by single spherical harmonics so that topographic length scales are easily defined. Thus, the topographic shape function corresponding to a spherical harmonic of degree $\ell$ and order $m$ is denoted as $\hlm{\ell }{m}$ and defined by
\begin{equation}
	\hlm{\ell}{m}\lb \theta,\phi\rb = \frac{1}{Q_{\ell}^{m}}P_{\ell}^{m} \lb \cos{\theta}\rb\; \cos\lb m \phi\rb,
	\label{e:hlm}
\end{equation}
where $P_{\ell}^{m}\lb \cos\theta\rb $ is the associated Legendre polynomial for degree $\ell$ and order $m,$ and $Q_{\ell}^{m}$ is the maximum value of $P_{\ell}^{m}\lb \cos\theta\rb.$
We study topographic amplitudes ranging from $0.01<\epsilon<0.2.$ All studied values of $\epsilon$ are larger than the estimated core value $\epsilon \approx 10^{-3},$ because we desire topography with amplitudes greater than the viscous Ekman boundary layer thickness $\delta_{E}=O\lb Ek^{1/2}\rb.$ In the core, a conservative estimate is $\epsilon /\delta_{E}=10^{4}$, and it is necessary to maintain this hierarchy of length scales. $\delta_{E}= O\lb 10^{-3}\rb $ in our simulations, so we study topographic amplitudes $\epsilon >\delta_{E}.$
We consider $\hlm{32}{16},\hlm{8}{4},$ and $\hlm{2}{1}$ across a sweep of $\epsilon$, and $\hlm{32}{17}$ for $\epsilon  = 0.1$. \TOr{We note that $\epsilon  = 0.1$ corresponds to a topographic amplitude of $226 \text{ km}$ when scaled to the parameter regime of the Earth's core.}

We also study a topographic shape informed by seismological investigations of the deep mantle. 
For this shape, which we refer to as the ``tomographic model'' and denote the corresponding shape function as $h_{t}$, we use the cluster analysis maps developed in \cite{vL12}, and consider only the areas where all of the included studies \citep{cH08,jR11,bK08,cM00,nS10} agree on the presence of an LLSVP.
\TOr{The LLSVPs are likely denser than the surrounding mantle and are expected to generate topographic intrusions into the core \citep{jT15}. However, identifying density variations and the resulting topography at the CMB is more complicated than simply locating LLSVPs. For example, many studies have identified ultralow velocity zones \citep[e.g.][]{sR06,sY18,sH23}, indicating heterogeneity within the LLSVPs. Furthermore, dynamical modeling in the mantle indicates that the thermal and compositional structure of the LLSVPs can generate a variety of topographic structures \citep{tL07}. Since our intention is to model a topography with a similar spatial distribution to deep mantle structure,
	we follow the conceptual model of \cite{jT15}, who were motivated by the occurrence of a prominent CMB reversed flux patch with the edge of the African LLSVP. \cite{jT15} noted that the LLSVP might affect core flow through differences in core-side CMB topography \citep{mC12}, and mantle-side differences in conductivity or heat flow \citep{bB96}. While recognizing the desire to understand the interplay between these factors, \cite{jT15} focused first on a potential topographic effect, with an intrusion into the core compensating the high density of the LLSVP.} Therefore, we
simply associate LLSVPs with \TOr{uniform} intrusions into the core.

\TOr{
Several studies have investigated the effects of heterogeneous heating at the CMB, which are believed to influence the morphology of the geomagnetic field \citep{cD08,pO10,jM23}. In this study we do not consider heterogeneous thermal boundary conditions except for a small sweep of low $\Rat$ simulations discussed in Section \ref{s:subc}; otherwise, the temperature is fixed at a constant value across the surface $r= r_{cmb}$.
}

Although CMB topography is likely broadband in structure, the voting map is essentially binary -- it shows where models agree and where they disagree concerning LLSVPs. \TOr{The shape of the LLSVPs is likely well represented by the voting map; however, the structure near the edges is artificial. The voting procedure yields a discrete distribution of votes which generates an abrupt transition between regions with and without LLSVPs.
Although we note that the boundaries of the LLSVPs are expected to be sharp \citep[e.g.][]{jR11}, a choice must be made with regards to representing the voting map as a continuous, smooth function on the CMB.} 
To smooth the edges we use an error function to build $h_{t}$. 
We fix the parameter determining the gradient of the topographic edges, so that $h_{t}$ depends on the value of the topographic amplitude according to
\begin{equation}
	h_{t}\lb \epsilon ,\theta,\phi\rb  = 0.5\lb\text{erf}\ls \delta \lb \theta,\phi\rb/\sigma \epsilon \rs -1\rb,
	\label{e:ht}
\end{equation}
where $\delta\lb \theta,\phi\rb $ gives the nearest distance to a cluster edge with the convention chosen such that $\delta\lb \theta,\phi\rb >0$ ($<0$) inside (outside) the clusters. The slope parameter is chosen such that $\sigma=28.2,$ and corresponds to an average gradient of 0.02 per degree (that is a topography of amplitude $\epsilon=0.02$ decays to $10^{-2} \times\epsilon $ over $1^{\circ}$). 
\TOr{
We note that $h_t \leq 0$ everywhere, with the most negative values occurring far away from the LLSVPs. Therefore, all of the modeled topography can be thought of as areas of relief where there are no LLSVPs or areas of intrusion at the LLSVPs.
}
\TOr{ The same sweep of amplitudes is performed for $h_t$ as for the single-harmonic topographies.
}

\begin{figure*}
	\begin{center}
         \subfloat[][]{\includegraphics[width=0.3\textwidth]{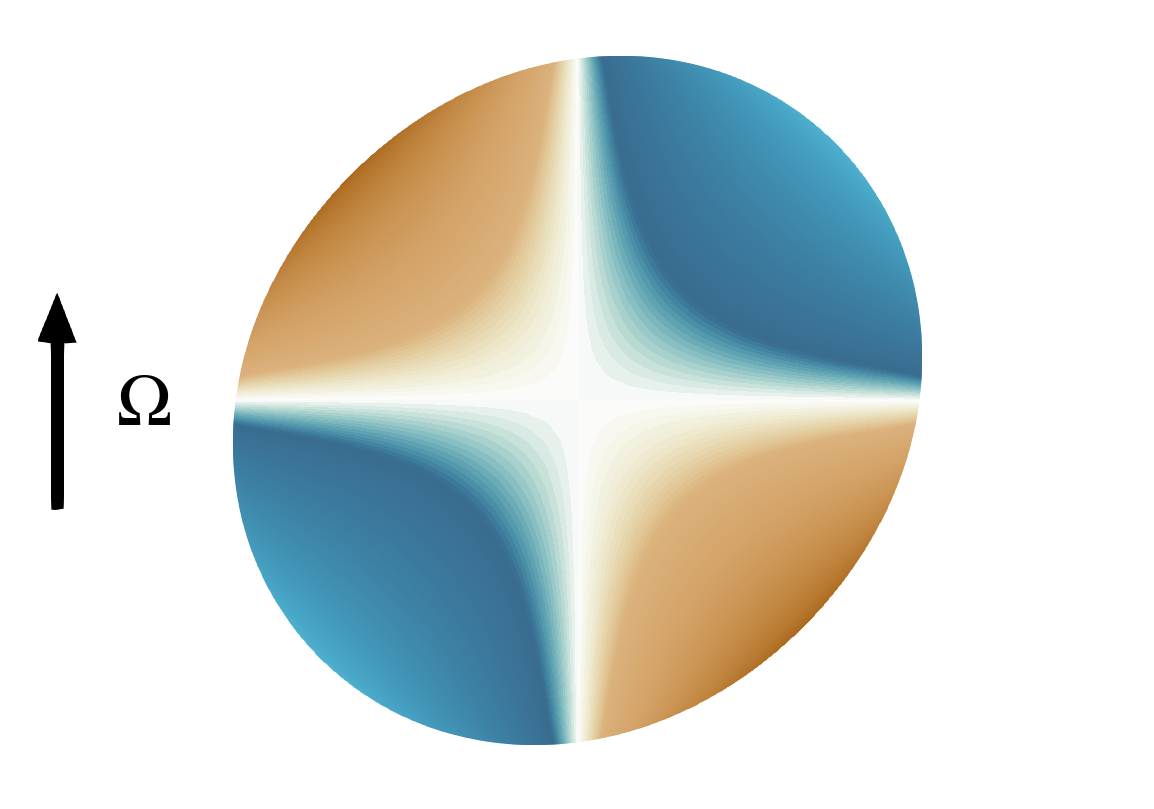}}
	\subfloat[][]{\includegraphics[width=0.3\textwidth]{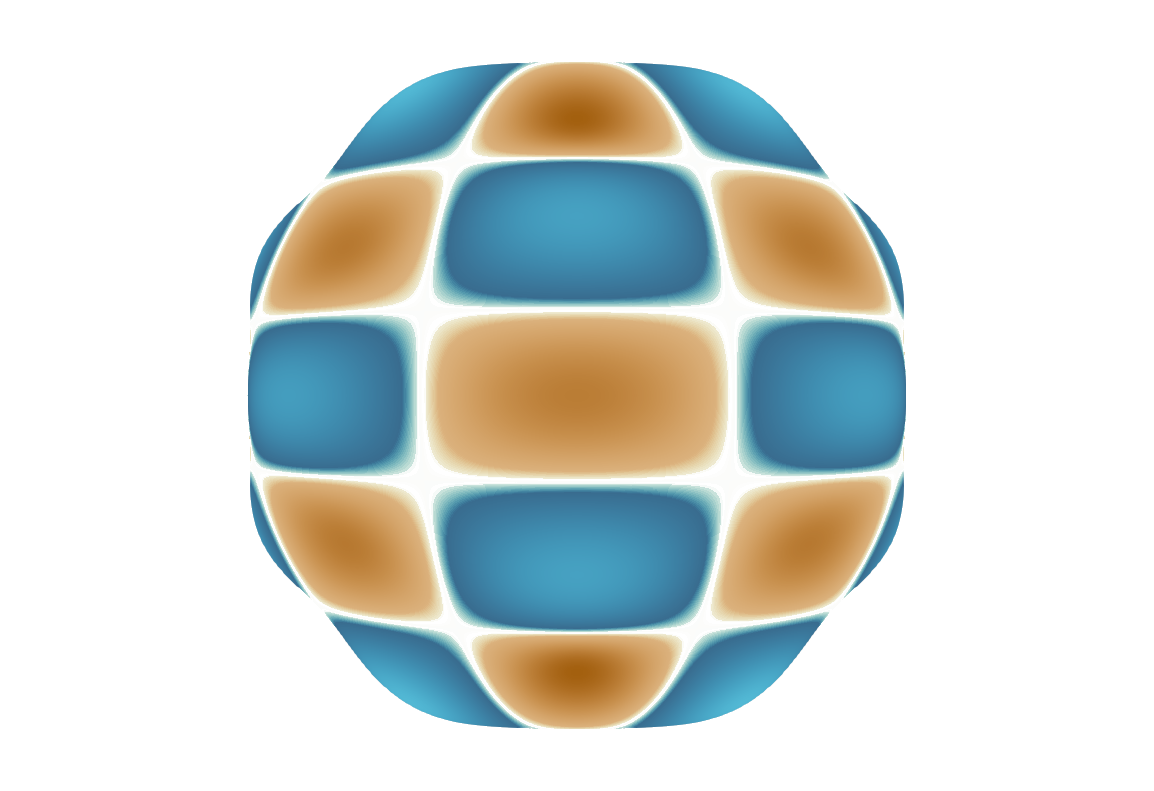}} 
	\subfloat[][]{\includegraphics[width=0.3\textwidth]{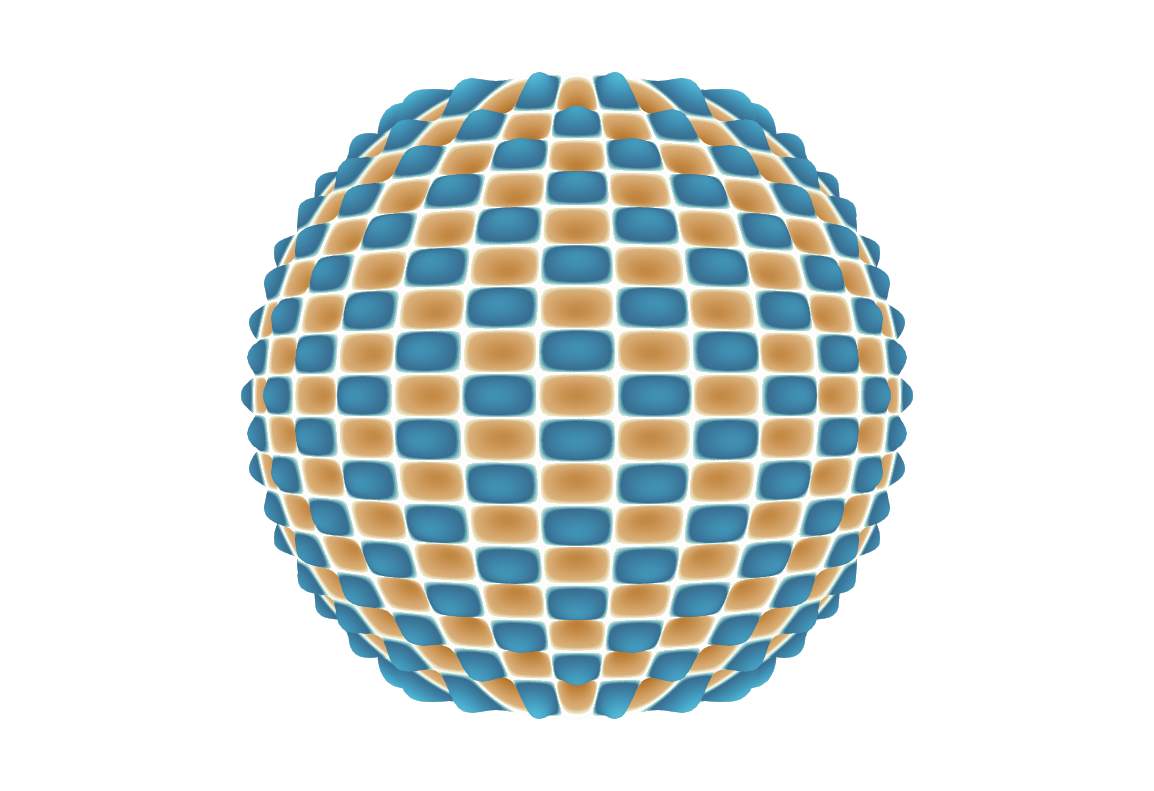}}\\
	\subfloat[][]{\includegraphics[width=0.3\textwidth]{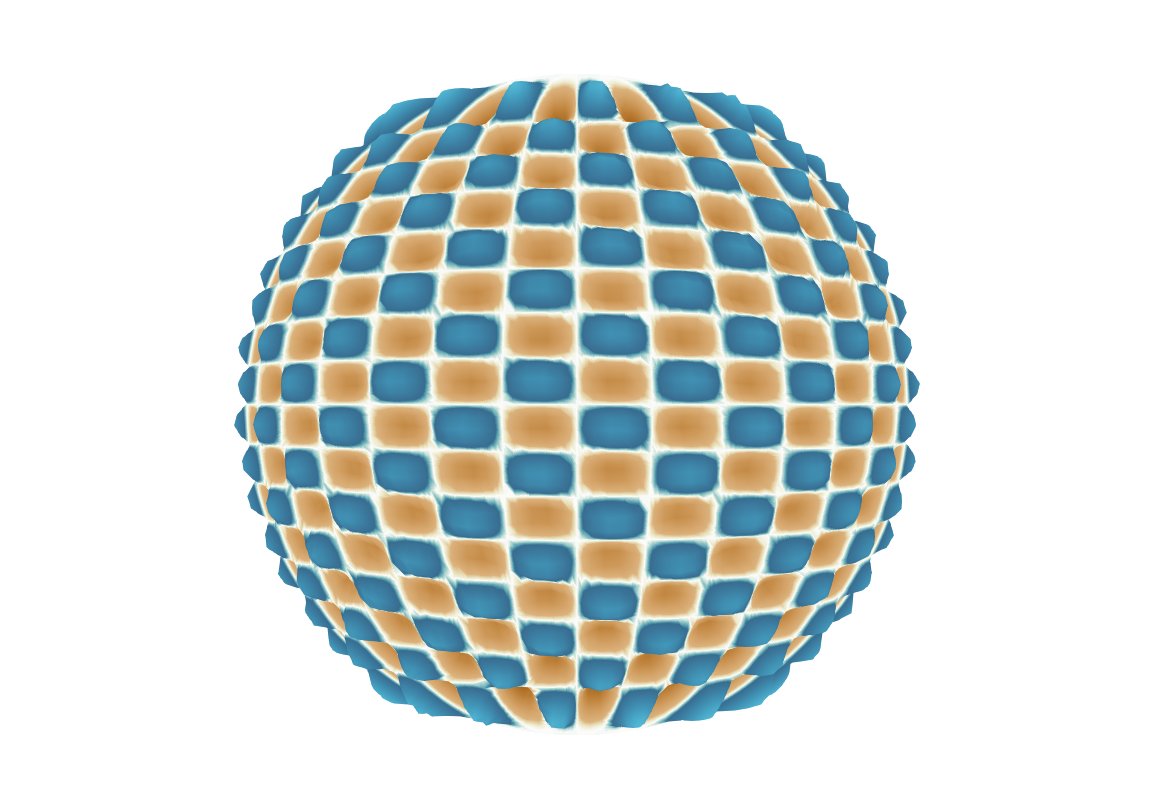}} 
	\subfloat[][]{\includegraphics[width=0.3\textwidth]{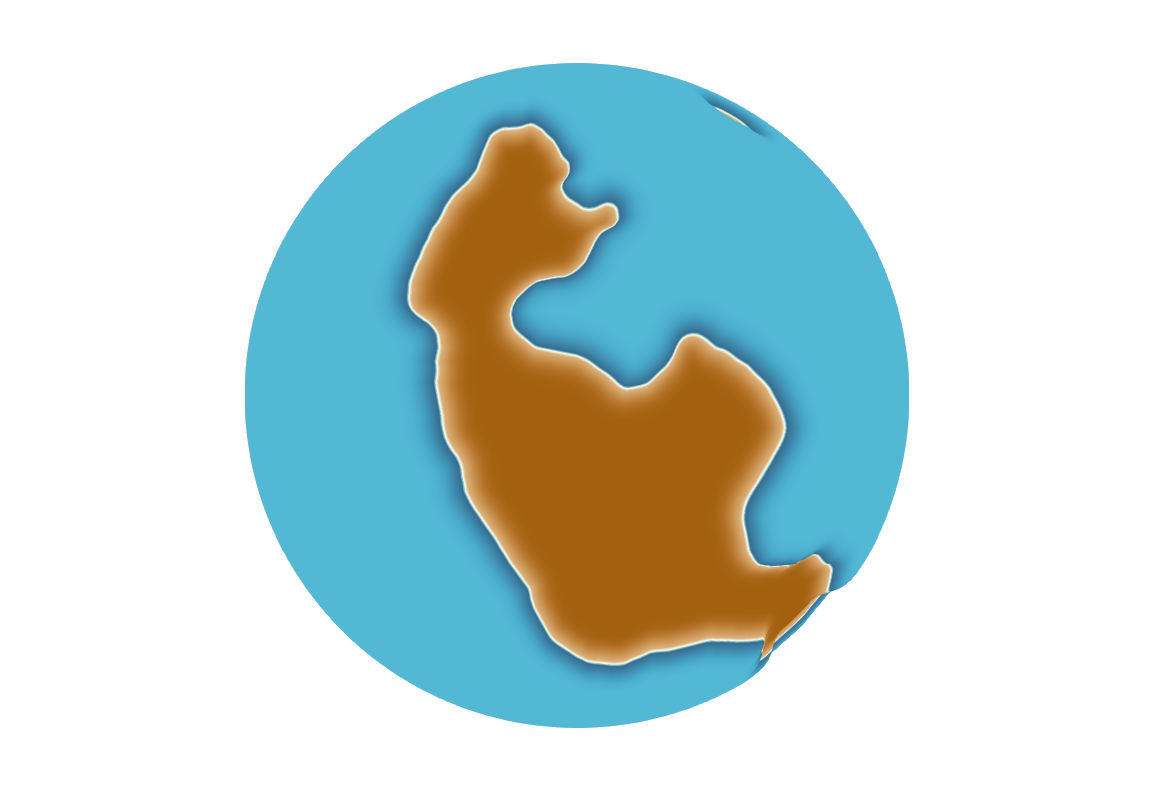}}
	\subfloat[][]{\includegraphics[width=0.3\textwidth]{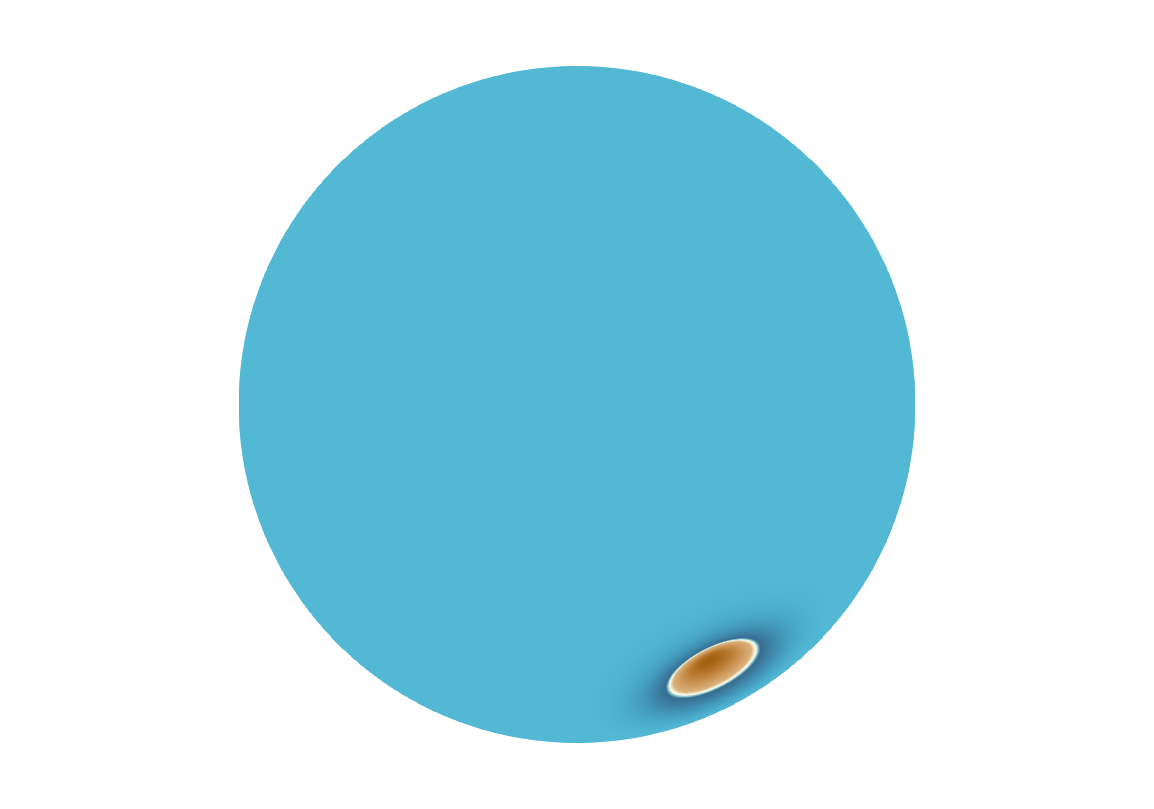}}\\
	\end{center}
	\caption{Visualizations of the topography used in the simulations: (a) $\hlm{2}{1}$;  (b) $\hlm{8}{4}$; (c) $\hlm{32}{16}$; (d) $\hlm{32}{17}$; (e) $h_t$; (f) localized topography used in the study of \cite{tO25}.
	}

	\label{f:topo}
\end{figure*}

Figures \ref{f:topo}(a)-(e) display the topographic shapes studied in the present work. For comparison, Figure \ref{f:topo}(f) displays the localized Gaussian bump investigated in \cite{tO25}.
In all cases $h\lb \theta,\phi\rb $ is normalized such that $\text{max}{\left|h\right|} = 1.0$. 
We note that this is not the same as the orthonormal or  ``4$\pi$'' normalizations used in many texts, and that the surface area of the outer boundary is not constant across the different topographic shapes.
Since $\hlm{\ell}{m}$ is composed of a single harmonic, the surface area of the CMB is only weakly dependent on the topographic amplitude ($O\lb \epsilon^{2}\rb $). This is not the case for the tomographic model, where $h_{t}$ has an $O(\ep)$ spherical mean. Table \ref{t:sa} provides the CMB surface areas for all geometries, demonstrating that only small changes in the surface area are observed over our range of parameters.

\subsubsection*{S\MakeLowercase{pectrum of the tomography model}}
\begin{figure}
	\begin{center}
		\subfloat[][]{\includegraphics[width=0.48\textwidth]{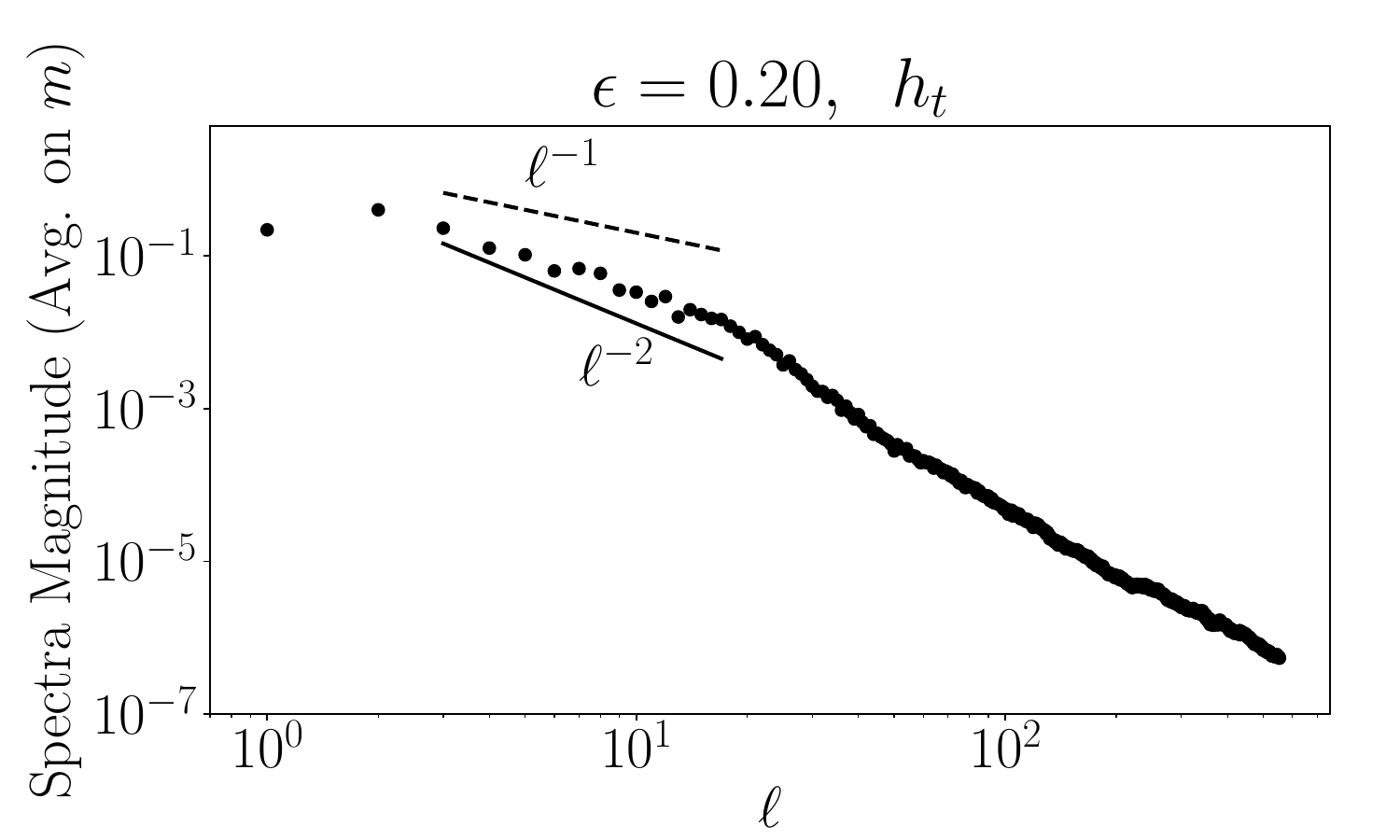}}
	\end{center}
	\caption{Spectral decomposition of the tomographic shape function, $h_{t}$ at $\epsilon  = 0.20$, averaged over $m$. The solid black line shows the scaling for Kaula's rule \citep{wK66}, which we use to estimate the torques. The dashed black line shows $\ell^{-1}$ for reference.}
	\label{f:ht_spec}
\end{figure}

During the discussion of topographic torques we present a scaling argument which requires a distribution rule for the size of the spherical harmonic coefficients of topography. 
\cite{mP23} and \cite{vD25} used Kaula's rule \citep{wK66} to extend the CMB topography spectrum when calculating topographic torques, and we adopt the same approach while developing the scaling argument.
Kaula's rule suggests that the spherical harmonic coefficients for geophysical topography scale as $\ell^{-2}$. 
The tomographic shape that we employ is artificial (with no extended spectrum), but it is helpful to compare it with Kaula's rule to assess the validity of the scaling argument for our particular dataset.
Figure \ref{f:ht_spec} shows the magnitudes of the spectral coefficients for $\epsilon  = 0.20,$ $h_{t},$ where we observe that the spectrum roughly follows Kaula's rule. This rough agreement is interpreted as coincidental although it may be that the coarse cluster analysis performed by \cite{vL12} is sufficient to reproduce the rule. 
Regardless, the plot in Figure \ref{f:ht_spec} does not reflect any physical result, it merely indicates that we expect our argument using Kaula's rule to apply to the tomographic model.  It is straightforward to adjust the scaling law according to any topography spectrum.

\subsection{The geostrophic height function}

\begin{figure}
	\begin{center}
		\subfloat[][$\hlm{2}{1}$]{\includegraphics[width=0.19\textwidth]{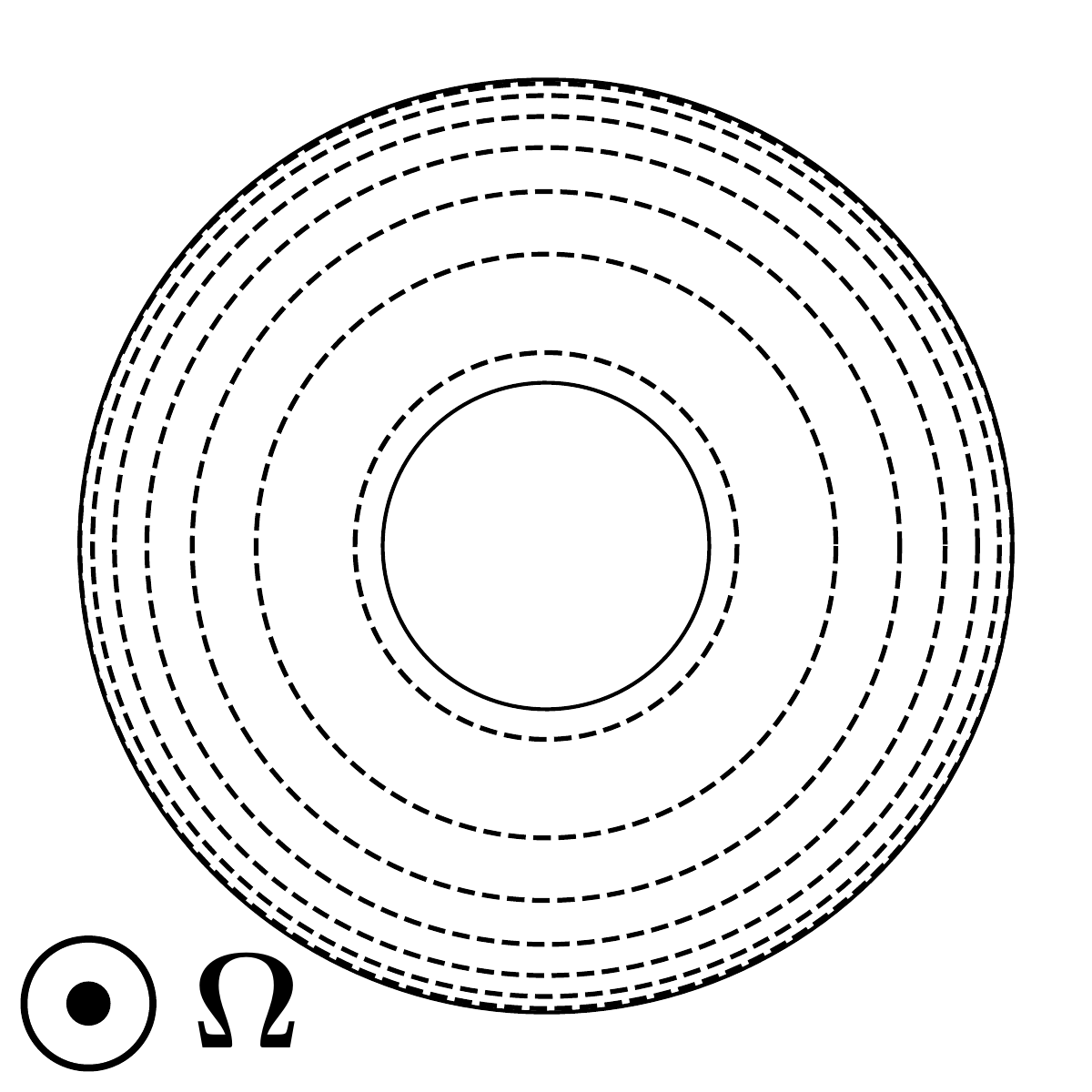}}
	\subfloat[][$\hlm{8}{4}$]{\includegraphics[width=0.19\textwidth]{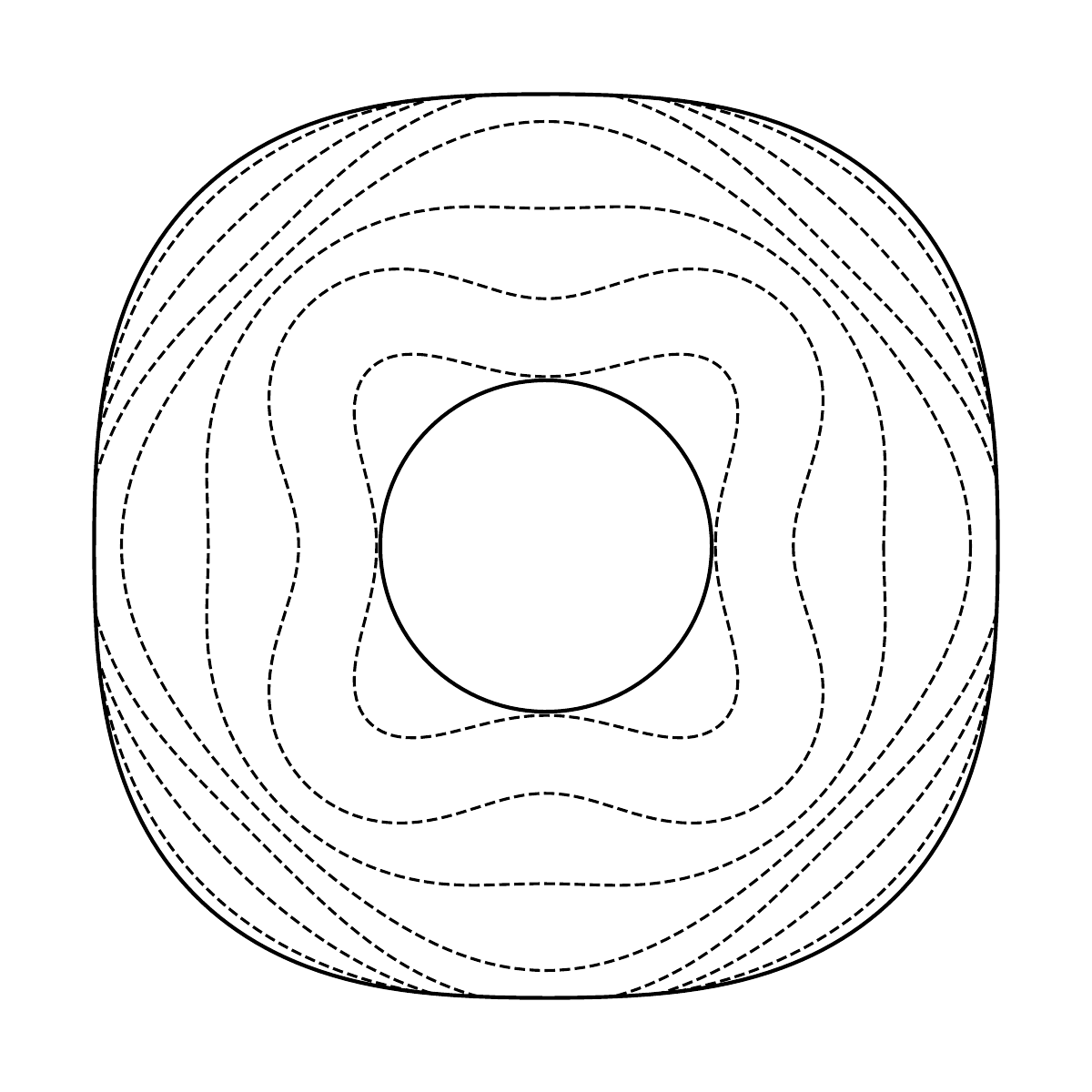}}
	\subfloat[][$\hlm{32}{16}$]{\includegraphics[width=0.19\textwidth]{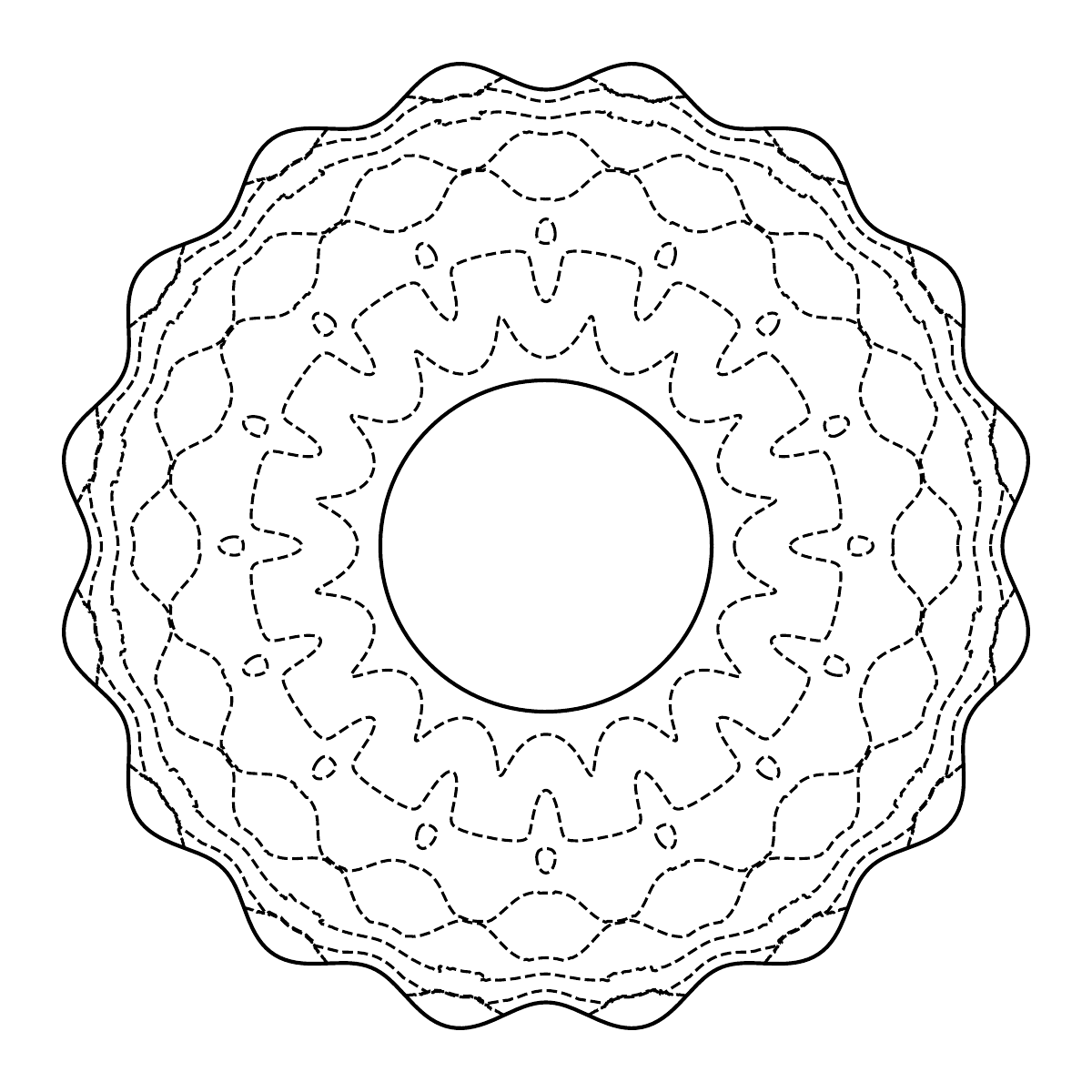}}
	\subfloat[][$\hlm{32}{17}$]{\includegraphics[width=0.19\textwidth]{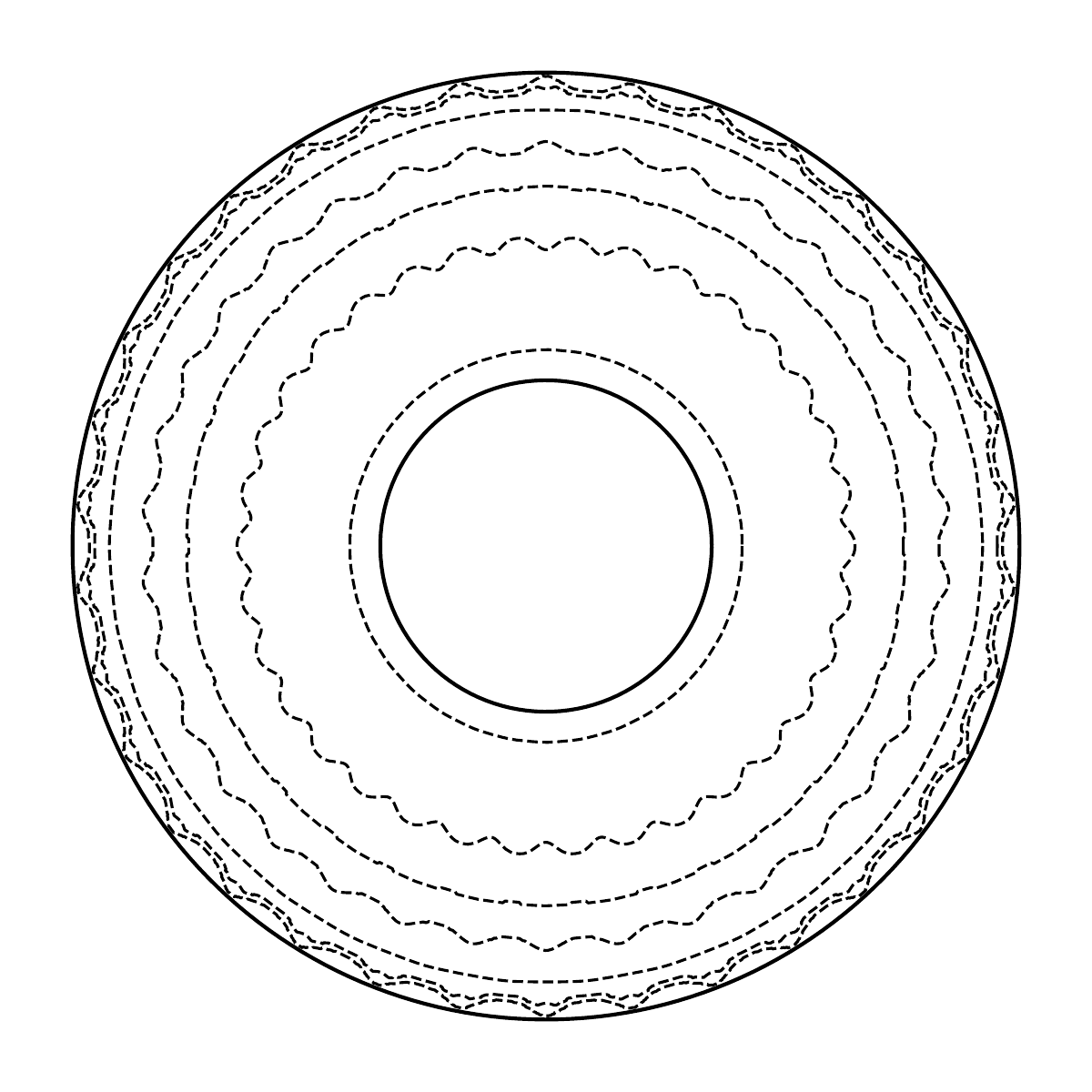}}
	\subfloat[][$h_{t}$]{\includegraphics[width=0.19\textwidth]{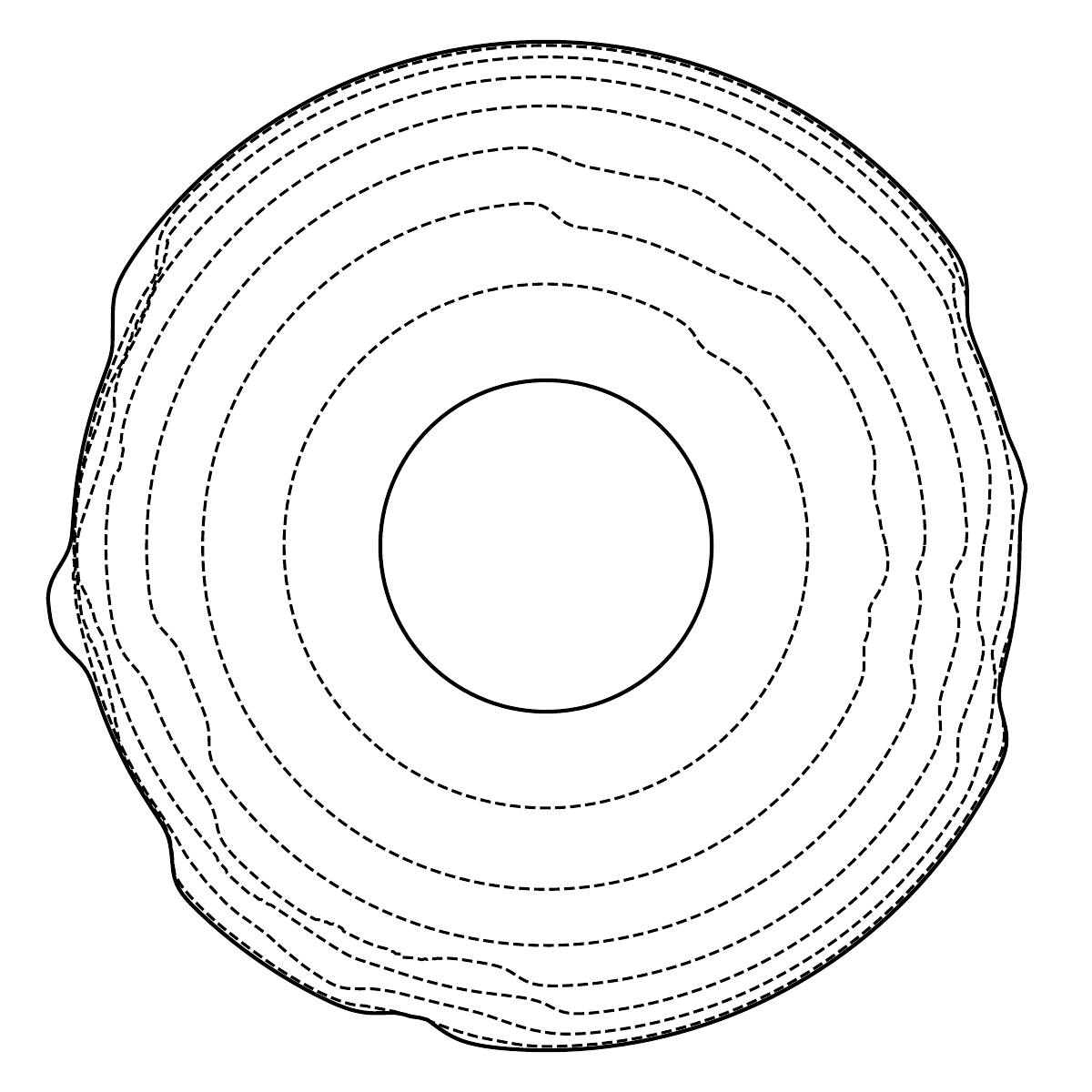}} \\
	\end{center}
	\caption{Geostrophic contours for (left to right) $\hlm{2}{1}$, $\hlm{8}{4}$, $\hlm{32}{16}$, $\hlm{32}{17}$, and $h_{t}$ for $\epsilon  = 0.1$. \TOr{Azimuthal oscillations of the contours correspond to the topographic order $m$. Topographies with odd $\ell-m$ (panels a, d) are equatorially anti-symmetric, so intrusions in the northern hemisphere are matched by extrusions in the southern hemisphere; the deformations of $H_{geo}$ partially cancel between hemispheres and scale as $O\lb \epsilon^{2}\rb$ with order $2m$. Topographies with even $\ell-m$ (panels b, c) are equatorially symmetric, the deformations add constructively, and $H_{geo}$ is perturbed at $O\lb \epsilon\rb$ with order $m$.} The derivation for this result is provided in \ref{a:oddveven}.}
	\label{f:t_geo_c}
\end{figure}

Using a cylindrical coordinate system ($s,\phi, z$), the geostrophic height function, $H_{geo}\lb s , \phi \rb$, measures the axial distance spanned by the fluid domain at $(s,\phi)$.  
Curves of constant $H_{geo}$ are known as the geostrophic contours since motions that are in a purely geostrophic balance follow these contours \citep{hG68} (see \ref{a:oddveven} and Figure \ref{f:sketch} for a more complete discussion). 
As an example, $H_{geo}$ for a full sphere of radius $r_{o}$ is
\[H_{geo}^{s phere}\lb \mathbf{r}\rb =  H_{geo}^{s phere}\lb s\rb = 2\sqrt{r_{o}^{2} - s^{2}}.\]
In a spherical shell (inner radius $r_{i}$), $H_{geo}^{shell}$ has a discontinuity at the tangent cylinder $s = r_i$. 
Nevertheless, the geostrophic contours are circles centered about the rotation axis.
The isosurfaces of $H_{geo}^{shell}$ are cylinders about the origin, and the term ``geostrophic cylinder'' commonly refers to these surfaces.
When topography is added to the outer spherical boundary, the geostrophic contours are, in general, no longer circular, nor do they necessarily form closed curves. Sufficiently steep topography can introduce discontinuities in $H_{geo},$ which has significant consequences for the resulting flow. 
In this study we will refer to the isosurfaces of $H_{geo}$ as geostrophic cylinders as long as the associated contours form closed loops.

Figure \ref{f:t_geo_c} displays geostrophic contours for the topographic shapes investigated in the present work when $\epsilon  = 0.1$. 
Figures \ref{f:t_geo_c}(b), corresponding to $\hlm{32}{16}$, and \ref{f:t_geo_c}(c), $\hlm{32}{17}$, illustrate an important distinction between topographies where $\ell - m$ is even and odd.
In our notation $\hlm{\ell}{m} (\theta,\phi) = \lb -1\rb ^{\ell - m} \hlm{\ell}{m}\lb \pi-\theta,\phi\rb $, which indicates that when $\ell - m$ is even the topography at antipodal points are identical. When $\ell - m$ is odd, the topography at antipodal points differs in sign.
An important consequence is that the perturbations to $H_{geo}$ are larger amplitude and have longer wavelength when $\ell-m $ is even. In particular, outside the tangent cylinder
\begin{equation}
	H_{geo}^{shell}- H_{\ell,m}^{\epsilon}\sim
	\begin{cases}
		 \epsilon  \cos m\phi\qquad &\ell - m \text{ even}\\
		 \epsilon^{2}  \cos 2m\phi\qquad &\ell - m \text{ odd,}
	\end{cases}
	\label{e:h_geo}
\end{equation}
where $H_{\ell,m}^{\epsilon}$ is $H_{geo}$ for the topography characterized by shape $\hLM$ and amplitude $\epsilon$.
A derivation of this result is included in \ref{a:oddveven}.

\subsection{Output/definitions}

To quantify flow speeds, we will report the Reynolds number $Re,$ which is calculated from the root mean square velocities according to
\be
Re = \sqrt{\left<\ub\cdot\ub\right>},
	\label{e:re}
	\ee
where the $\left< \cdot\right>$ indicates a full volume average. Unless otherwise noted, all reported values of $Re$ will also be time averaged. 
We also report the Reynolds number for particular components of the velocity field. 
We denote the included components in the subscript so that (for example) $Re_{u,v}= \sqrt{\left<u^{2} + v^{2}\right>}$.
We discuss the geostrophic and ageostrophic Reynolds numbers ($Re_{\gchar}$ and $Re_{\achar}$ respectively) determined from the components of the velocity field tangent and perpendicular to the geostrophic contours. For these two quantities the spatial average is performed over only the equatorial plane. All of the turbulent simulations are performed at the same value of $Ek$ and $\Rat;$ however, the values of $Re$ vary. In the case with no topography (hereafter referred to as the control) $Re \approx 1280.$ Our most turbulent simulation ($\epsilon=  0.2, \;\hlm{8}{4}$) has a Reynolds number of $Re \approx 2480$.

The Nusselt number ($Nu$) compares the ratio of convective to conductive heat transport over a closed surface enclosing the inner core. 
The conductive heat profile differs between topographies, but the effect is negligible for small amplitudes. We therefore approximate the conductive heat profile using $\vartheta_c\lb r\rb$ the temperature field that solves Laplace's equation in a sphere (see equation \eqref{e:tec}).

 $Nu $ can be calculated over any surface, but for a statistically steady flow, the time-averaged value is the same for any surface. 
We therefore present a Nusselt number calculated over the inner core boundary (ICB) surface according to
\[Nu = r_{i}\lb 1-r_{i}/r_{o}\rb\;\oint_{ICB}\frac{1}{r} \frac{\partial }{\partial r} \lb r\vartheta\rb da .\]

One-dimensional (azimuthal) kinetic energy spectra are computed for various components of the flow in the equatorial plane. 
In general, we define the radially averaged one dimensional energy spectra over fields $v_{i}$ as
\begin{equation}
	\mathcal{E}_{m}\lb v_{0},v_{1}...,v_{N}\rb = \lb r_c -r_{i}\rb ^{-1} 
	\int_{r_i}^{r_{c}} \sum_{i}^{N}\hat{v}_{i}^{m}\lb r\rb \hat{v}_{i}^{m^*}\lb r\rb  d r ,
\label{e:spectra_def}
\end{equation}
where $\hat{\lb\cdot\rb}^{m} $ is the $m$ Fourier component of the field $\lb \cdot\rb $, and $r_c$ is the largest radial value within the equatorial plane for all $\phi$.    
The Fourier transform is performed over the azimuthal coordinate $\phi$. 
In particular, the kinetic energy spectra in a cylindrical coordinate system is $\mathcal{E}_{m} \lb u_{s},u_{\phi},w\rb.$  
We use the shorthand $\mathcal{E}_{m}=\mathcal{E}_{m}\lb  u_{s},u_{\phi},w\rb $ for notational simplicity. 

To quantify the flow morphology we report a weighted wavenumber $m_{eq}$, based on the kinetic energy spectra in the equatorial plane. It is defined as 
\begin{equation}
	m_{eq} = \frac{\sum_{m=1}m \mathcal{E}_{m}}{\sum_{m=1}\mathcal{E}_{m}}.
	\label{e:meq}
\end{equation}
The weighted wavenumber is similar to the mean spherical harmonic degree used in \cite{uC06} (also \cite{jN24}); however, it is defined for the order rather than the degree.
We found $m_{eq}$ was a useful quantity because the topography significantly alters the azimuthal structure of the flow.


The topographic torque is defined as
\be
\mathbf{\Gamma} = \oint_{CMB} \mathbf{r}\times p \; d\mathbf{a},
\label{e:ttorque}
\ee
where $d\mathbf{a}$ is a surface area element on the CMB. We use the convention that the normal points away from the core. As in previous work \citep{tO25}, we will limit our analysis to the axial torque, which can generate $\Delta$LOD. For small topographic amplitude ($\ep \ll 1$) an approximate expression for the axial torque is 
\be
\Gamma_z \approx \oint \epsilon h\lb\theta,\phi\rb \partial_\phi p\; da.
\label{e:ax_torque}
\ee
All torques presented in this study are calculated with the exact expression (\ref{e:ttorque}); however, we will appeal to equation (\ref{e:ax_torque}) for the purposes of analysis. 
The approximation in equation \eqref{e:ax_torque} is accurate up to $O\lb \epsilon^{2}\rb$ \citep{pR12}.

Motivated by equation \eqref{e:ax_torque}, we find it useful to investigate the spherical harmonic decomposition of the pressure field to better understand which components of the pressure contribute to the torque. For the $\hlm{\ell}{m}$ topographies, only a single harmonic contributes to the torque, whereas for the tomographic model the entire spectrum of $p$ is convolved with $h_{t}.$ 
We report fluctuations in $\Gamma_{z}$ so rather than present the spectra of $p,$ we present the rms spectra of $p^{\prime}=p-\overline{p}$ on the outer surface. The overline $\overline{\cdot}$ denotes a time average.
We define the quantity
\be \phatLM = \sqrt{\overline {c_{\ell,m}^{2} + s_{\ell,m}^{2} } },
\label{e:phat}
\ee
where $c_{\ell,m}$ and $s_{\ell,m}$ satisfy
\be 
p(r_{cmb})-\overline{p}\lb r_{cmb}\rb = \sum_{\ell=0}\sum_{m = 0}^{\ell}P_{\ell}^{m}\lb \cos\theta\rb\; (c_{\ell,m}\cos(m\phi) + s_{\ell,m}\sin(m\phi)).
\label{e:pfs}
\ee
We refer $\phatLM$ as the fluctuating pressure spectrum.
For the $\hlm{\ell}{m}$ topographies, which are proportional to $\cos \lb m\phi\rb$, only the power in the corresponding $s_{\ell,m}$ mode contributes to the torque, so we find it useful to separate the sine and cosine modes. The data in Figure \ref{f:p_disp} reports 
$C_{\ell,m}=\sqrt{\overline{c_{\ell,m}^{2} } }$ and
$S_{\ell,m} =\sqrt{\overline{s_{\ell,m}^{2} } }$ separately.

We define the viscous torque as  
\[\mathbf{\Gamma}^{v} = \oint_{CMB} [(d\mathbf{a}\cdot\nabla) \mathbf{u} ]\times \mathbf{r},\]
although we only consider the axial component $\Gamma_{z}^{v}$. 
We report  the ratio of the standard deviations of the viscous and topographic axial torques
\[\gamma = \frac{\sqrt{\overline{(\Gamma_{z}^{v}-\overline{\Gamma_{z}^{v}})^{2}}}}{\sqrt{\overline{(\Gamma_{z}-\overline{\Gamma_{z}})^{2}}}}.\]
We found that $\gamma<1$ for all of the investigated parameters except the $\hlm{2}{1}$ topographies at $\epsilon \leq0.02.$ As in previous studies \citep[e.g.][]{bB07}, we conclude that viscous torques do not significantly contribute to $\Delta$LOD. Our results regarding $\gamma$ are provided in \ref{a:visc_t}.


\section{Results}
\label{s:results}

\subsection{Subcritical flows: comparison with Bell \& Soward theory}
\label{s:subc}
\subsubsection{The instability}

\begin{figure}
	\begin{center}
		\subfloat[][]{\includegraphics[width=0.45\textwidth]{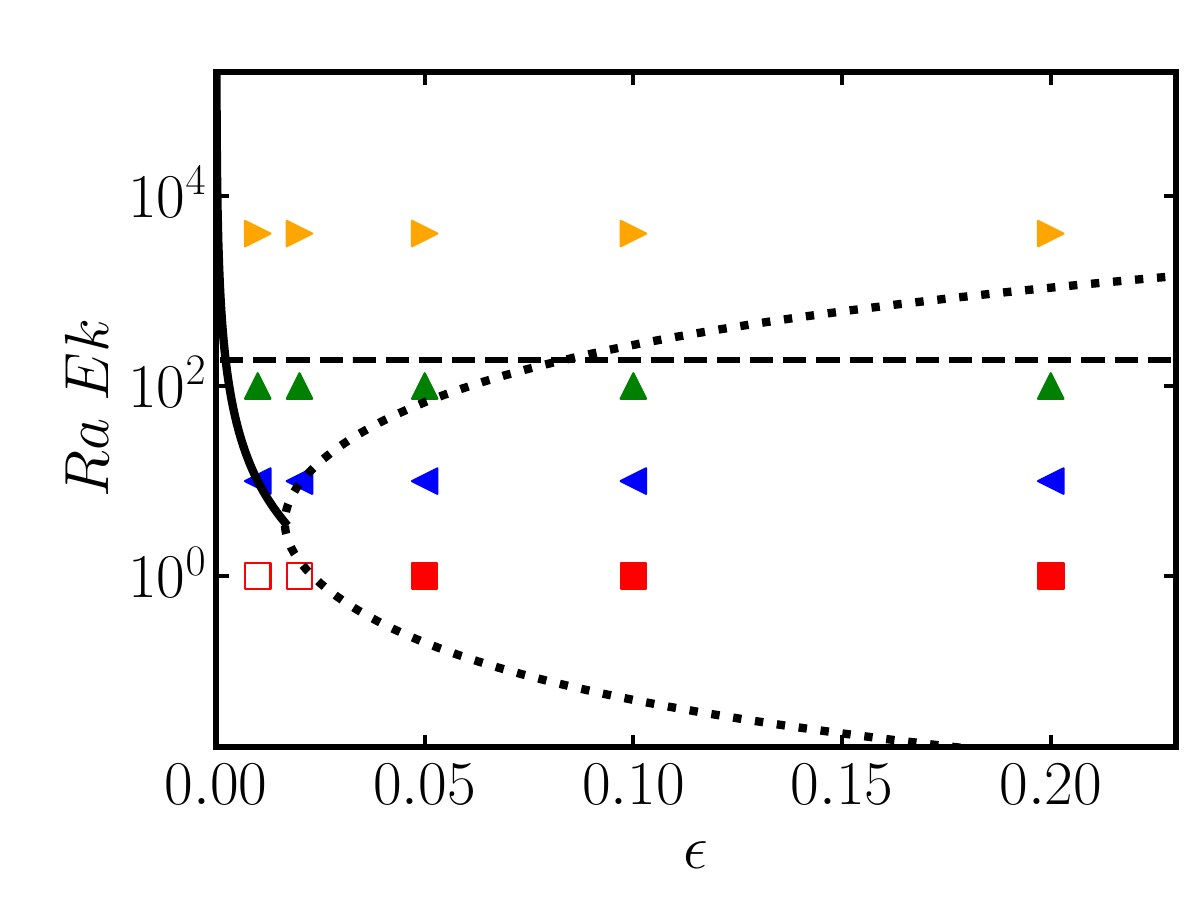}}
		\subfloat[][]{\includegraphics[width=0.45\textwidth]{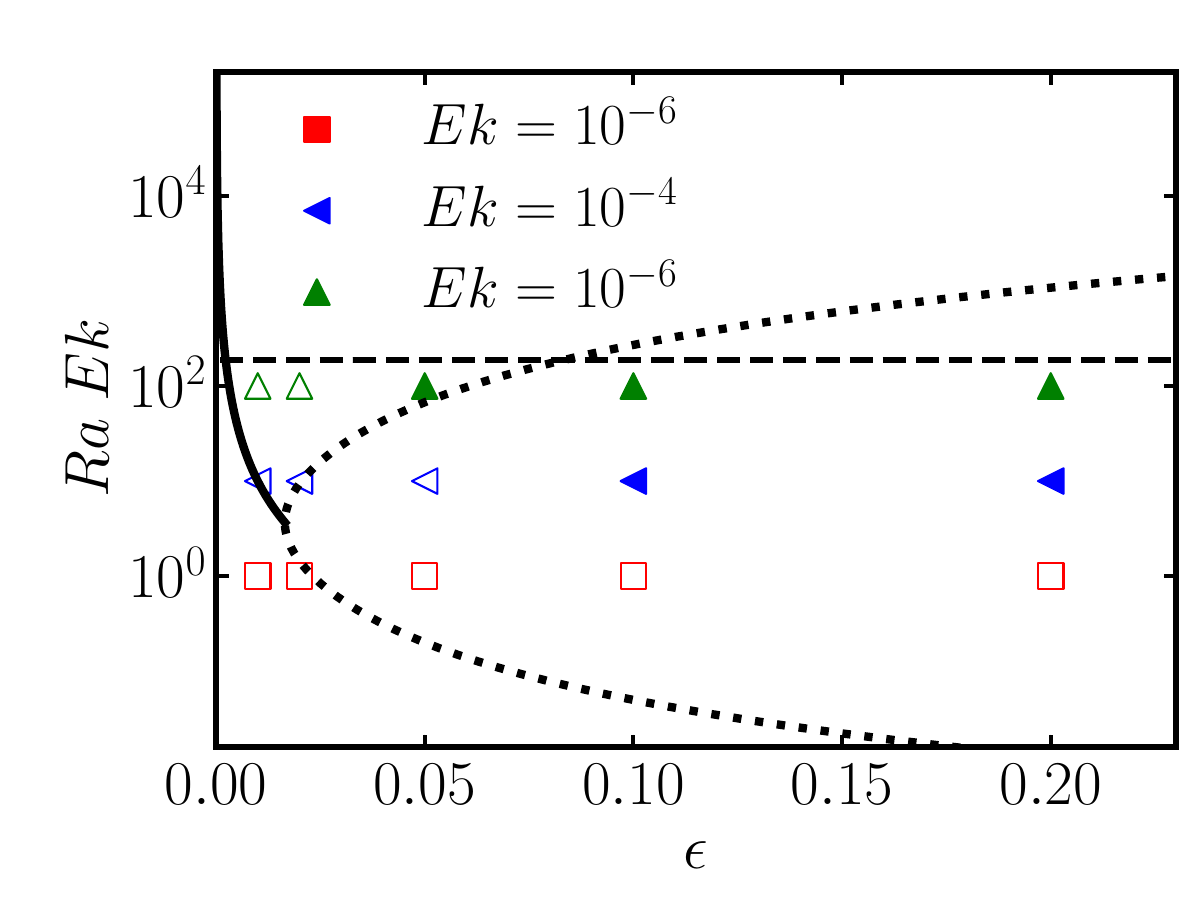}}
	\end{center}
	\caption{
		(Both panels) Linear stability diagram for $\hlm{8}{4}$ and simulated parameter values. The dashed line marks $Ra_{c}Ek,$ above which convection in a pure spherical shell is unstable. The solid and dotted lines mark the stability curve from BS96. BS96 predicts that at fixed $\epsilon$ the system is susceptible to an oscillatory (stationary) instability above the solid (dotted) line. 
		(a) The open (closed) symbols are stable (unstable) according to the BS96 theory. 
		(b) Stability results for subcritical cases with a boundary condition that prevents baroclinic flows $\vartheta\lb r_{cmb}\rb =\vartheta_{c}\lb r_{cmb}\rb$.
	$RaEk = 10$ data (blue) is performed at $Ek = 10^{-4}$ rather than $Ek = 10^{-6}$. Open (closed) symbols indicate stable (unstable) configurations based on the growth of kinetic energy. Compared to panel (a), the flows are more stable than BS96 would suggest; however, the flow is still unstable below $Ra_{c}$ for a spherical shell \citep[e.g.][]{aB22}.
	}		
	\label{f:bs_stab}
\end{figure}

As previously explained, the presence of topography allows the buoyancy force to enter the leading-order force balance, and
dissipation or other small perturbative forces are not necessary for instability. 
\cite{pB96} (hereafter referred to as BS96) developed theory that identified a distinct linear instability due to this topographic effect. 
The instability is characterized by a critical Rayleigh number that scales as $Ra\sim\Ek^{-1}$, which is less than $Ra_c$ by a factor of $Ek^{-1/3}$. 
 We performed a sweep of simulations with $Ra <Ra_{c}$ (referred to as ``subcritical'') for the $\hlm{8}{4}$ topography with varying topographic amplitude $\epsilon$. 
The $\hlm{8}{4}$ topography was chosen because it is both simple and characterized by a wavenumber that is significantly smaller than the critical wavenumber for convection, i.e.~$m = 4 < m_{c} = 32$. 
Figure \ref{f:bs_stab}(a) shows the simulated parameter regime and the linear stability curves (solid and dotted) from BS96.
The solid line is the boundary for an oscillatory instability and the dotted line is the boundary for a stationary instability.
 A detailed explanation of the generation of these curves is provided in \ref{a:bs_stab}. 
 The dashed line is placed at $Ra_{c}Ek,$ marking the boundary between stable and unstable flows in a spherical shell with the same aspect ratio $\chi$ and no topography \citep{aB22}.
 According to the BS96 theory, the closed (open) symbols denote unstable (stable) cases. The yellow, right-pointing triangles mark the parameter space for our turbulent simulations and are shown for reference only.

The instability can be identified by a positive-valued kinetic energy growth rate, but only if a static state solves equations (\ref{e:mom})-(\ref{e:incom}).
 The homogeneous temperature boundary condition on $r_{cmb}$ generates non-zero baroclinic flows which make the identification of the instability difficult. Therefore, we use the boundary condition $\vartheta\lb r_{cmb}\rb = \vartheta_{c}\lb r_{cmb}\rb$ (hereafter referred to as the heterogeneous boundary condition), to test the instability. The heterogeneous boundary condition permits a static, conductive solution to equations (\ref{e:mom})-(\ref{e:incom}).  The heterogeneous boundary condition is consistent with the BS96 theory because in the topographic annulus model used in that study, the fixed temperature boundaries are separate from the topographic boundaries such that the system supports a static, conductive solution.
 Figure \ref{f:bs_stab}(b) displays the same parameter space as Figure \ref{f:bs_stab}(a) and each symbol identifies the stability based on the growth rate of the kinetic energy. 
 The closed (open) symbols in Figure \ref{f:bs_stab}(b) denote unstable (stable) cases. We note that the $RaEk = 10$ simulations were performed at $Ek = 10^{-4}.$ 
 We find that the BS96 stability curves only provide qualitative agreement with the data. 
 In general, the flows with topography are \textit{more} stable in the spherical geometry than in the annulus. 
 Nevertheless, the flows are unstable below $Ra_{c}$ which means that topography reduces the critical Rayleigh number.

\subsubsection{Saturated flow speed predictions}
 The BS96 analysis predicts that flow speeds should saturate when buoyant forces balance the ageostrophic component of the Coriolis force. 
 The analysis requires that the velocity field be decomposed into components parallel and perpendicular to the geostrophic contours.
 We adopt the notation
\begin{equation}
	\mathbf{u} = u_{\gchar}\gvec + u_{\achar}\avec + w \hat{\mathbf{z}},
	\label{e:geo_dec}
\end{equation}
where $\gvec$ and $\avec$ are unit vectors tangential and perpendicular to the geostrophic contours. Our chosen handedness is such that in a sphere $\gvec$ and $\avec $ coincide with $\hat{\boldsymbol{\phi}}$ and $\hat{\mathbf{s}}$ respectively. 
The cross-product in the Coriolis term is expressed as
\[\hat{\mathbf{z}}\times\mathbf{u} = -u_{\gchar}\avec + u_{\achar}\gvec,\]
and the steady, low-amplitude momentum equation is
\begin{equation}
	\frac{2}{Ek}\lb -u_{\gchar}\avec + u_{\achar}\gvec\rb   = - \nabla p + \frac{Ra}{Pr} \frac{\mathbf{r}}{r_{o}} \vartheta + \nabla^{2}\mathbf{u}. 
	\label{e:m_s}
\end{equation}
Integrating equation (\ref{e:m_s}) along a closed geostrophic contour and multiplying by $Ek$ yields
\begin{equation}
2\oint u_{\achar}\;d\gchar =\frac{Ra Ek}{Pr}\frac{1}{r_{o}}\oint \vartheta \mathbf{r} \cdot d\mathbf{\gchar} 
+ Ek\oint \nabla^{2}u_{\gchar}\;d\gchar.
\label{e:m_avg}
\end{equation}

The incompressibility condition requires that $\partial w/\partial z$ is proportional to the left-hand side of equation (\ref{e:m_avg}).
When $\gvec$ is everywhere perpendicular to buoyancy (as is the case in the sphere), the right-hand side of equation (\ref{e:m_avg}) indicates that internal viscosity is necessary for a non-zero axial derivative of $w$. 
Provided that the topography is sufficiently large,
\begin{equation}
	\frac{\partial w}{\partial z} \sim\frac{Ra Ek }{Pr}\frac{1}{r_{o}}\oint \vartheta \mathbf{r}\cdot d\gvec.
	\label{e:m_avg2}
\end{equation}
The flows are quasigeostrophic so  axial derivatives are $O\lb 1\rb.$ To determine a scaling for $w$, we only need to determine the magnitude of the integral in equation \eqref{e:m_avg2}. In the absence of topography, the dot product in equation \eqref{e:m_avg2} is zero, but topography deforms the contours and $\mathbf{r}\cdot d\gvec\sim\epsilon .$ 
The azimuthal dependence of the dot product has the same order and opposite phase compared to the geostrophic contours.
For equation \eqref{e:m_avg2} to have a non-zero right-hand side, temperature fluctuations in phase with the dot product are necessary. BS96 determines that these fluctuations are $O\lb \epsilon\rb $.
The axial velocity therefore scales as
\begin{equation}
w \sim Ra Ek\; \epsilon^{2}/ Pr. 
\label{e:u_bs}
\end{equation}We note the similarity between equation (\ref{e:u_bs}) and the Coriolis-inertial-Archimedean (CIA) scaling $Re\sim RaEk/Pr$ observed in turbulent rotating flows \citep[e.g.][]{cG19}. 
Both results arise from a balance between Coriolis and buoyancy terms, although we stress that equation (\ref{e:u_bs}) is for laminar flows and non-linear effects are so far ignored.
The Ekman pumping condition \citep[e.g.][]{hG68} yields a scaling for the geostrophic velocity based on $w.$
\be
u_{\gchar} \sim Ra Ek^{1/2}\epsilon^{2}/Pr.
\label{e:g_bs}
\ee

With the heterogeneous temperature boundary condition (data in Figure \ref{f:bs_stab}(b)), we were unable to adequately test the scalings predicted by equations \eqref{e:u_bs} and \eqref{e:g_bs}. The saturation time is on the order of a viscous diffusion time, which was computationally prohibitive for $Ek=10^{-6}.$  
The simulations at $RaEk=10$ were performed at the more moderate value of $Ek=10^{-4}$ so that we could saturate the flow speeds, although only two data points were unstable. We will discuss these two points in the following section which otherwise focuses on simulations with the homogeneous temperature boundary condition on $r_{cmb}$. 

\subsubsection{Saturated flow speeds with $\vartheta\lb r_{cmb}\rb = 0 $ }
We found that it was much easier to saturate the flow speeds when $\vartheta\lb r_{cmb} \rb =0$.
We attribute this effect to the baroclinic flows at the outer boundary. 
\TOr{In light of equations \eqref{e:u_bs} and \eqref{e:g_bs}, we define rescaled axial and geostrophic Reynolds numbers as
\begin{equation}
	Re^{*}_{w} = \frac{Pr}{Ra Ek \;\epsilon^{2}}Re_{w}\qquad Re^{*}_{\gchar} = \frac{Pr}{Ra Ek^{1/2} \;\epsilon^{2}}Re_{\gchar}.
	\label{e:rew_r}
\end{equation}
For simplicity, we calculate the geostrophic Reynolds number $Re_{\gchar}$ in the equatorial plane only (see equation \eqref{e:re} and ensuing discussion for a definition).
}

\begin{figure}
	\begin{center}
		\subfloat[][]{\includegraphics[width=0.30\textwidth]{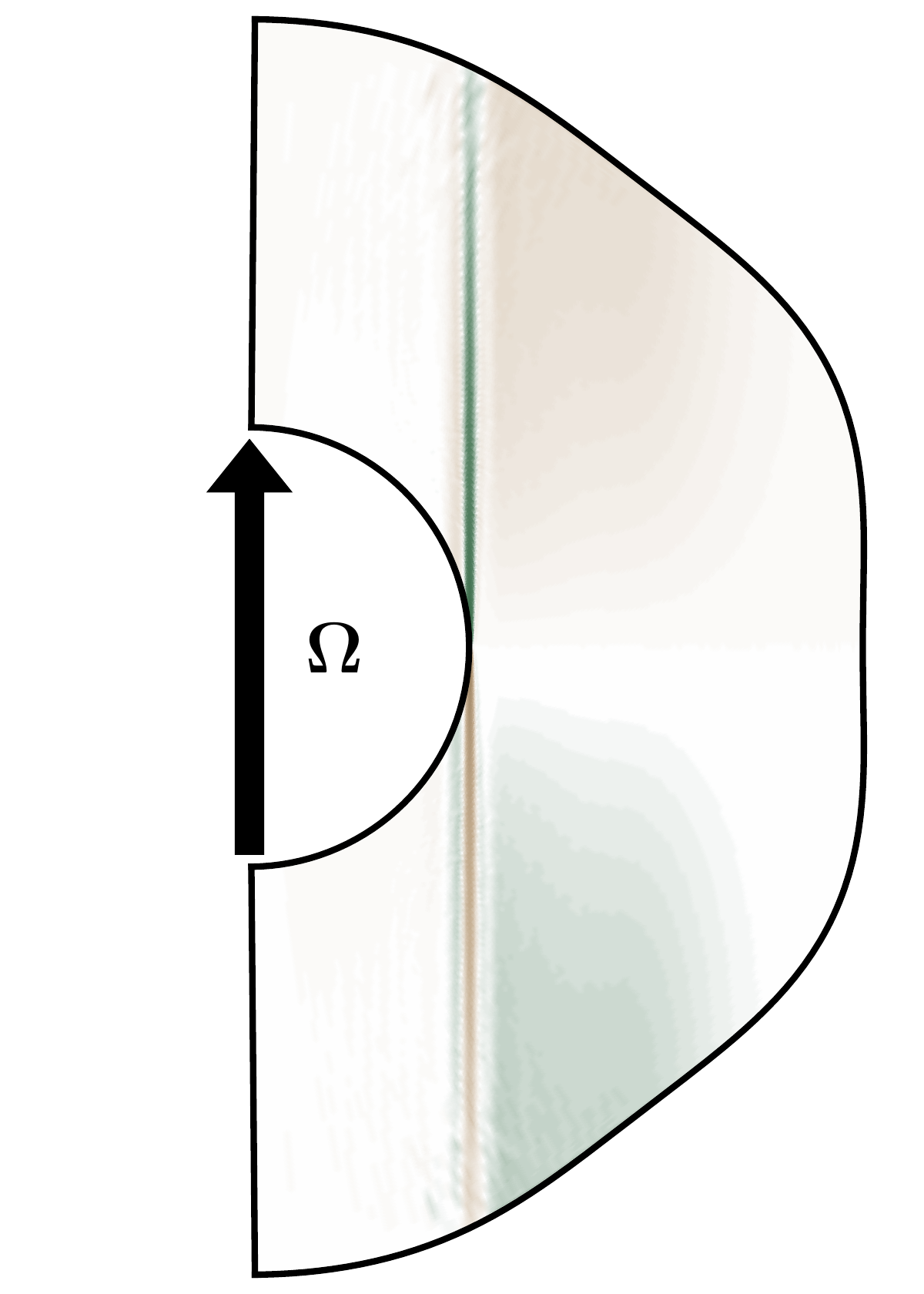}} \hspace{1.5cm}
		\subfloat[][]{\includegraphics[width=0.41\textwidth]{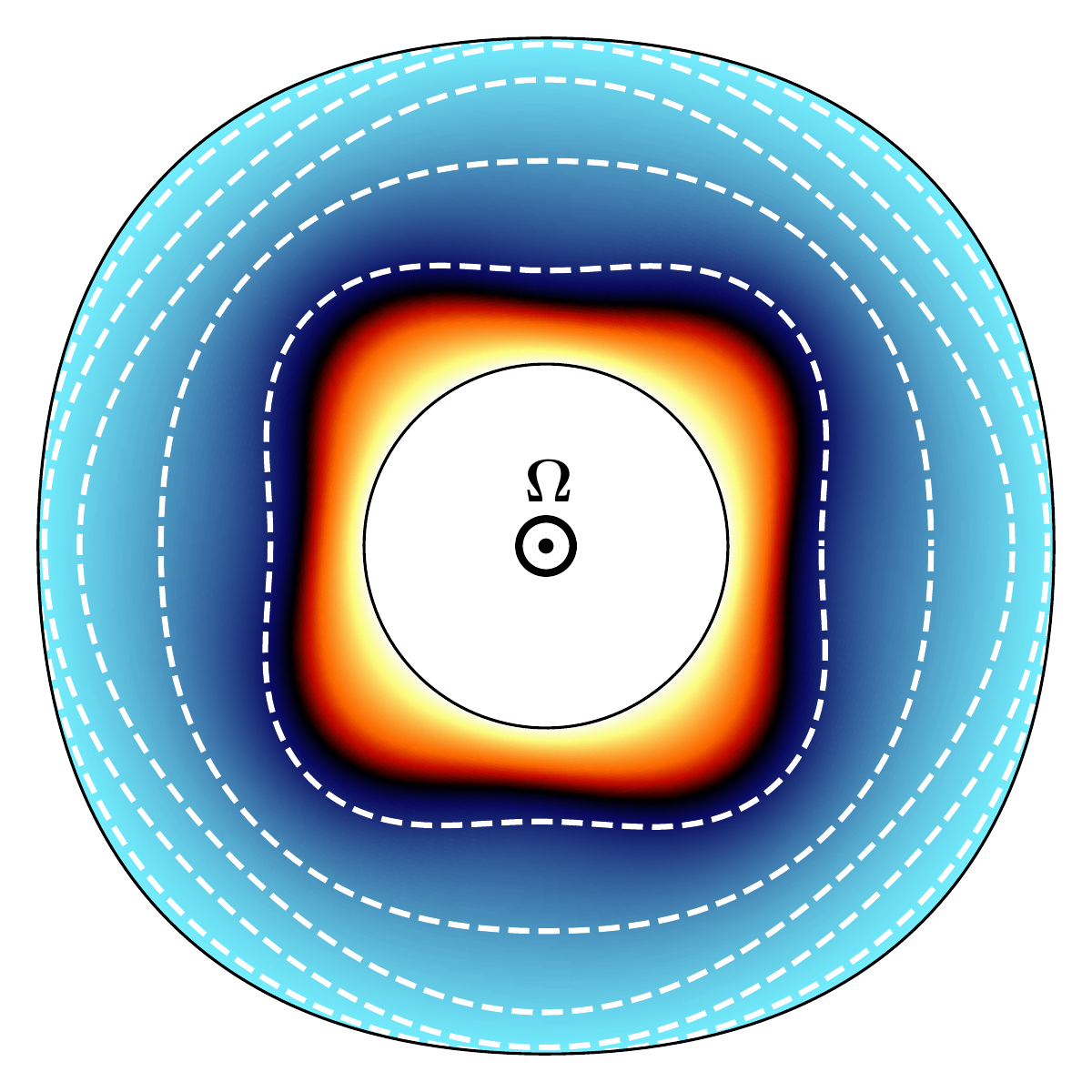}} 
	\end{center}
	\caption{\textcolor{black}{Visualizations of the subcritical flow for $\hlm{8}{4}$ at $RaEk = 10,$ $Ek = 10^{-6},$ and $\epsilon  =0.05.$
		(a) Axial velocity $w$ in the meridional plane. 
	 	(b) Temperature $\vartheta$ in the equatorial plane. Although difficult to see, the isotherms are slightly misaligned with the geostrophic contours. The misalignment facilitates buoyancy work on the flow.}
	 The homogeneous boundary condition $\vartheta\lb r_{cmb}\rb =0 $ is used.}
	\label{f:lin_vis}
\end{figure}
Figure \ref{f:lin_vis} displays a snapshot of the flow for $RaEk =10,$ $Ek = 10^{-6},$ and $\epsilon  = 0.05.$ 
The axial velocity $w$ is shown in the meridional plane in Figure \ref{f:lin_vis}(a). It is characterized by a columnar structure that varies slowly across the entire axial extent of the domain. $w$ crosses zero in the equatorial plane. 
The temperature in the equatorial plane is shown in Figure \ref{f:lin_vis}(b), and we see that it is significantly warped by the topography. In the absence of topography, the temperature would follow a conductive, axisymmetric profile.
Instead, the temperature takes on a structure similar to the geostrophic contours (dashed lines) although we note that it is slightly misaligned. As previously shown, a component of the temperature out of phase with the contours is necessary for buoyancy to do work on the geostrophic flow.

\begin{figure}
	\begin{center}
		\subfloat[][]{\includegraphics[width=0.45\textwidth]{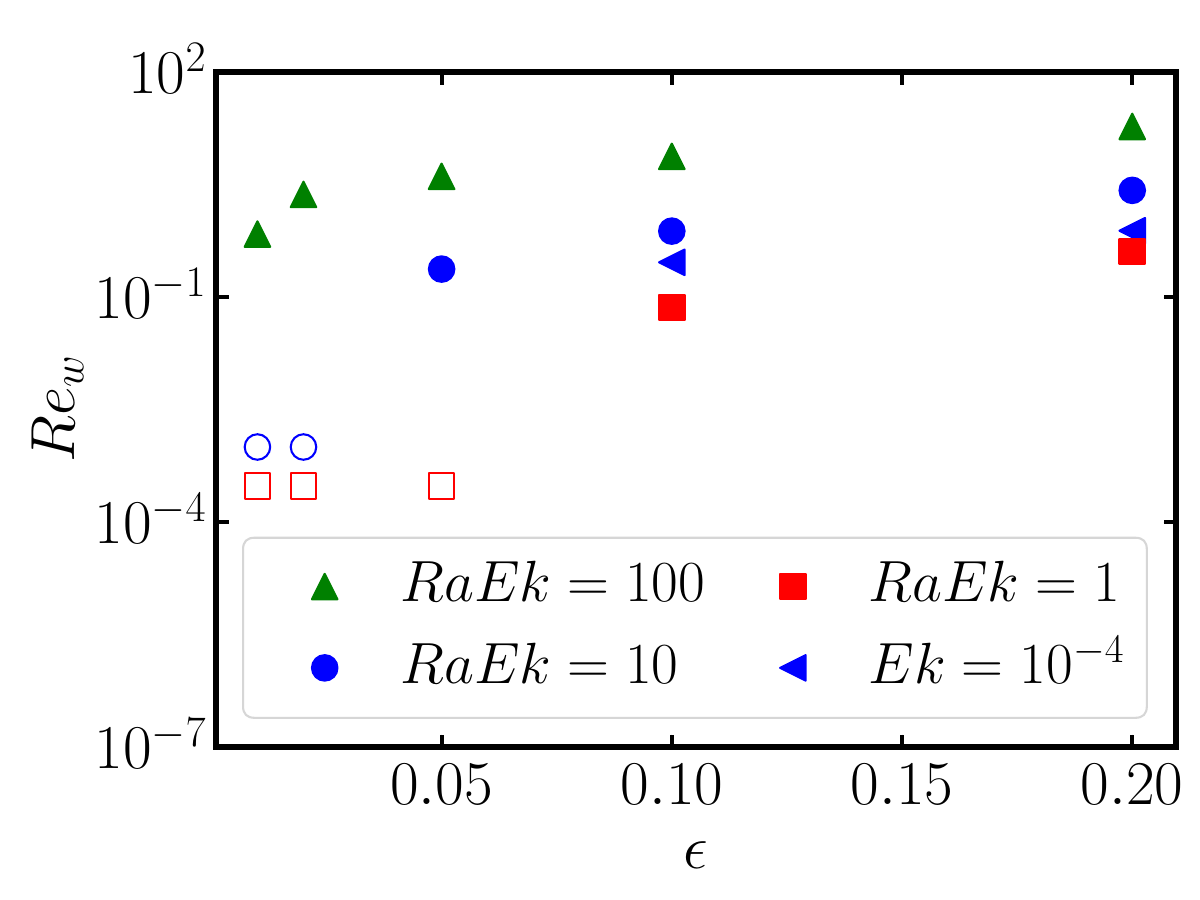}}
		\hspace{1cm}
		\subfloat[][]{\includegraphics[width=0.45\textwidth]{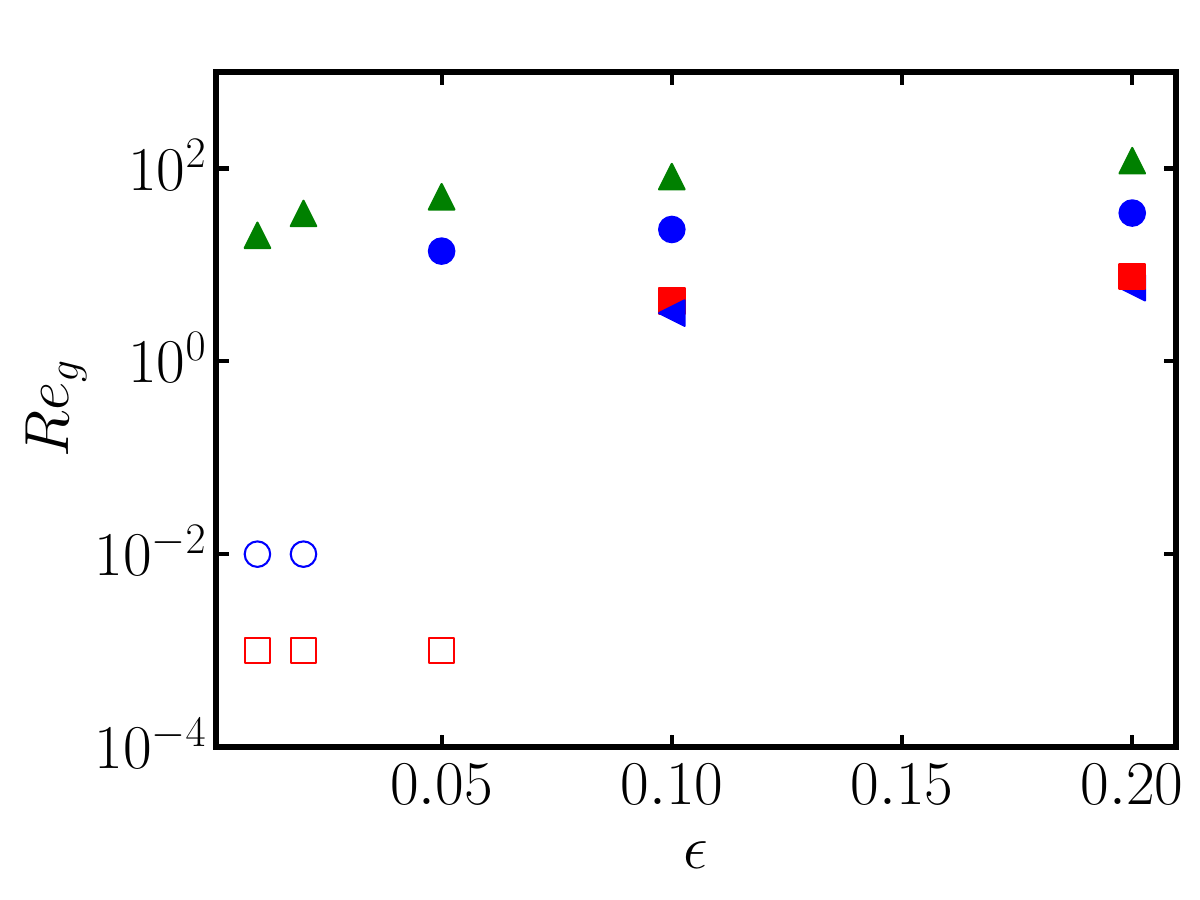}}\\
		\subfloat[][]{\includegraphics[width=0.45\textwidth]{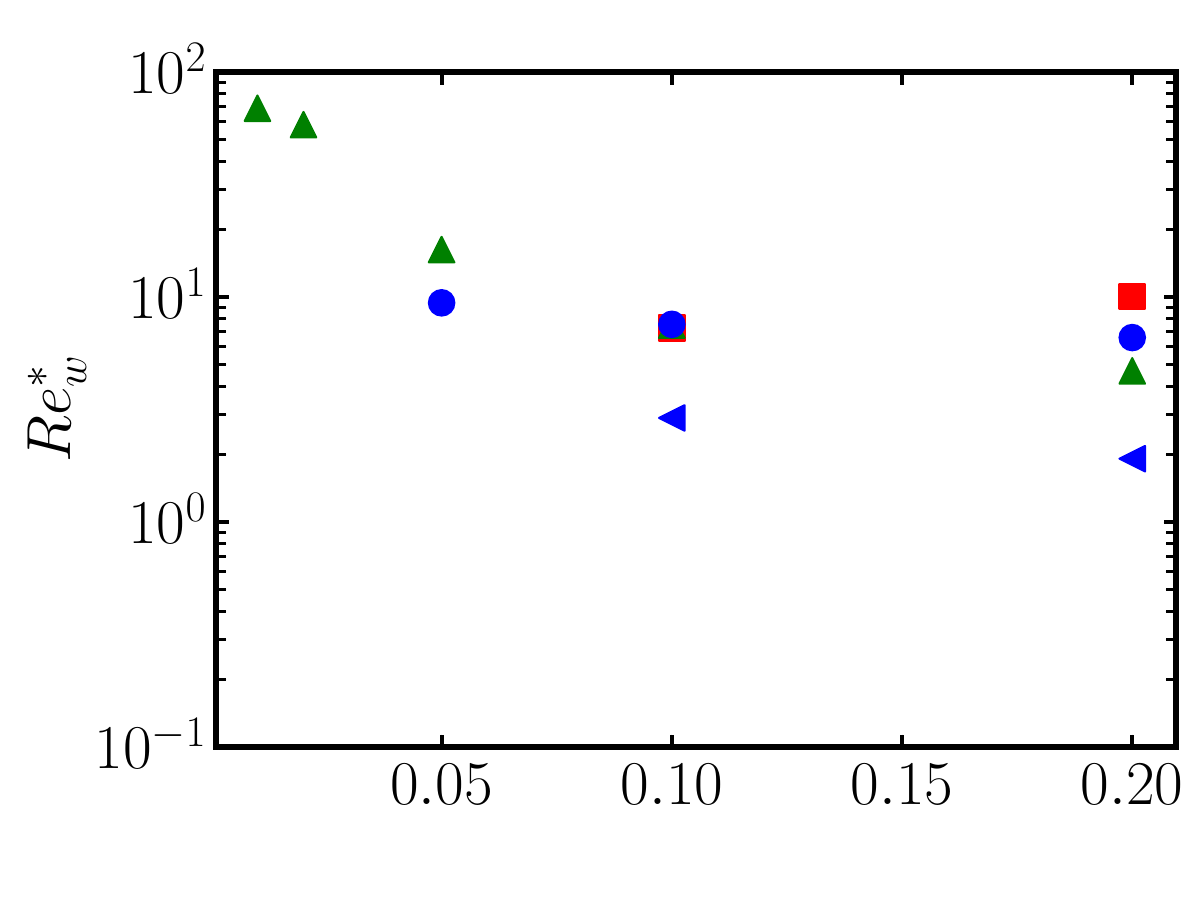}}
		\hspace{1cm}
		\subfloat[][]{\includegraphics[width=0.45\textwidth]{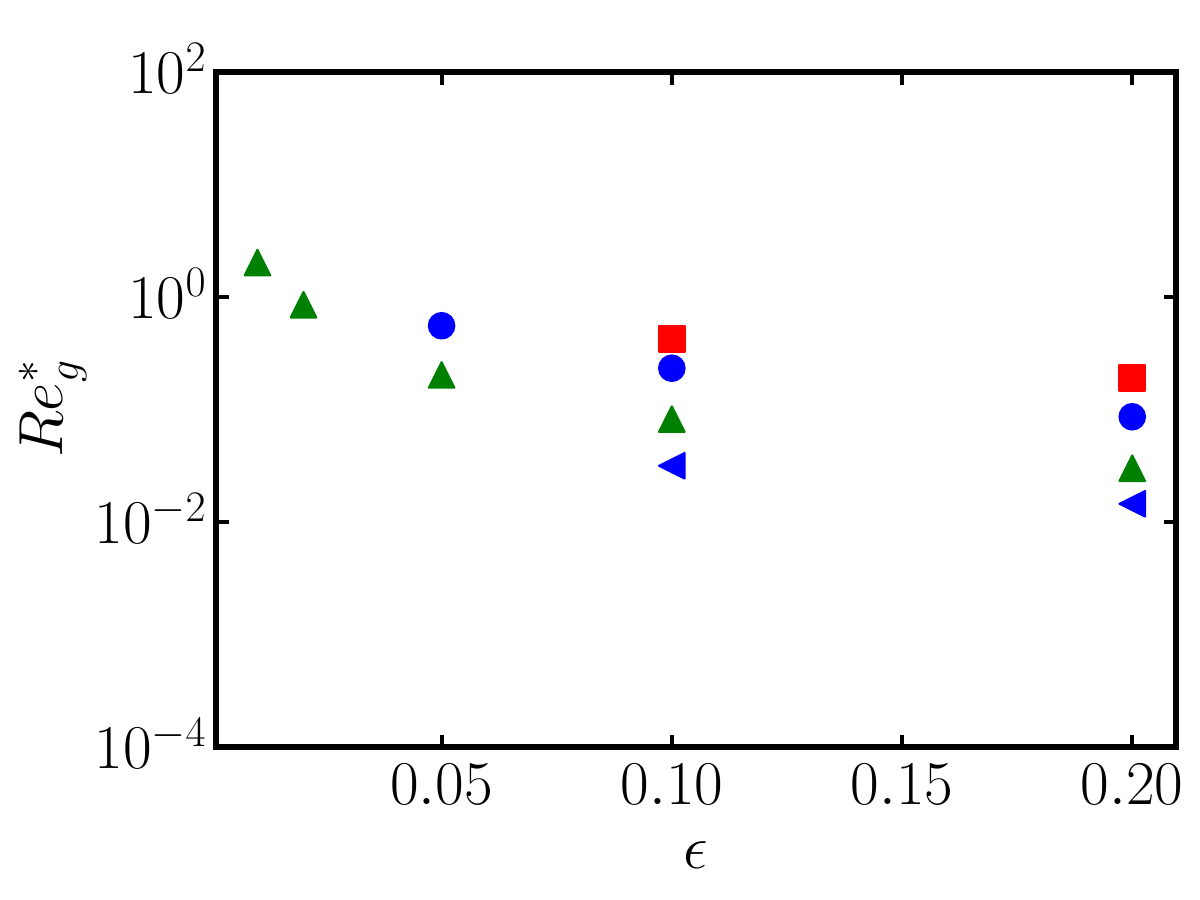}}
		
	\end{center}
	\caption{Saturated flow speeds for the subcritical parameters. The open markers denote simulations that had monotonically decreasing speeds with time and could not be saturated; they are omitted from panels (c) and (d). All data are for the homogeneous temperature boundary condition except for the two points denoted by blue, left-pointing triangles, which correspond to the heterogeneous boundary condition simulations shown in Figure \ref{f:bs_stab}(b) ($RaEk = 10$). We note that the empty symbols all correspond to parameter regions below or near the stability curves shown in Figure \ref{f:bs_stab}(a) suggesting qualitative agreement with BS96. The rescaled axial and geostrophic Reynolds numbers are shown in (c) and (d) respectively, and we find that the scaling proposed in BS96 is inadequate for the data.}
	\label{f:bs_stab_vel}
\end{figure}

Figure \ref{f:bs_stab_vel} displays the Reynolds numbers for subcritical parameter values with the homogeneous temperature boundary condition $\vartheta\lb r_{cmb}\rb =0$. 
The axial and geostrophic Reynolds numbers are shown in Figures \ref{f:bs_stab_vel}(a) and \ref{f:bs_stab_vel}(b) respectively. 
The closed symbols correspond to saturated flows. 
The open symbols distinguish cases where the Reynolds numbers monotonically decrease with time (after initial fluctuations at initialization). We could not simulate the flows marked with open symbols long enough to equilibrate the decaying speeds.
As such, the Reynolds numbers are not plotted, and instead we simply place the markers on the plot to indicate that the speeds monotonically decrease over our simulation window, which was around $0.1$ viscous diffusion times for each. 

The axial and geostrophic Reynolds numbers both increase with topographic amplitude and $Ra.$
To test the BS96 scaling, Figures \ref{f:bs_stab_vel}(c) and \ref{f:bs_stab_vel}(d) display $Re_{w}^{*}$ and $Re_{\gchar}^{*}$ respectively.
We observe that $Re_{w}^{*}$ is $O\lb 10\rb $ and decreases with $\epsilon$, which indicates that the $\epsilon^{2}$ scaling presented in BS96 is too strong to fit the data.
The rescaled geostrophic Reynolds number $Re_{\gchar}^{*}$ is $O\lb 1\rb $, but also decreases with $\epsilon $.
 The blue, left-pointing triangles indicate Reynolds numbers for the two unstable heterogeneous boundary condition cases at $RaEk=10$. Although the Reynolds numbers are less than for the homogeneous boundary condition counterparts shown by blue circles, the behavior is similar and the scalings in equation \eqref{e:u_bs} are too strong.

Overall, we find that the BS96 theory correctly identifies the instability mechanism though the details appear to be geometry-dependent.
Based on the flow speeds, we conclude that there is significantly more forcing for the configurations denoted by the closed symbols in Figure \ref{f:bs_stab_vel} than the open symbols. 
In the annulus, $O\lb \epsilon\rb $ buoyancy fluctuations generate $O\lb \epsilon^2\rb$ buoyant forcing in unstable flows (see equations \eqref{e:m_avg2}-\eqref{e:g_bs}). 
Although the data in Figure \ref{f:bs_stab}(b) indicates that the spherical shell is more stable than the annulus, the flow speeds associated with the unstable flows are nevertheless larger than $O\lb \epsilon^{2}\rb.$  
We expect that this is due to the differences in the background temperature profiles of the sphere and annulus. 
In the annulus, the isotherms of the conductive profile run parallel to the rotation axis. 
Axial velocity is perpendicular to the conductive temperature gradient. This is not the case in a spherical geometry, and axial velocity can transport cold fluid from the boundaries towards the centre of the shell. 
This effect yields larger temperature variations than in the annulus, and therefore larger velocities.

The homogeneous boundary condition creates an additional source of axial velocity via the Ekman pumping of the baroclinic flows. We propose that this generates an additional source of temperature variations, which promote the instability and explain why we observe a jump in the flow speeds at values of $RaEk$ and $\epsilon$ that were stable in the absence of the boundary baroclinic flows.
Finally, we point out that the temperature fluctuations are visible in the visualization shown in Figure \ref{f:lin_vis}(b).
A geostrophic flow coincides with the contours, and the visible misalignment between the isotherms and the contours indicates the presence of the large fluctuation which is able to do work on the geostrophic flow.
We stress the importance of buoyancy forcing the geostrophic flow. This source of mechanical work underpins the BS96 analysis, and in the next section we will show how in turbulent flows this mechanism can generate large variations in the momentum and heat transport.

A final comment on the linear theory is that BS96  assumes that topography  on the upper cap of the annulus is mirrored onto the lower cap. The spherical equivalent is topography composed entirely of spherical harmonics where $\ell - m$ is even (we refer to this as symmetric topography). 
Special care needs to be taken when considering topography generated exclusively by spherical harmonics corresponding to odd values of $\ell - m$. 
In the annulus, the effect of an asymmetric topography (for example an intrusion on the lower cap matched with an extrusion on the upper cap) would be to remove any deformations to the geostrophic contours. 
In the sphere, the geostrophic contours are still deformed; however, the size of the deformations is $O\lb \epsilon^{2}\rb $ rather than $O\lb \epsilon\rb $. 
Furthermore, the wavenumber of the geostrophic contours is doubled from $m$ to $2m.$ 
As such, the effects of topography on the flow morphology can be significantly reduced if the topography is composed primarily of harmonics with odd  $\ell - m$ values.
The difference between odd and even values of $\ell - m$ comes down to geometry, and we provide a more thorough explanation in \ref{a:oddveven}.


%
%
%
%
%
%

\subsection{Flow structure in the turbulent regime}

\begin{figure}
	\begin{center}
        \subfloat[][]{\includegraphics[width=0.29\textwidth]{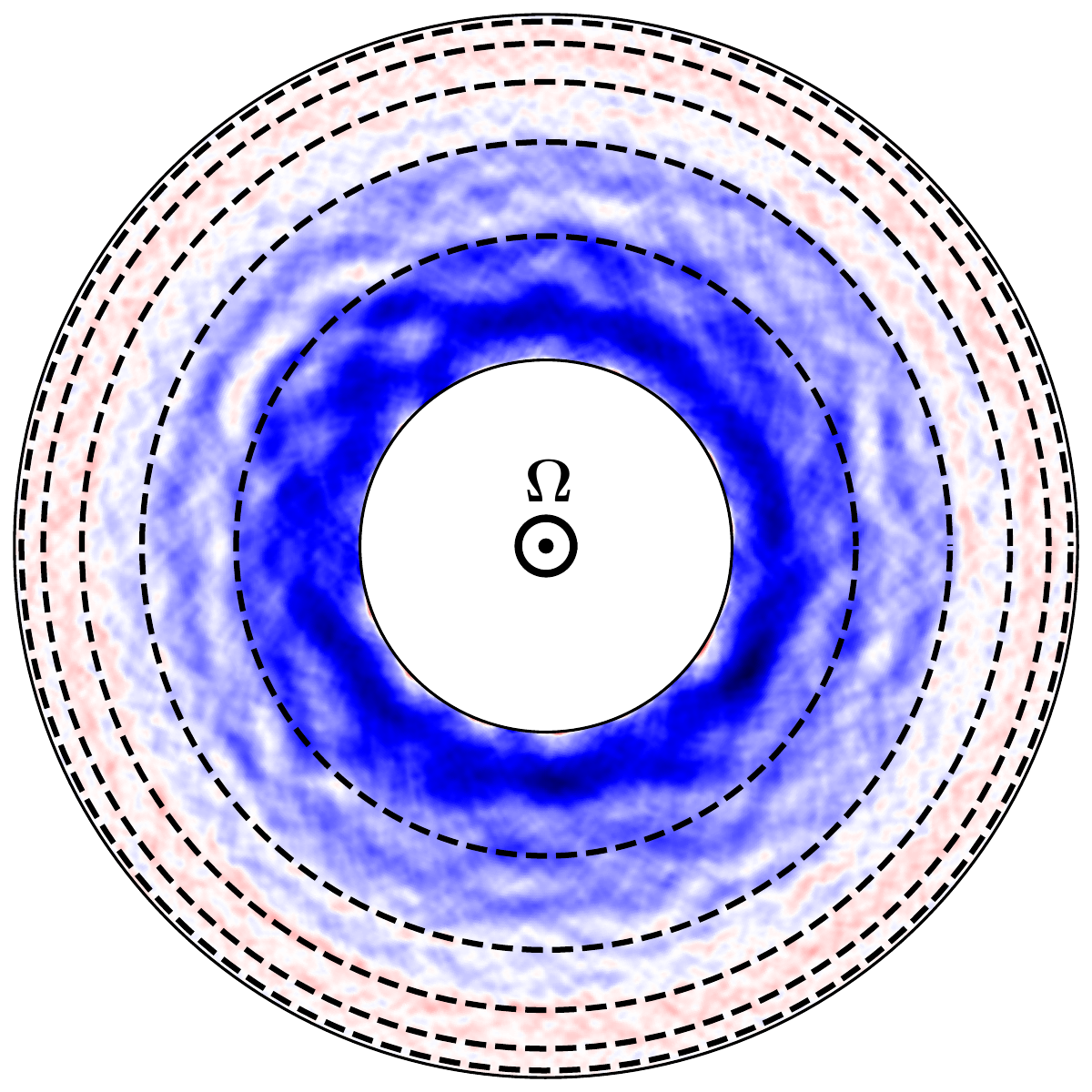}}
        \subfloat[][]{\includegraphics[width=0.29\textwidth]{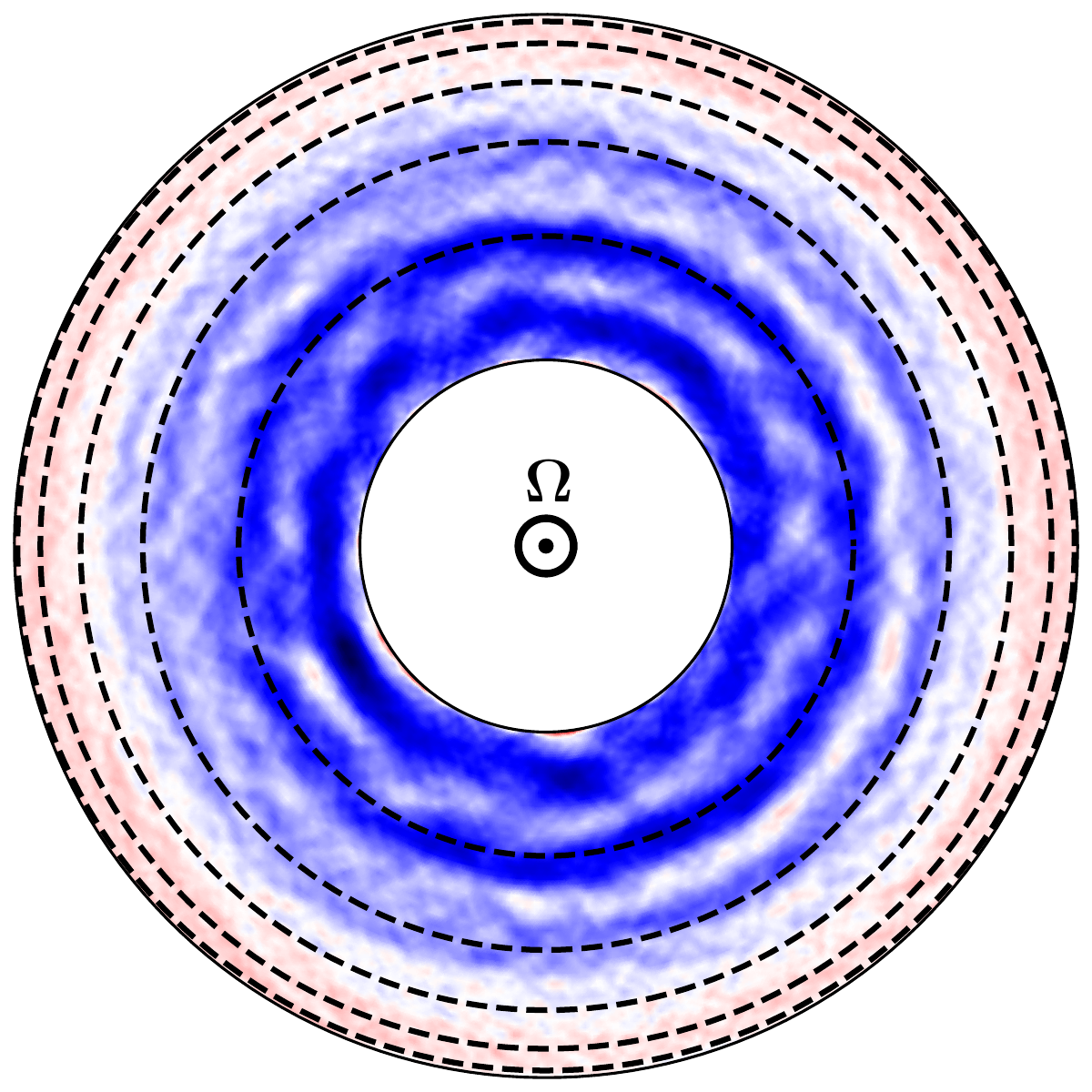}}
        \subfloat[][]{\includegraphics[width=0.29\textwidth]{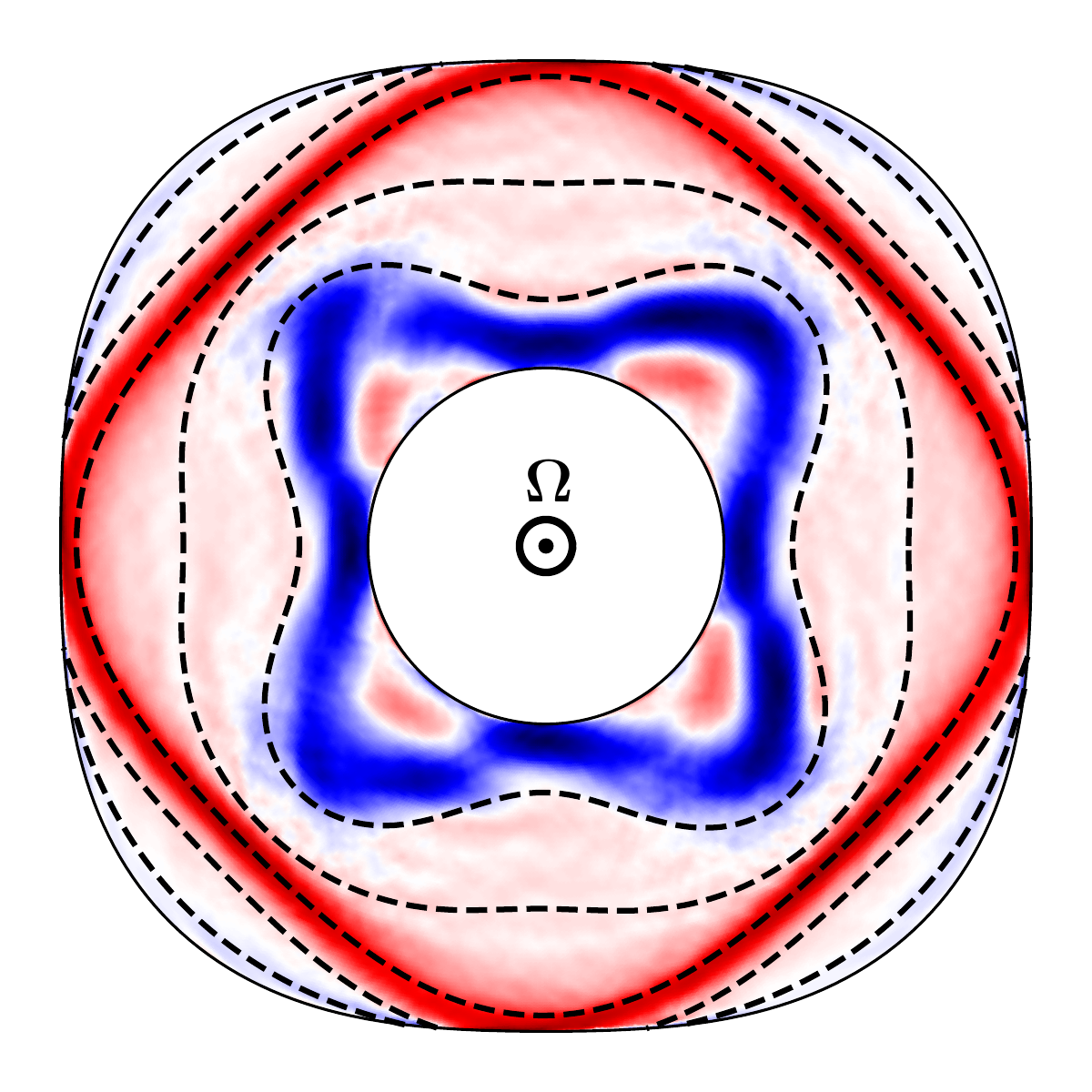}}
	\\
	\subfloat[][]{\includegraphics[width=0.29\textwidth]{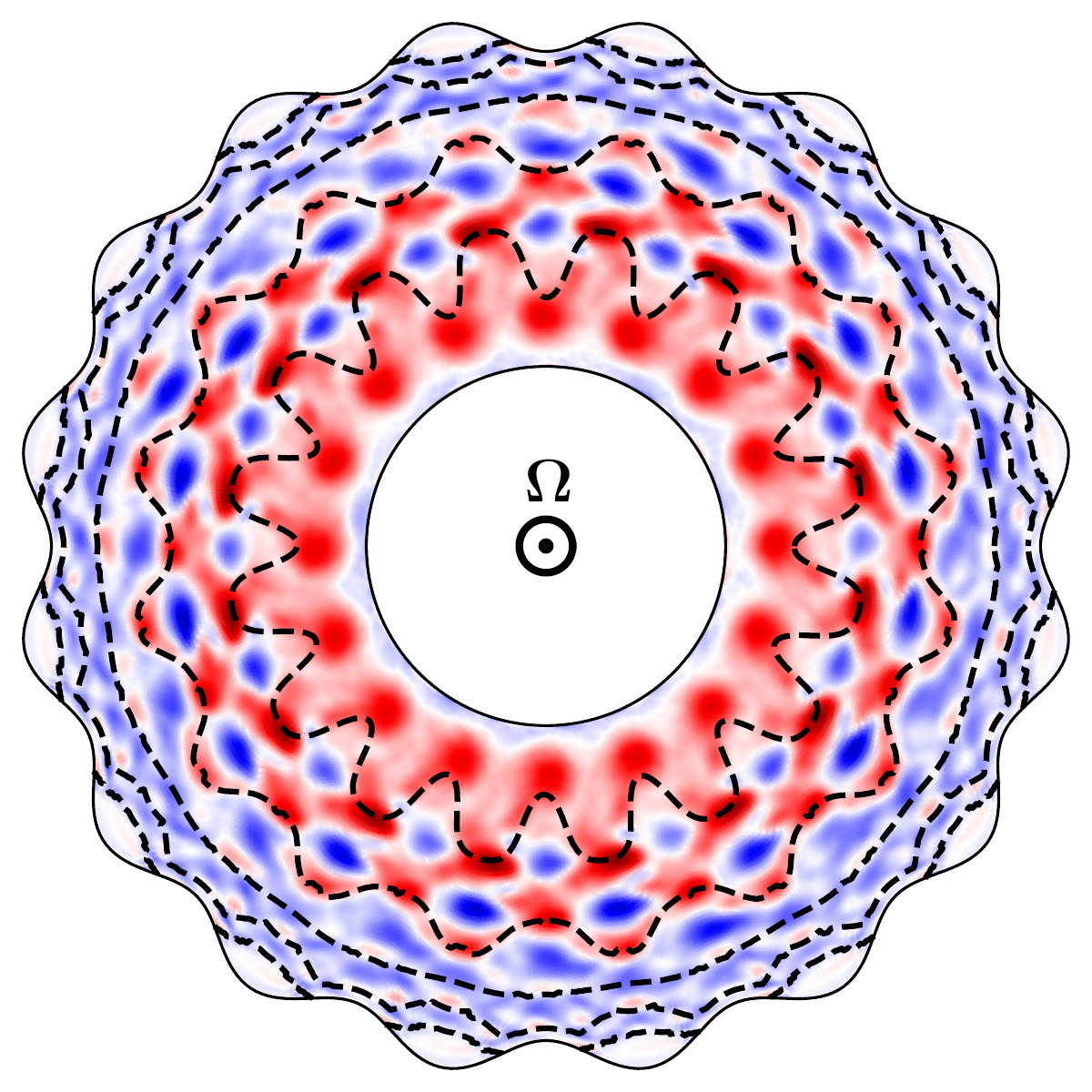}}
	\subfloat[][]{\includegraphics[width=0.29\textwidth]{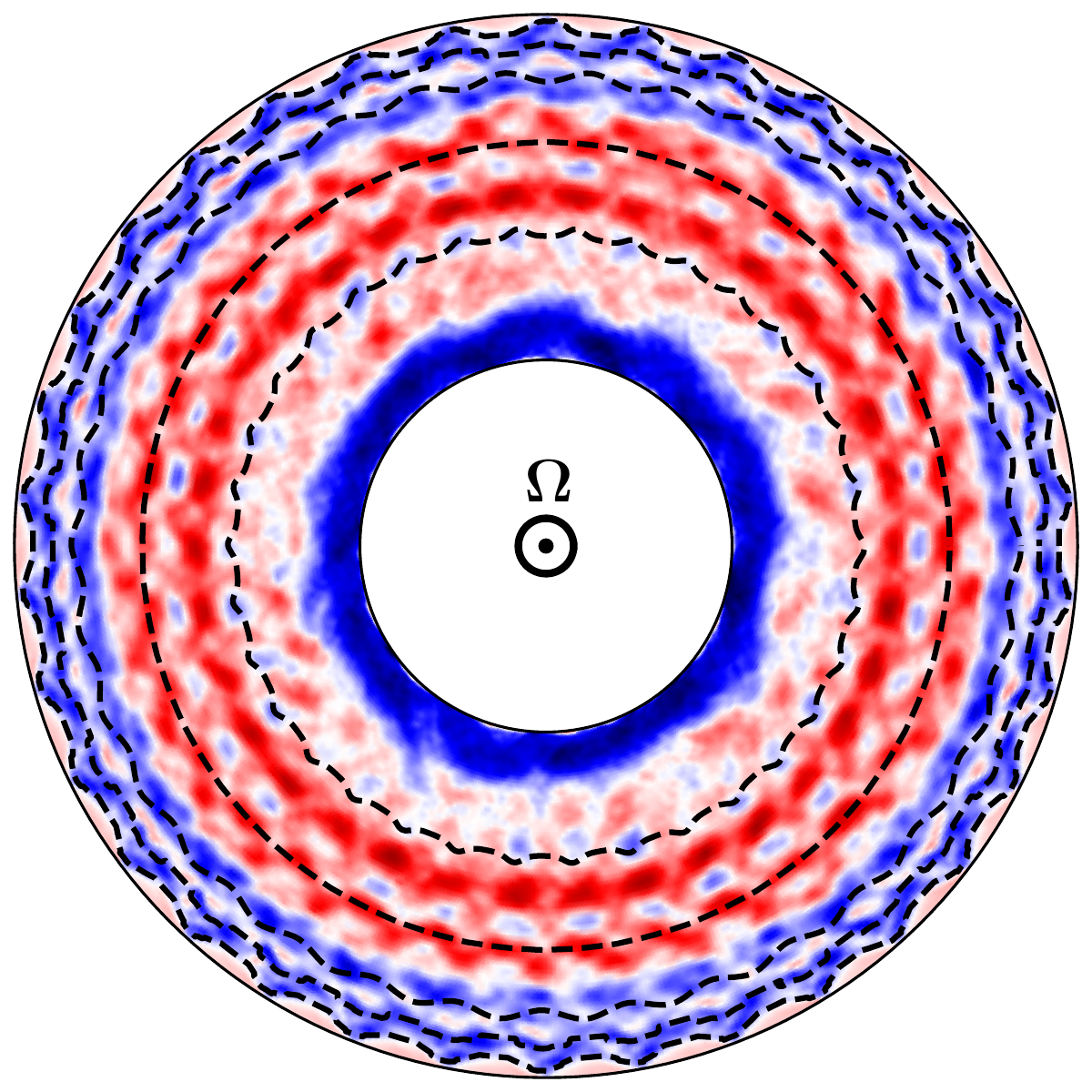}}
	\subfloat[][]{\includegraphics[width=0.29\textwidth]{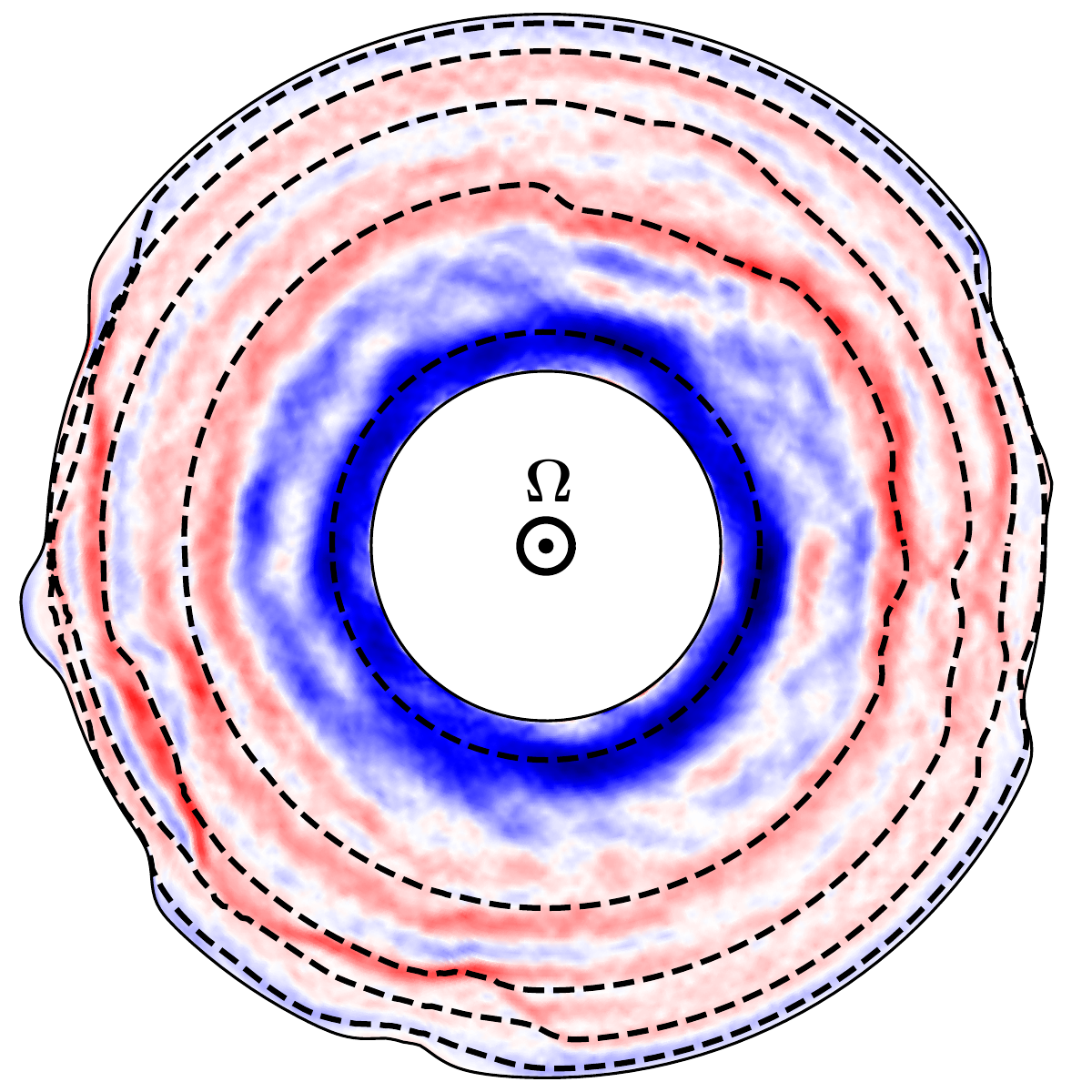}} 	
	\end{center}
	\caption{Time averaged azimuthal velocity ($u_{\phi}$) in the equatorial plane. All simulations were performed at $\Rat = 40, Ek = 10^{-6}.$ (a) The control case with no topography ($\epsilon =0$). The topographic amplitude for (b)-(f) is $\epsilon =0.1$. (b) $\hlm{2}{1}$ (c) $\hlm{8}{4}$ (d) $\hlm{32}{16}$ (e) $\hlm{32}{17}$ (f) $h_{t}$.
	}
	\label{f:t_geo_f}
\end{figure}

\begin{figure}
	\begin{center}
		\subfloat[][]{\includegraphics[width=0.45\textwidth]{ 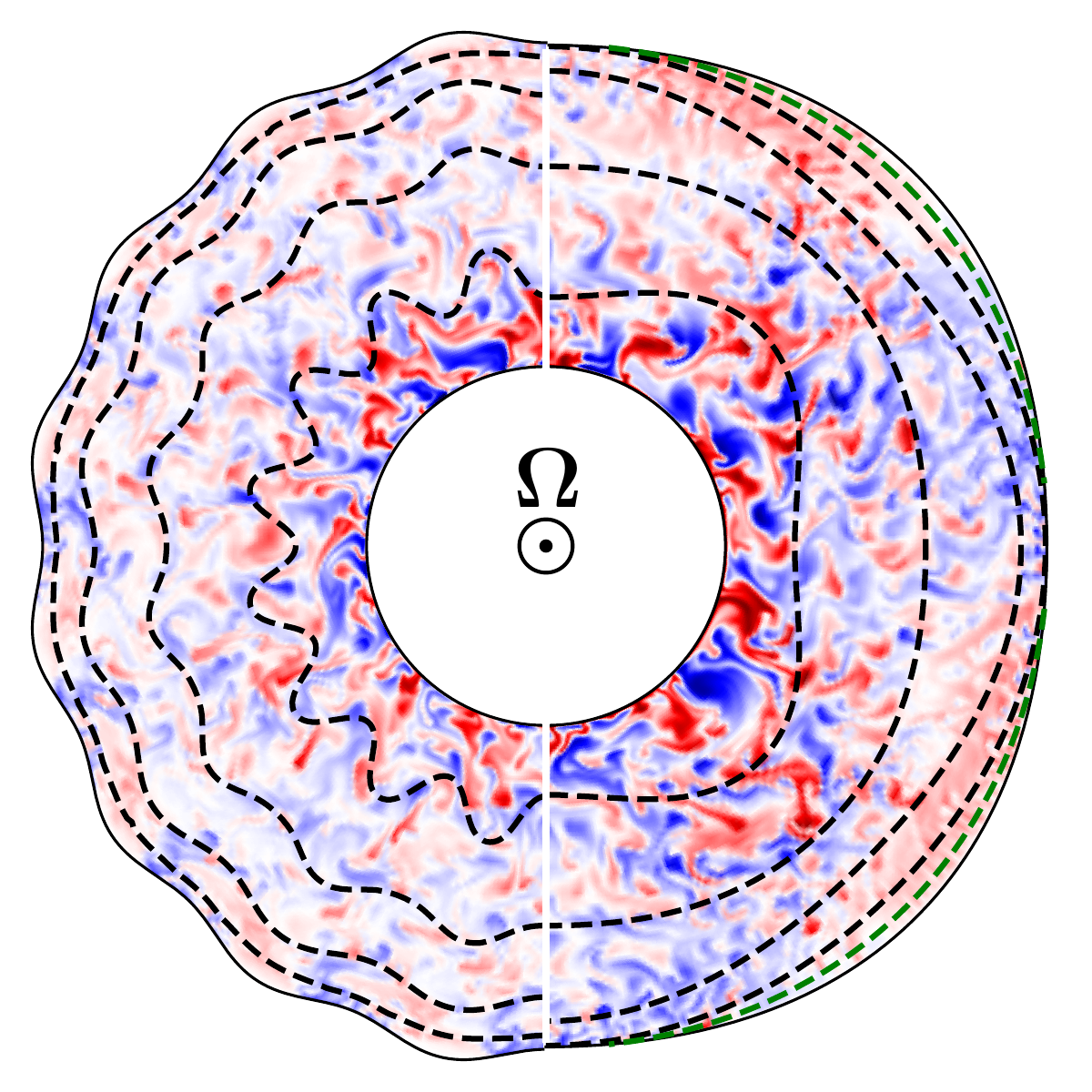}
		}
		\subfloat[][]{\includegraphics[width=0.45\textwidth]{ 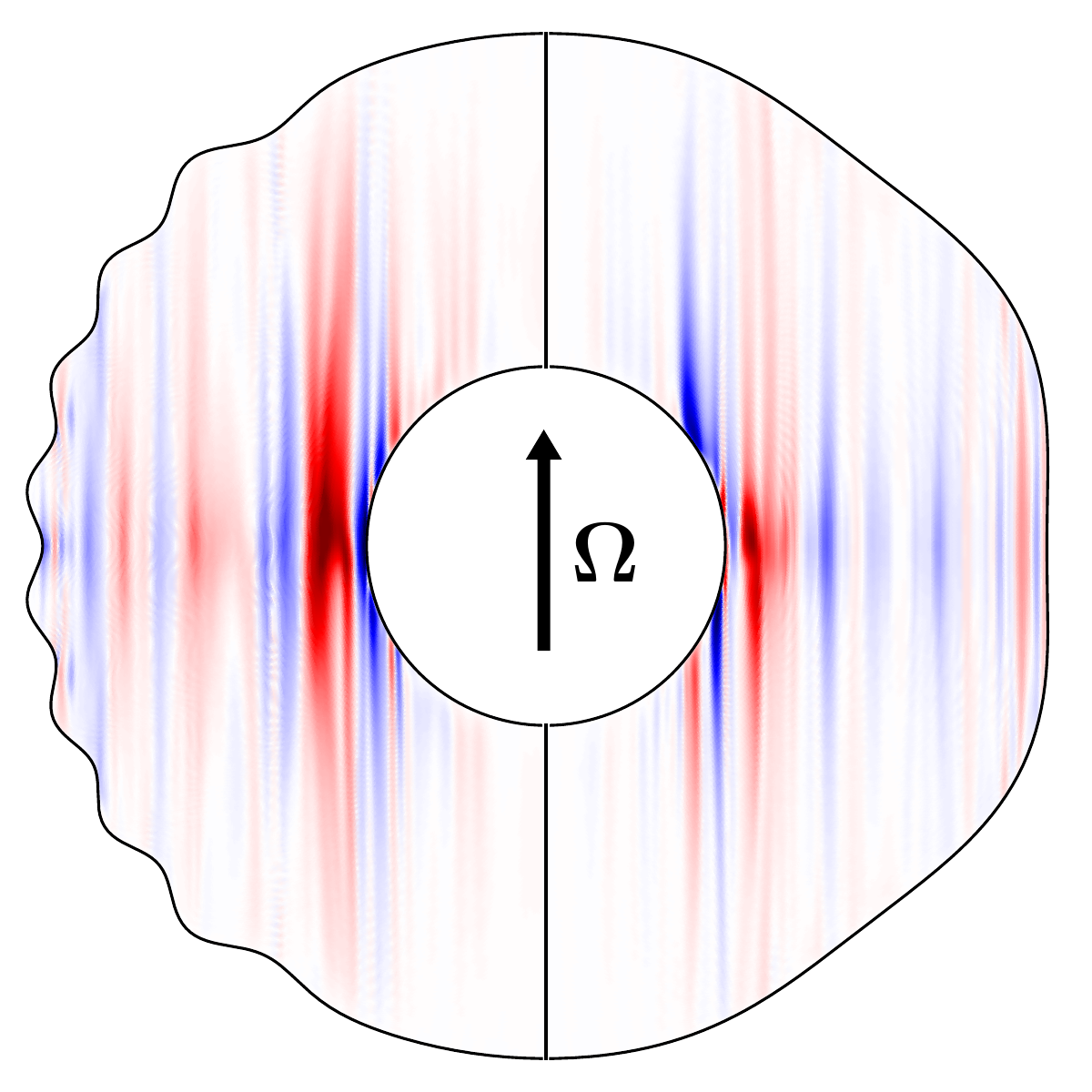}
		}
	\end{center}
	\caption{Instantaneous visualizations for temperature field fluctuations in the (a) equatorial and (b) meridional planes. The left-hand side of either visualization corresponds to $\hlm{32}{16}$ and the right-hand side to $\hlm{8}{4}$. (a) $\vartheta_{a}$ the temperature with the time and geostrophic contour average removed. The green contour depicts a contour that terminates on the CMB rather than completing a closed loop. (b) The fluctuating temperature field $\vartheta^{\prime}=\vartheta -\overline{\vartheta}$.}
	\label{f:t_vis}
\end{figure}

The supercritical flows can be modified significantly by the presence of topography. 
Here we investigate the behavior for $\Rat = 40$ and $Ek = 10^{-6}$. Previous work showed that for the same parameter range as investigated here, a localized Gaussian bump caused measurable changes to the flow and temperature field in the form of a counterrotating, cold region that was axially aligned with the bump \citep{tO25}. However, bulk quantities such as the Reynolds number and Nusselt number were not significantly modified, likely as a result of the localized nature of the topography in that study.

The changes in flow morphology are much more striking in models with global topography. 
The shape of the topography has a significant impact on the structure of the convective flow. 
Figure \ref{f:t_geo_f}(a) displays the time averaged azimuthal velocity for the control case where $\epsilon =0$. The remaining panels (b-f) show the same quantity across all topographies at $\epsilon  = 0.1$.
We note that the colorbars are normalized to maximum values in each visualization and are not equivalent across the six panels (the center of the colormap (white) is always at $u_{\phi} = 0$). In the studied parameter regime the topography can increase $Re$ by nearly $100\%$.
The structure of the large scale flow is primarily dictated by the geostrophic contours. In particular, consider the difference between Figures \ref{f:t_geo_f}(d) and \ref{f:t_geo_f}(f) and the topographies ($\hlm{32}{16}$ and $\hlm{32}{17}$) shown in Figures \ref{f:topo}(c) and \ref{f:topo}(d). Compared to the topographies, the flow structures are easily distinguishable. The $\hlm{32}{17}$ topography has an odd value of $\ell-m$, which means that the geostrophic contours have much smaller amplitude and wavelength. As a result, the flow appears much more azimuthally symmetric and the azimuthal scale is twice as small.

In regions near the CMB, where the topography bulges away from the core, the contours do not always form closed loops and the flow is inhibited. For example, in Figure \ref{f:t_geo_f}(c) the regions in the corners have reduced azimuthal velocity.
Contours that terminate at the boundaries (rather than complete closed loops) cannot support geostrophic flow, although they can support depth-invariant Rossby waves \citep{hG68}.
As a result, fluid motion into and out of regions with terminated contours is inhibited.

Figure \ref{f:t_vis} shows a comparison of fluctuating temperatures for two different topographic shapes. 
The visualizations on the left-hand sides of Figures \ref{f:t_vis}(a) and \ref{f:t_vis}(b) are for $ \hlm{32}{16}$ and the right-hand sides for $\hlm{8}{4}.$
Both visualizations correspond to a topographic amplitude of $\epsilon  = 0.05$.
In the equatorial plane we have plotted $\vartheta_{a}$, the temperature fluctuation about the geostrophic average.
\[
\vartheta_{a} = \vartheta-
	\begin{aligned}
		\oint \overline{\vartheta} dq\\[1pt]
		\hline
\oint dq
\end{aligned}\] 
The fluctuations are influenced by the contours, although the spatial structure is smaller. 
For example, in both topographic models the intensity of the fluctuations is greatest near the ICB. The region of high intensity is roughly traced by the contours, but the fluctuations within are smaller scale.

The green dashed line in Figure \ref{f:t_vis}(a) traces a contour which terminates at the boundary. The fluctuations in the region of this contour are larger scale. 
The fluctuations are colder than the average on one half of the contour and warmer on the other half.
Visualizations of the full temperature field (not shown) indicate that these regions are anomalously cold.
The reduced velocities associated with these regions inhibit advection from the interior. The effect is less striking in the $\hlm{32}{16}$ topography because the regions where geostrophic contours terminate at the boundaries are quite small.

Figure \ref{f:t_vis}(b) shows meridional slices of the fluctuating temperature field $\vartheta^{\prime} = \vartheta -\overline{\vartheta}$.
Except for near the boundaries, the two temperature fields are much less distinguishable between the different topographies than in the equatorial slices. 
The radial structure of the flow is visibly less dependent on the topographic scale than the azimuthal structure. 
The geostrophic contours run in planes perpendicular to meridional slices, which explains why the flows are not as distinguishable in this visualization.

\subsection{Global transport quantities}
\label{s:gs}
\begin{figure}
	\begin{center}
		\subfloat[][]{\includegraphics[width=0.45\textwidth]{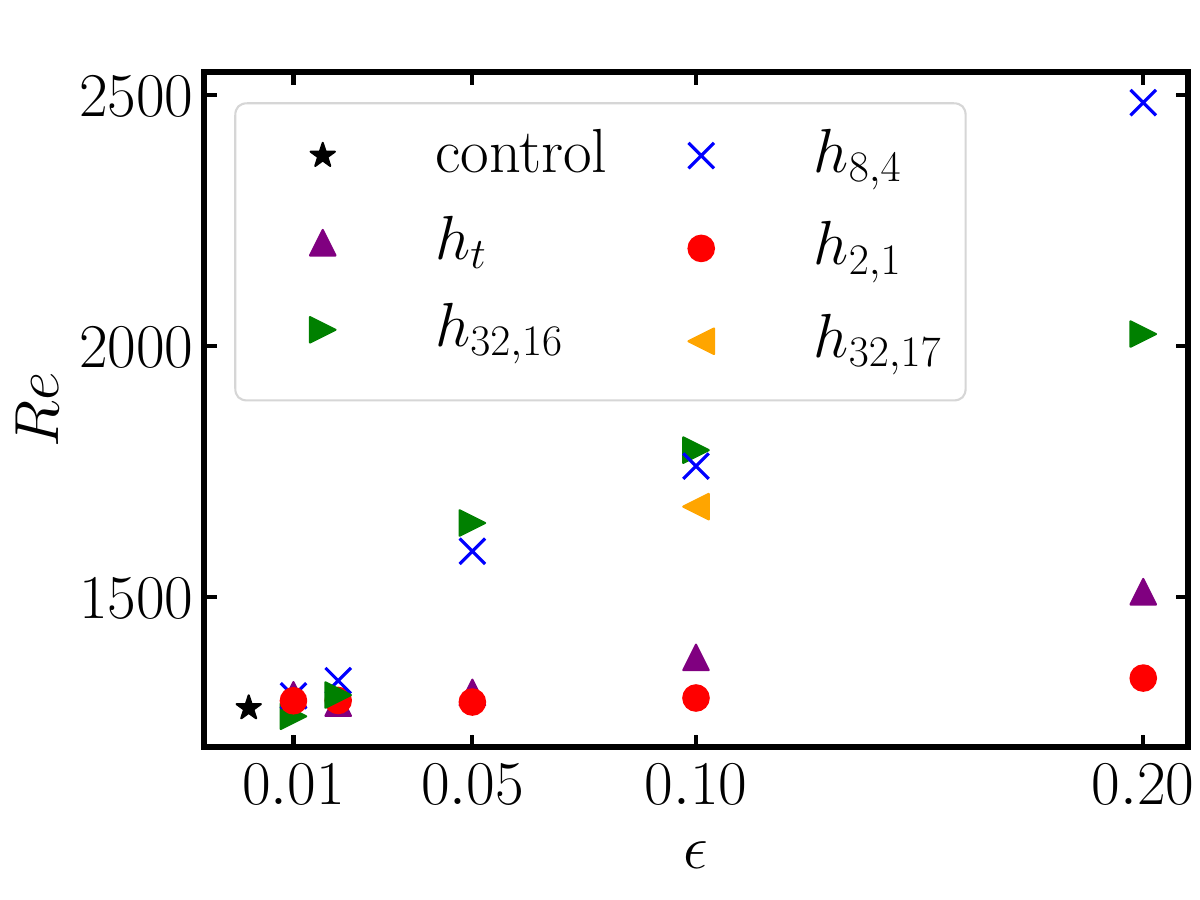} 
		}
		\subfloat[][]{\includegraphics[width=0.45\textwidth]{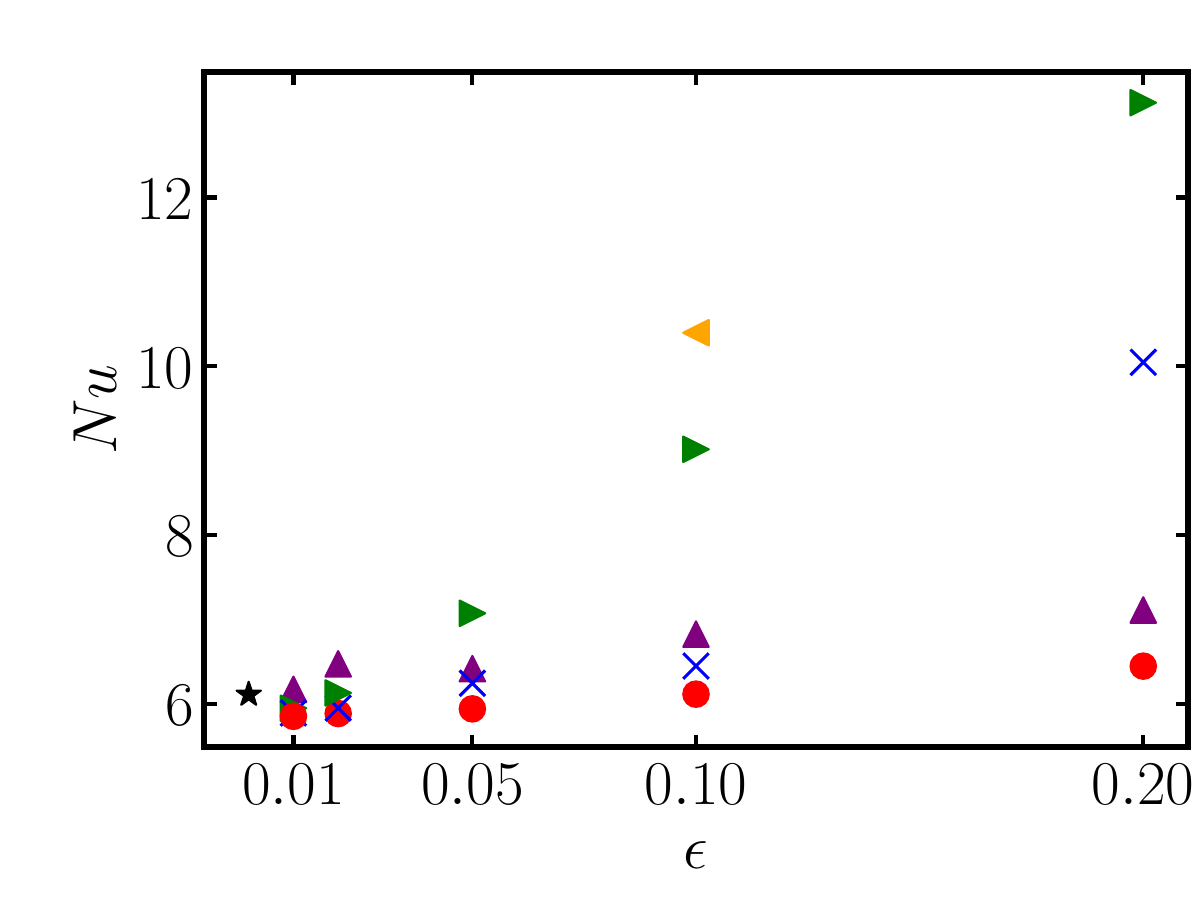} 
		}
	\end{center}
	\caption{Global transport quantities: (a) Reynolds and (b) Nusselt numbers plotted against topographic amplitude $\epsilon$. Both quantities tend to increase with $\epsilon$ and the effect is most pronounced in the $\hlm{8}{4},\hlm{32}{16},$ and $\hlm{32}{17}$ topographies. }
	\label{f:gs}
\end{figure}
The top row of Figure \ref{f:gs} displays the Reynolds number and Nusselt number for all turbulent simulations.
The black datum gives values for the control case ($\epsilon  = 0.0$). 
We first note that for the smoothest topography ($\hlm{2}{1}$) there is little variation in both $Re$ and $Nu$. Otherwise, there tends to be an increase in both quantities as the topographic amplitude is increased, although no obvious scaling law exists. 
\TOr{
	The increase in $Re$ and $Nu$ is most easily explained by recognizing that the buoyancy forcing increases with topographic amplitude, which we primarily attribute to the mechanism described by equation \eqref{e:m_avg2}: when geostrophic contours are no longer circular, buoyancy is able to exert work on the component of the flow parallel to the contours. 
	In \ref{a:b_work} we provide a quantitative analysis of the buoyancy work and demonstrate that this mechanism is responsible for elevated $Re$ and $Nu$ in the $\hlm{8}{4}$ and $\hlm{32}{16}$ topographies.
}

\begin{figure}
	\begin{center}
		\subfloat[][]{\includegraphics[width=0.45\textwidth]{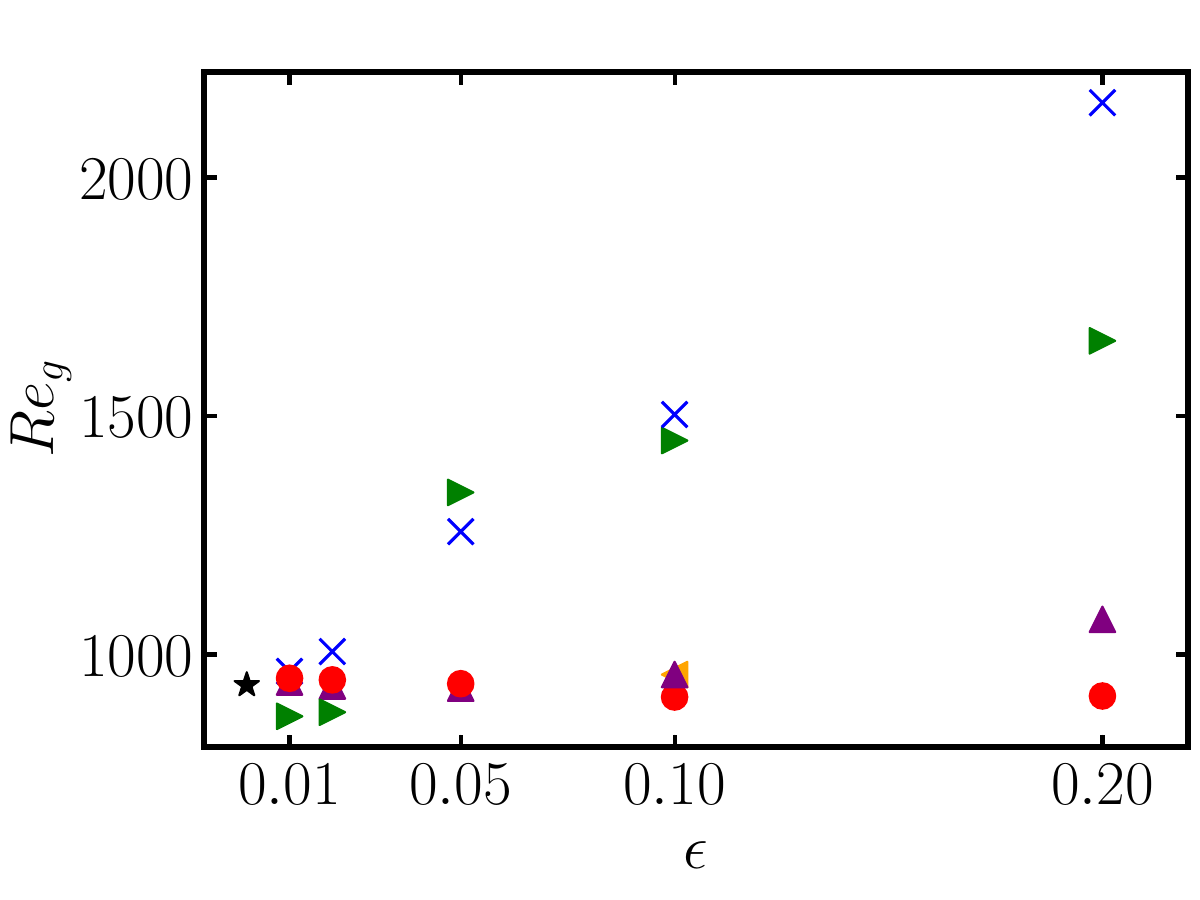}
		}
			\subfloat[][]{\includegraphics[width=0.45\textwidth]{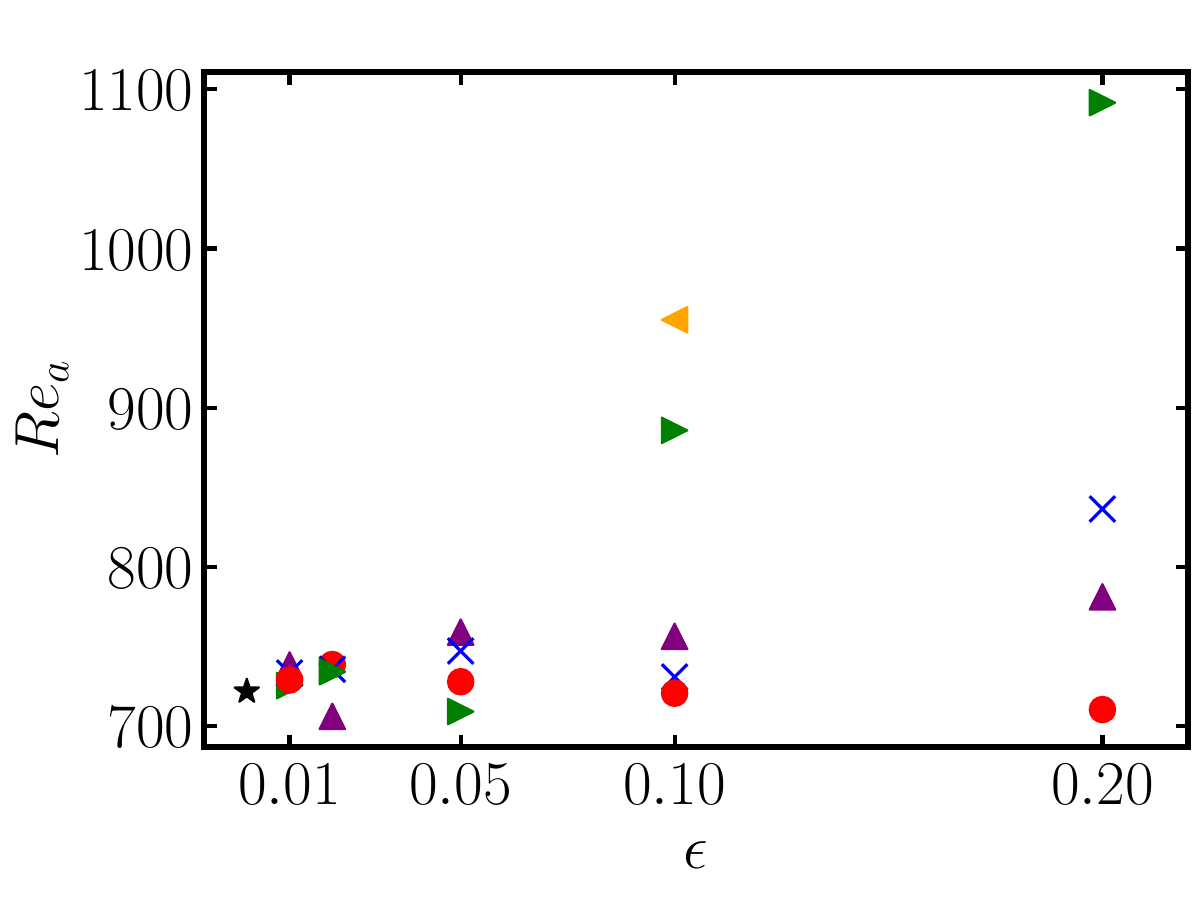} 
			} \\
			\subfloat[][]{\includegraphics[width=0.45\textwidth]{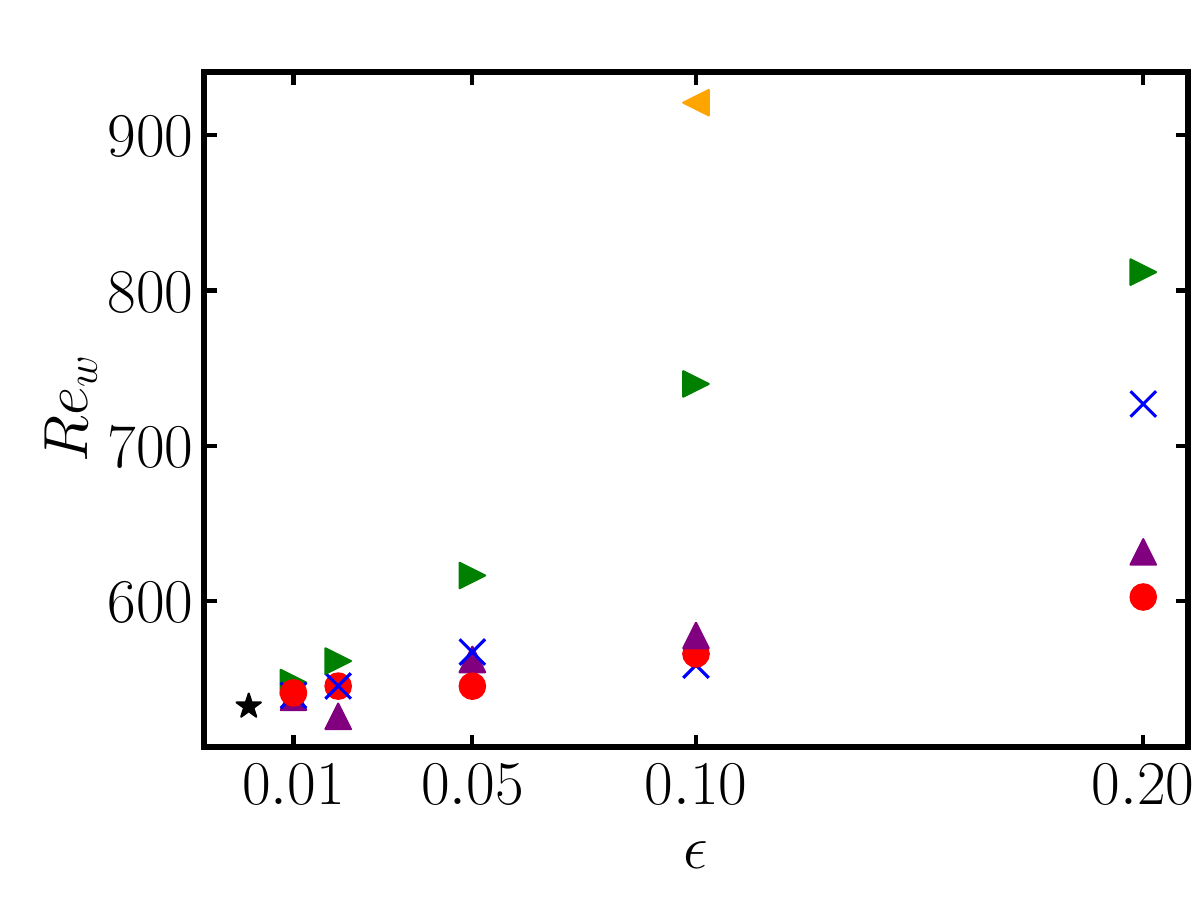}
			}
			\subfloat[][]{\includegraphics[width=0.35\textwidth]{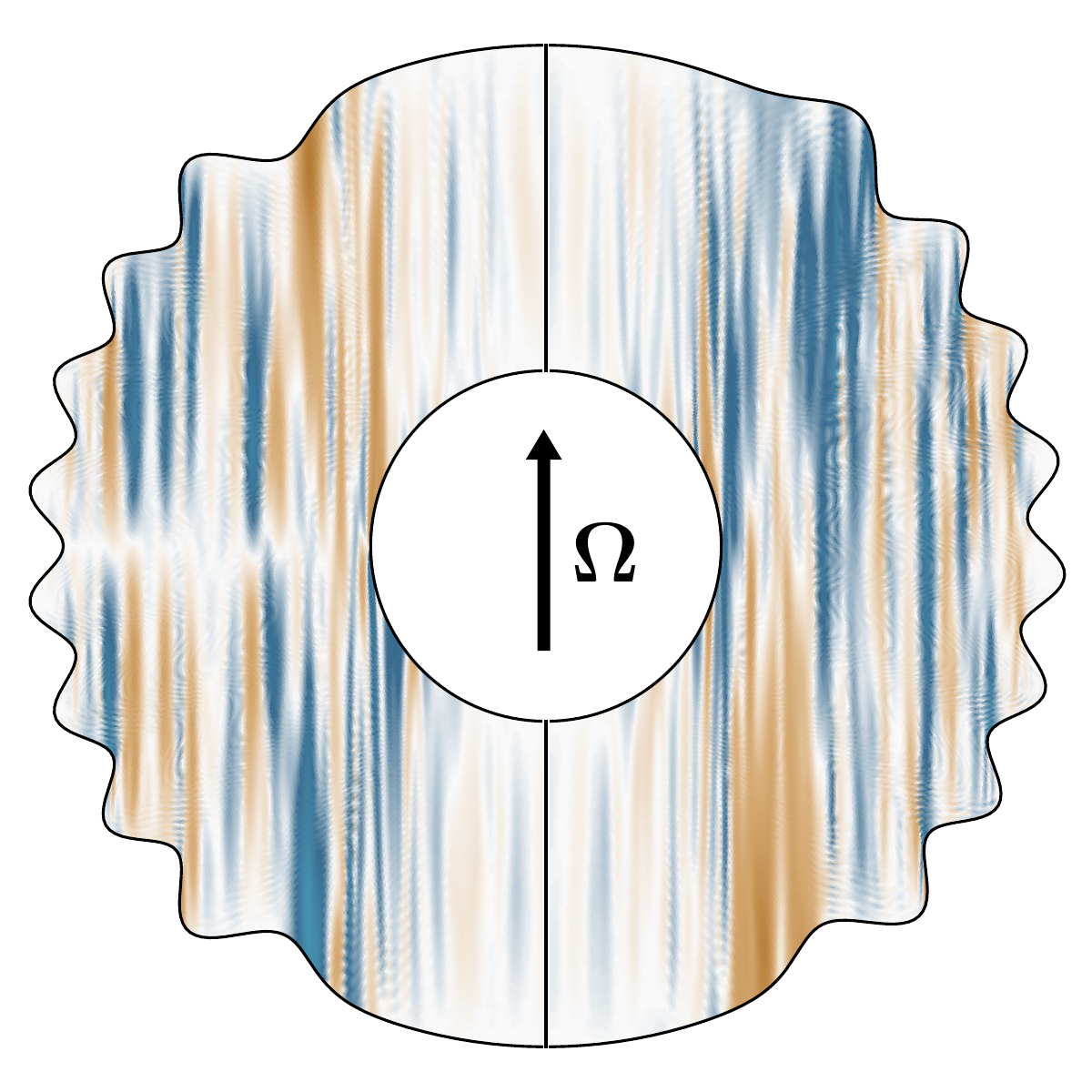}
			}
	\end{center}
	\caption{Reynolds numbers for all supercritical/turbulent simulations. Symbols are the same as in Figure \ref{f:gs}. (a) Reynolds from geostrophic velocity. (b) Reynolds from ageostrophic velocity. Both $Re_{\gchar}$ and $Re_{\achar}$ are calculated in the equatorial plane only. (c) Axial Reynolds number calculated over the entire domain. (d) Meridional slice of the axial velocity, for (left) $\hlm{32}{16}$ and (right) $\hlm{32}{17}$ topographies at $\epsilon =0.1.$
When $\ell - m$ is even the convection prefers axial modes that are odd across the equatorial plane. The opposite is true for odd values of $\ell - m$. 
}
	\label{f:re_geo_a}
\end{figure}

Motivated by the approach of BS96, we discuss Reynolds numbers calculated from individual velocity components, specifically $u_{\achar},$ $u_{\gchar},$ and $w.$
We note that this decomposition is analogous to the zonal decomposition often employed in spherical studies \citep[e.g.][]{pG78,uC02,jN24}.
In spheres and spherical shells, the zonal flow is coincidentally the flow that follows the circular geostrophic contours.  
Figure \ref{f:re_geo_a} displays Reynolds number calculated from three different components of the velocity field. Panels (a), (b), and (c) correspond to $u_{\gchar},u_{\achar},$ and $w$, respectively. 

For the $\hlm{32}{17}$ topography we find that the elevated values of $Re$ and $Nu$ are associated with elevated ageostrophic and axial velocities.  
The geostrophic Reynolds number is unchanged compared to the control case ($\ep=0$). Since the geostrophic contours are relatively unchanged compared to the control, buoyancy is able to perform only minimal work on the geostrophic component.
\TOr{Our analysis of the buoyancy work in \ref{a:b_work} corroborates this result}.
For the $\hlm{8}{4}$ and $\hlm{32}{16}$ topographies\TOr{ --- in which the geostrophic contours are significantly deformed --- }the variation in speeds is greatest in the geostrophic flow. $Re_{\gchar}$ increases by $130\%$ and $80\%$ respectively in the $\epsilon  = 0.2$ case compared to the control.


Figure \ref{f:re_geo_a}(d) displays meridional snapshots of $w$ for $\hlm{32}{16}$ (left) and  $\hlm{32}{17}$ (right). 
We noticed that $\hlm{32}{16}$ prefers an axial structure that flips sign about the equatorial plane, whereas $\hlm{32}{17}$ prefers a mode constant across the equatorial plane.
Inspection of snapshots from other topographies at super- and sub-critical $\Rat$ (not shown) follow a similar pattern. 
When $\ell - m$ is even (odd), an axial mode that is odd (even) across the equatorial plane is preferred. 
This may explain the elevated values of $Re_{w},$ in which we also see an increase in the $\hlm{2}{1}$ data.

%


\subsection{Energy spectra and length scales}
\label{s:ke}

\begin{figure}
	\begin{center}
	\subfloat[][]{\includegraphics[width=0.47\textwidth]{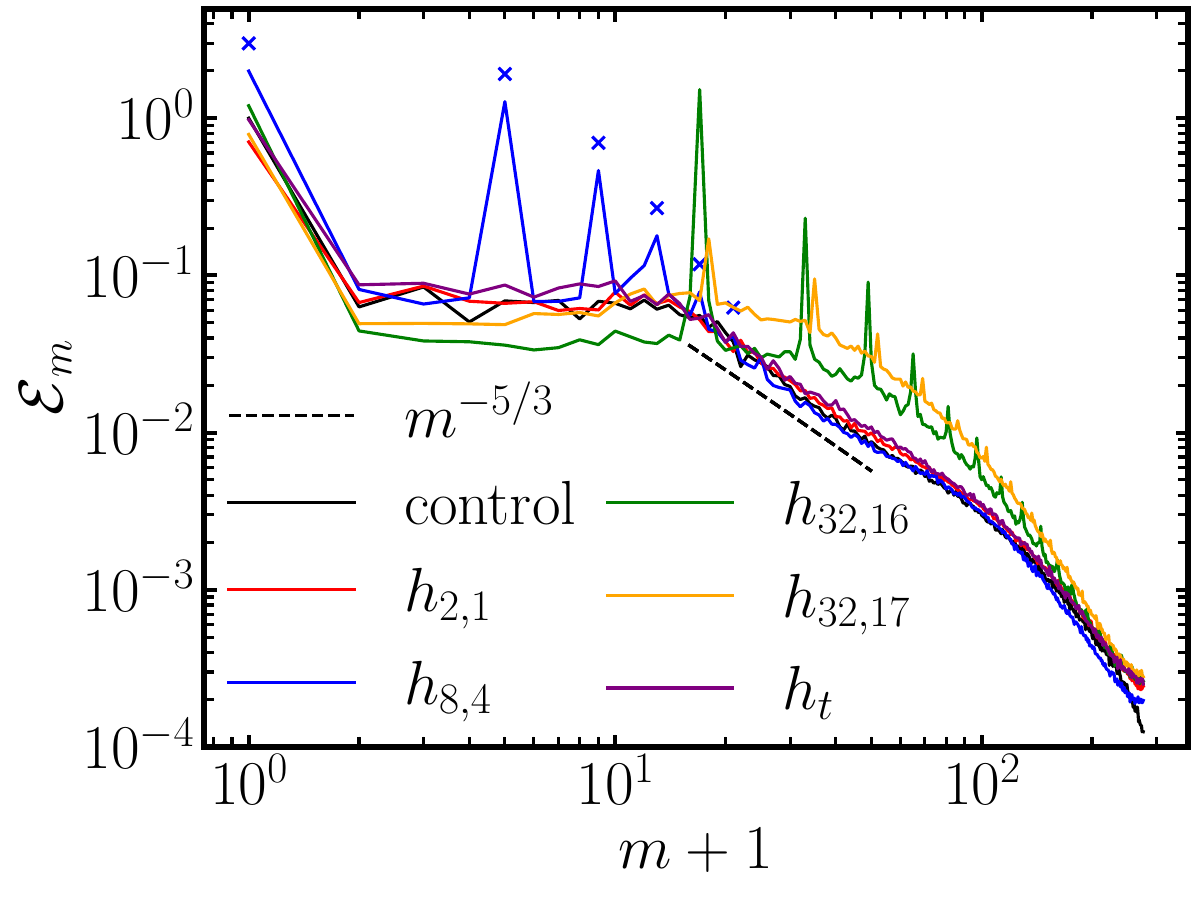}}\quad
	\subfloat[][]{\includegraphics[width=0.47\textwidth]{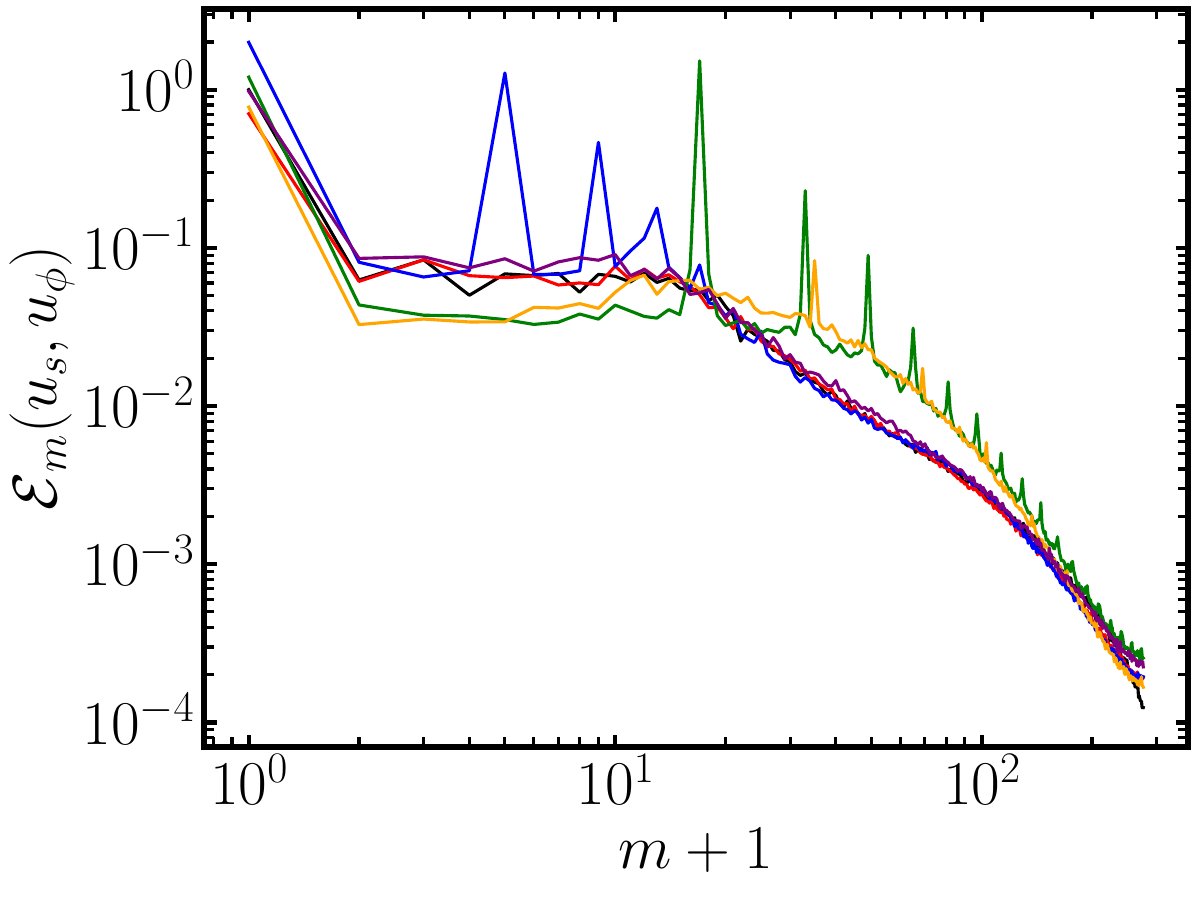}}\\
	\subfloat[][]{\includegraphics[width=0.47\textwidth]{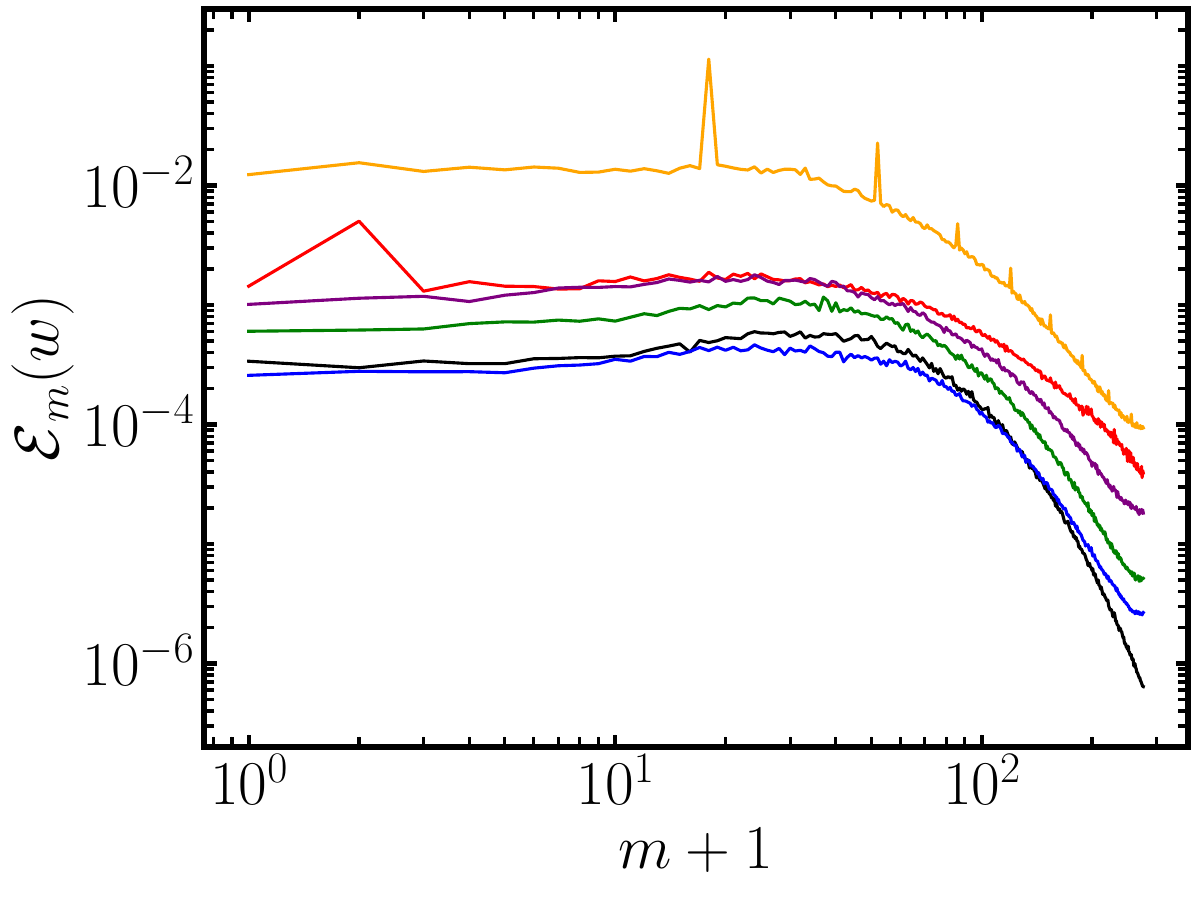}}\quad
	\subfloat[][]{\includegraphics[width=0.47\textwidth]{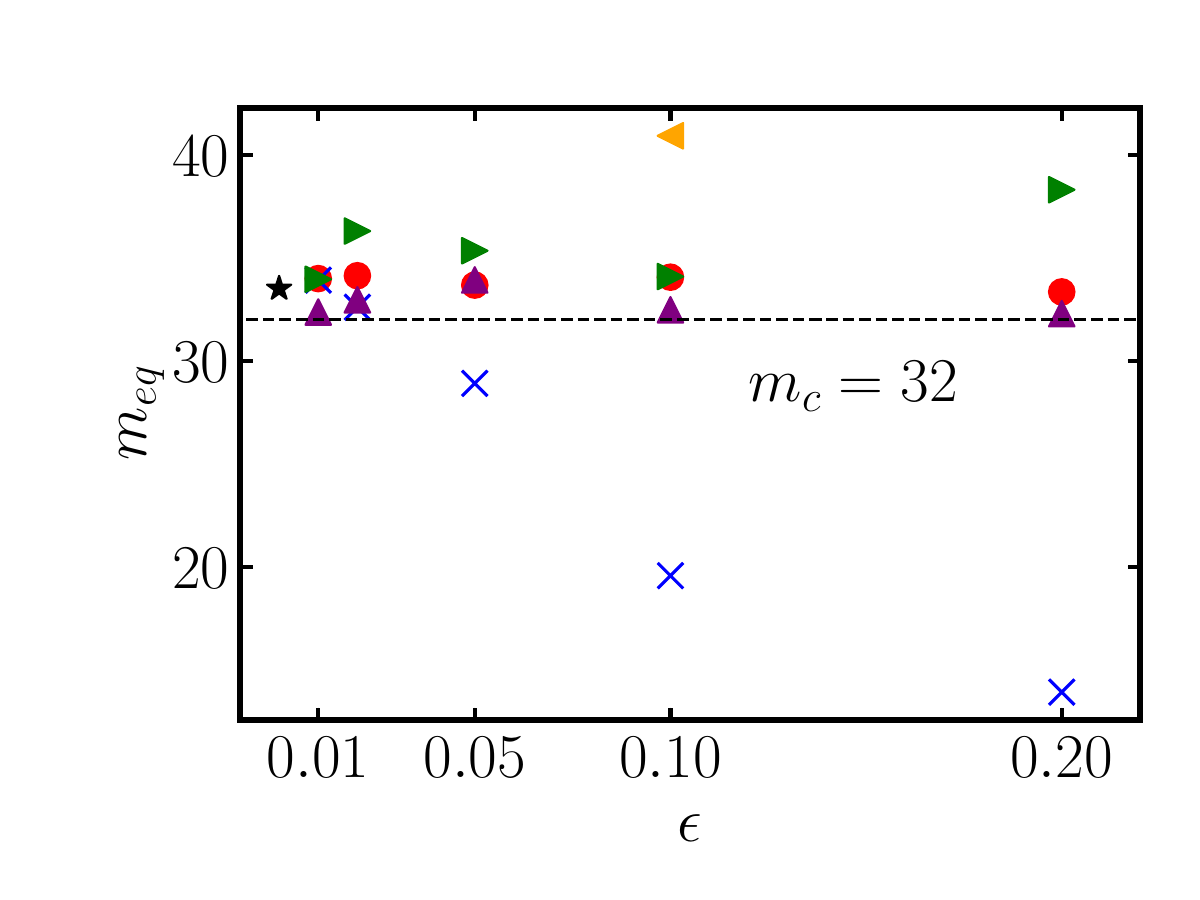}}
	\end{center}
	\caption{One dimensional kinetic energy spectra for control ($\epsilon  = 0$) and all $\epsilon  = 0.1$ cases. Spectra in panels (a-c) are normalized by the value $\mathcal{E}_0$ for the control. (a) $\mathcal{E}_{m}$, the spectra calculated from cylindrical velocity fields. Crosses denote locations of first six spectral peaks in the geostrophic height function for $\hlm{8}{4}$. They coincide with the spectral peaks of the kinetic energy.  (b) Same as (a) without contribution from axial velocity field $w$. (c) Same as (a) for only axial velocity field.
	(d) Weighted wavenumber versus topographic amplitude. The dashed line marks the critical wavenumber for convection with no topography ($\ep=0$).}

	\label{f:spectra}
\end{figure}

Kinetic energy spectra are useful for characterizing topographically-induced changes in the flow field. 
Given that the flows are primarily axially aligned, we restrict computation of the spectra to within the equatorial plane, averaging over radius from the ICB out to the minimum radius for which $r = r_{cmb}$ (that is we integrate over the largest annular disc which can fit within the boundaries at the equator).  
The resulting spectra are shown in Figure \ref{f:spectra} for the control and all cases where $\epsilon  = 0.1$.

The signature of the topography is evident in the $m\geq 4$ topographies, where repeated peaks at integer multiples of the topographic $m$ occur. 
Figure \ref{f:spectra}(a) displays the total kinetic energy spectra. The flow is highly structured and coordinated with the topography. 
Using the $\hlm{8}{4}$ case as an example, we find spectral peaks at integer multiples of 4. Although the topography is composed of only a single $m=4$ harmonic, the resultant geostrophic contours contain higher order harmonics (see \ref{a:oddveven}). 
The $m$ values corresponding to these harmonics are marked by the blue crosses in Figure \ref{f:spectra}(a). 
Figure \ref{f:spectra}(b) and Figure \ref{f:spectra}(c) display spectra calculated from the equatorial and axial velocity components respectively.
For the antisymmetric cases ($\hlm{2}{1}$ and $\hlm{32}{17}$),
the equatorial velocity components have increased power in even integer multiples of $m$ ($34,68, ...$ for  $\hlm{32}{17}$). 
The axial velocity peaks at odd integer multiples of $m.$

\TOr{
For the antisymmetric topographies, the spectral peaks at even and odd integer multiples of $m$ are} due to two separate effects. The equatorial spectra are largely constrained to the contours, which only contain $2nm$ harmonics (where $n$ is an integer). 
There is also a baroclinic flow influenced by the Ekman pumping at the boundaries. 
The odd integer multiples of $m$ do not conform to the contours, so the equatorial components of the flow associated with these modes are negligible; however, in order to satisfy the incompressibility condition
\TOr{the axial mode must be symmetric about the equator so that $\partial w/\partial z\approx 0$. This constraint explains $\mathcal{E}_{m}\lb w\rb $ peaks at odd integer multiples of $m$ for the antisymmetric topographies.}
We note that, in general, the flow in the $2nm$ harmonics does contain vertical motion, just not in the equatorial plane. Axial spectra calculated at latitudes other than $0^{\circ}$ (not shown) have peaks in both even and odd multiples of $m.$
This is also true for the symmetric cases, which have no peaks in the axial spectra at the equator (Figure \ref{f:spectra}(c)).

The picture in the symmetric cases ($\hlm{8}{4}$ and $\hlm{32}{16}$) is quite different, although the interpretation is the same. The equatorial velocity is primarily constrained to the contours, which have power in all integer multiples of $m.$ 
The lack of structure in the axial energy spectra is because the axial velocity is largely anti-symmetric about the equatorial plane, whereas it is symmetric for the odd $\ell-m$ topographies.
The highest wavenumber topographies ($\hlm{32}{16}$ and $\hlm{32}{17}$) display a ``belly'' in the spectra. 
For the $h_{t}$, $\hlm{2}{1}$, and $\hlm{8}{4}$ topographies, the spectral power at $m>16$ collapses to a constant power law which is nearly the $5/3$-slope commonly associated with isotropic turbulence \citep{sP00}. 
However, the power in the high wavenumbers for $\hlm{32}{16}$ and $\hlm{32}{17}$ bow upwards. 
We observe this effect for $\epsilon \geq0.05$, and expect that it may arise because the topographic wavelength is very near the critical wavelength. 
Indeed, the spectral slope in the control case changes from near flat to $5/3$ right around $m=16.$ \TOr{We expect that the convection is resonating with the topography, which may also explain the significantly elevated values of $Re$, $Nu$, $Re_{a}$, and $Re_{w}$.}

Figure \ref{f:spectra}(d) shows the weighted wavenumber from equation \eqref{e:meq} plotted against topographic amplitude. For low $\epsilon$, $m_{eq}$ is near $32,$ which is the critical wavelength in a sphere \citep{aB22}. 
As $\epsilon$ increases, $m_{eq}$ does not accurately reflect the topographic wavenumber. 
Although it decreases for $\hlm{8}{4},$ it is much larger than $4$, even at the largest $\epsilon,$ and the data for $\hlm{32}{16}$ is larger than the critical wavelength for the largest topography. 
Since $m_{eq}$ is a weighted average over $\mathcal{E}_{m},$ we attribute this behavior to the peaks at integer multiples of $m$ and, for $\hlm{32}{16}$ and  $\hlm{32}{17}$, the belly at high wavenumber, which shifts $m_{eq}$ to values higher than the topographic wavenumber.

\subsection{Topographic torques}
\label{s:top_torque}

\begin{figure*}
	\subfloat[][]{\includegraphics[width=0.45\textwidth]{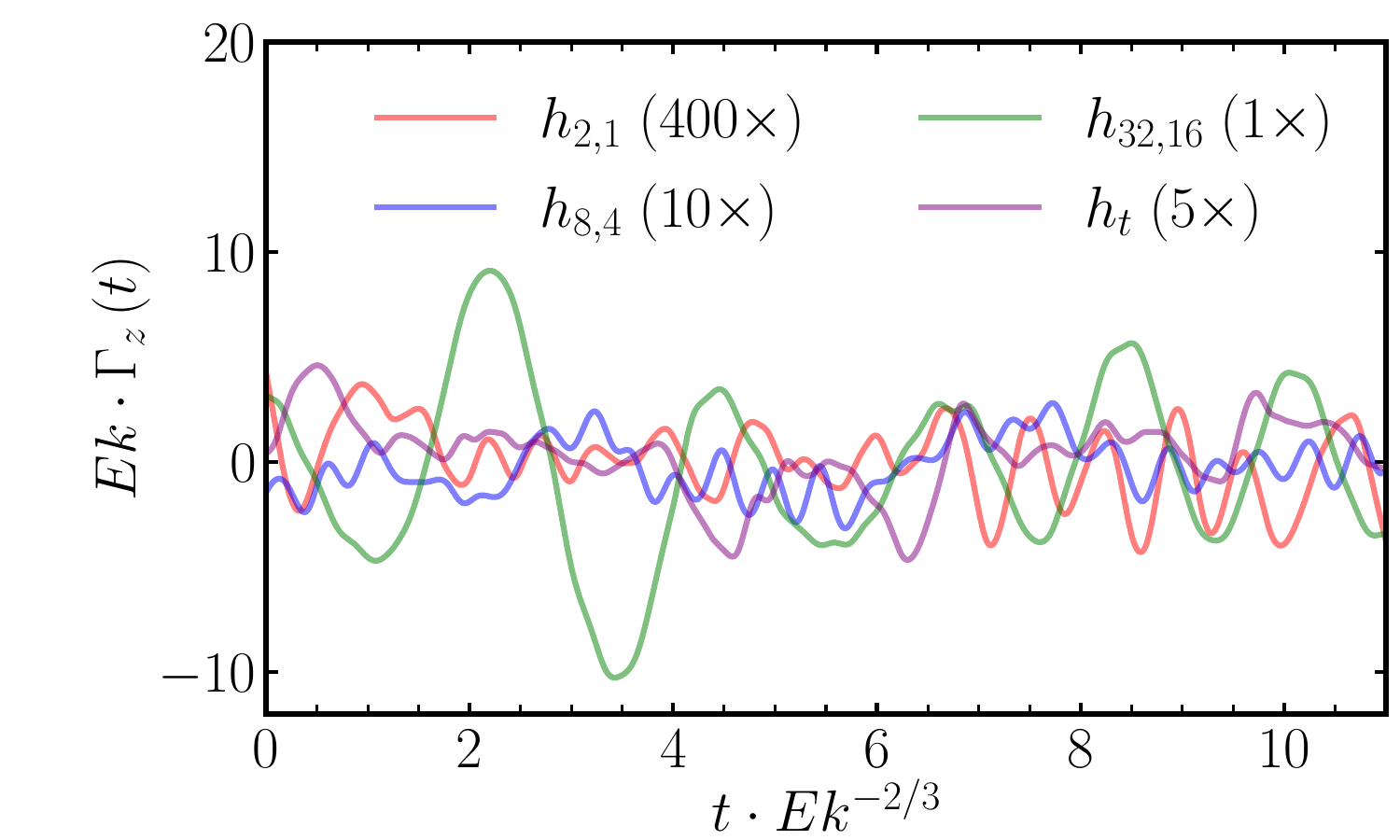}\label{f:ts_tq0x01}}
	\subfloat[][]{\includegraphics[width=0.45\textwidth]{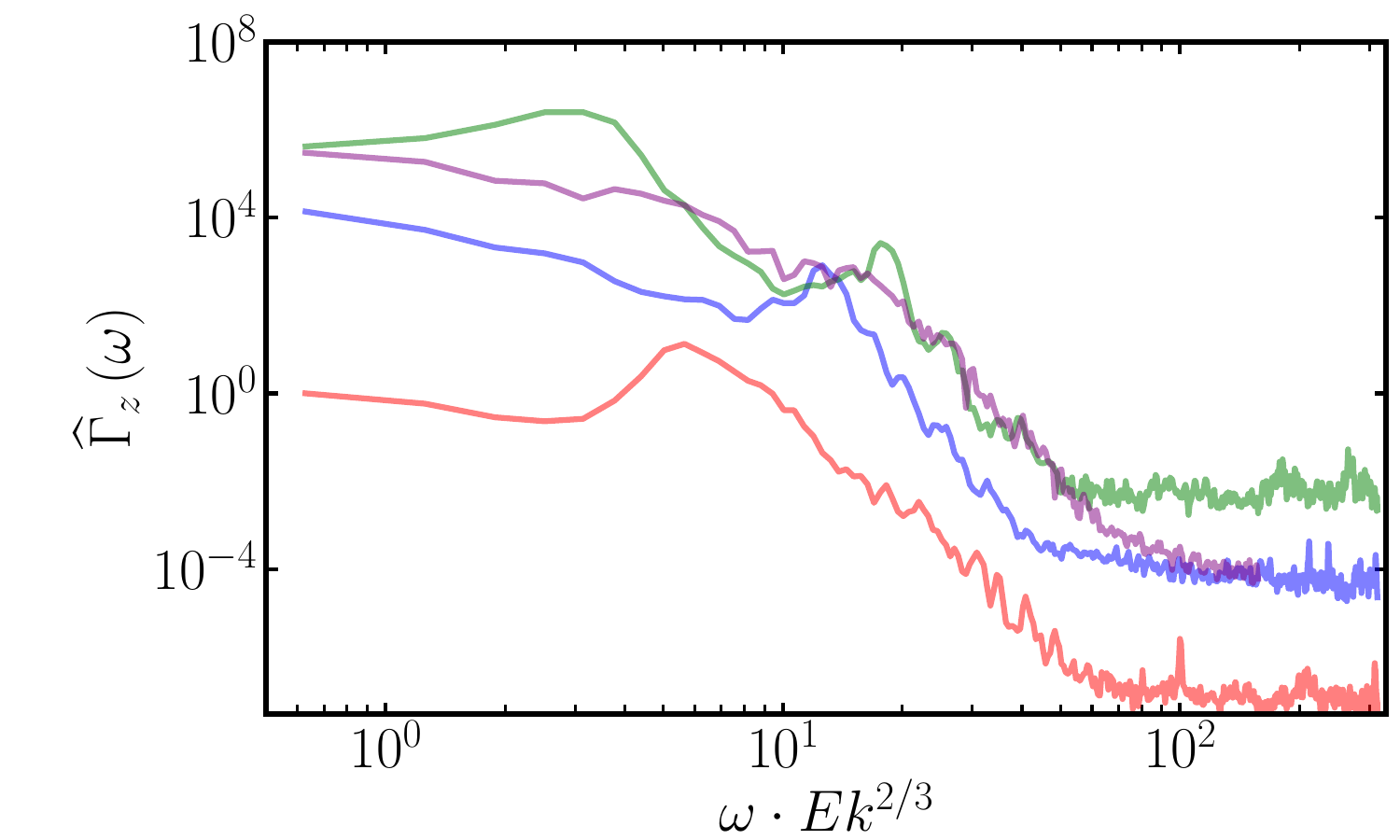}\label{f:ts_nu0x01}}
	\\
	\subfloat[][]{\includegraphics[width=0.45\textwidth]{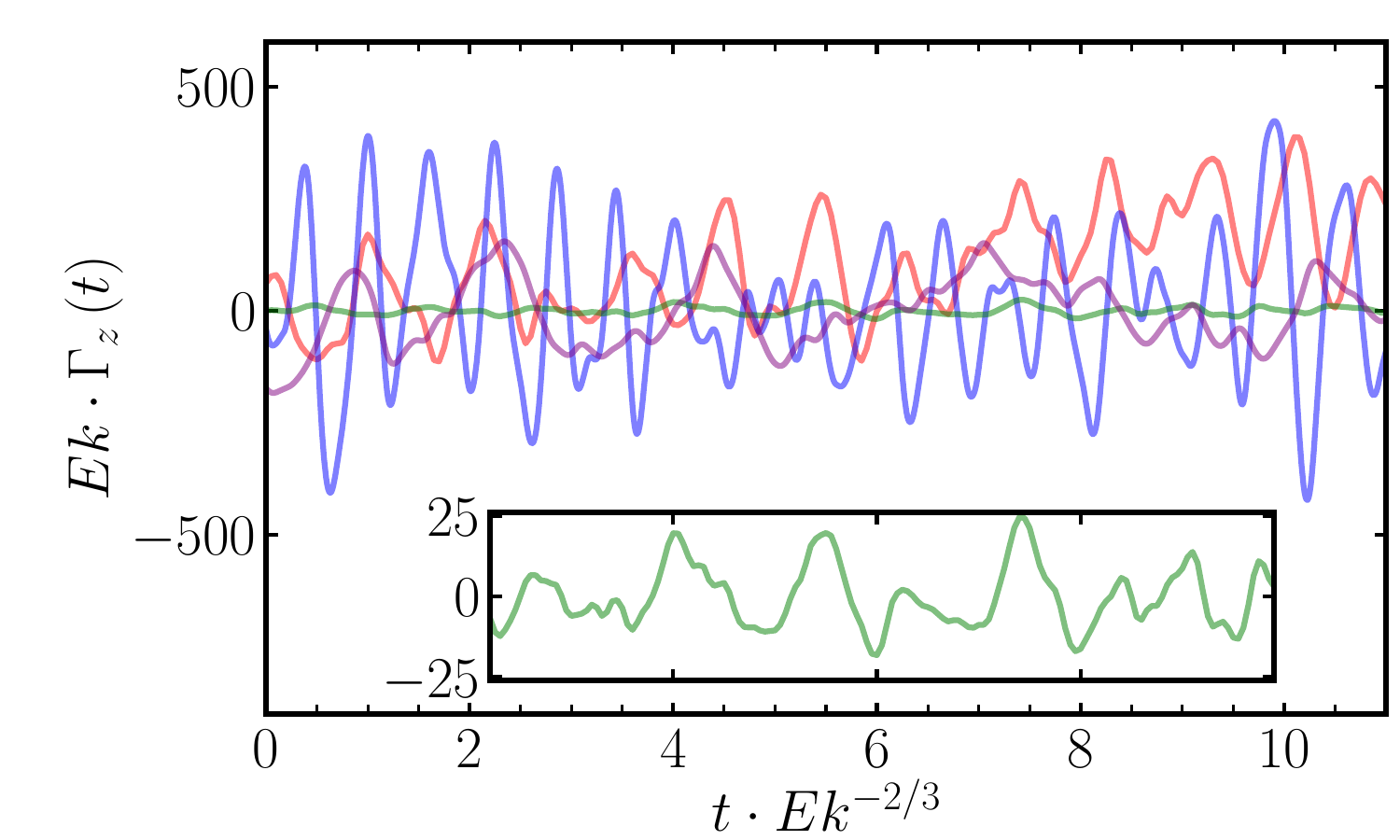}\label{f:ts_tq0x1}}
	\subfloat[][]{\includegraphics[width=0.45\textwidth]{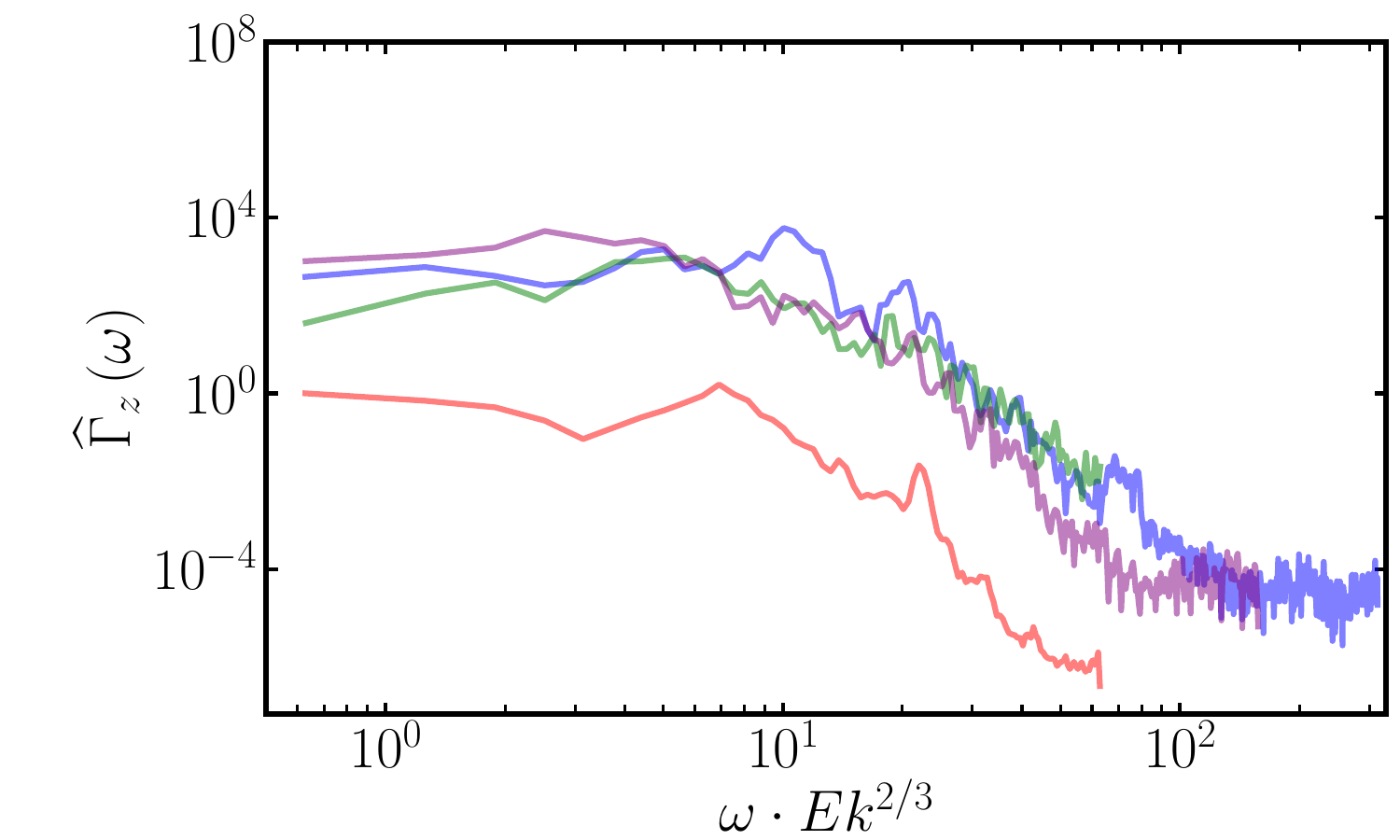}\label{f:ts_nu0x1}}
	\caption{Time dependence of topographic torque for (top) $\epsilon  = 0.01$ and (bottom) $\epsilon  = 0.2$. (Left) time series of the reduced torques and (right) periodograms of the torque. Frequencies and times are re-scaled by the small scale diffusion time $Ek^{2/3}$. The inset in (c) shows the same data as the main panel, but the vertical axis is re-scaled to see the smaller torque for $\hlm{32}{16}$.}
\label{f:torque}
\end{figure*}

Temporal fluctuations of the topographic torque are shown in Figure \ref{f:torque} for the different topographic shapes. The top row corresponds to small topographic amplitude $(\epsilon = 0.01)$ and the bottom row to large topographic amplitude ($\epsilon  = 0.20$). Figures \ref{f:torque}(a) and \ref{f:torque}(c) display time-series of the axial torque. The magnitudes of the torque can vary significantly across the different topographies. 
We therefore normalize each series by the factor indicated in the legend. 
With the exception of $\hlm{32}{16},$ the torque amplitudes are much smaller in Figure \ref{f:torque}(a) than in Figure \ref{f:torque}(c). 
Given a linear dependence on  $\epsilon$, we would expect $\Gamma_{z}\lb t\rb $ to have an amplitude $20\times$ smaller, although this is clearly not always the case.
We also note the coherence of the $\hlm{32}{16}$ signal in Figure \ref{f:torque}(a). Despite a Reynolds number of $Re \approx 1300$, the signal is smooth. 

Figures \ref{f:torque}(b) and \ref{f:torque}(d) present the periodograms corresponding to Figures \ref{f:torque}(a) and \ref{f:torque}(c) respectively. The frequency spectra are plotted against the normalized frequencies $\omega Ek^{2/3}$. We normalize by $Ek^{2/3}$ because the critical frequencies in rotating convection are $O\lb Ek^{-2/3}\rb.$
We use Welch's method with a Hann window of ten small scale diffusion times corresponding to $\tau = t\cdot Ek^{-2/3}$. In each plot the spectra are normalized by the lowest frequency power for the $\hlm{2}{1}$ topography. This choice allows us to compare the magnitudes of the torques across topographies.
The periodogram in Figure \ref{f:torque}(b) confirms that the majority of the power resides in frequencies around $O\lb Ek^{-2/3}\rb$ (that is $O\lb 1\rb$ values of $\omega Ek^{2/3}$). 
Equation \eqref{e:ax_torque} indicates that for the single harmonic topographies, only the components of the pressure field with the same $m$ value as the topography can contribute to the torques. Indeed the same $\ell$ is required as well, however, because the flow is axially aligned, we find it is sufficient and simpler to only discuss $m$. 
For small topography, the $\hlm{32}{16}$ torques are largest because the scale of convection is set by the instability occurring at $m= m_c = O(Ek^{-1/3}).$  
Of the topographies tested at $\epsilon  = 0.01$, the order of $\hlm{32}{16}$ is closest to $m_c$.
We conclude that the smoothness of the torque signal for $\hlm{32}{16},$ $ \epsilon = 0.01$ is due to the interaction of the topography with a few traveling waves at $O(Ek^{-2/3})$ frequencies characteristic of standard rotating convection. 
We note that the periodograms for the $h_{t}$ topography do not display noticeable peaks. Since $h_{t}$ is spectrally broad, we expect that it is interacting with a variety of convective modes with different frequencies.

\begin{figure*}
	\begin{center}
		\subfloat[][]{\includegraphics[width=0.49\textwidth]{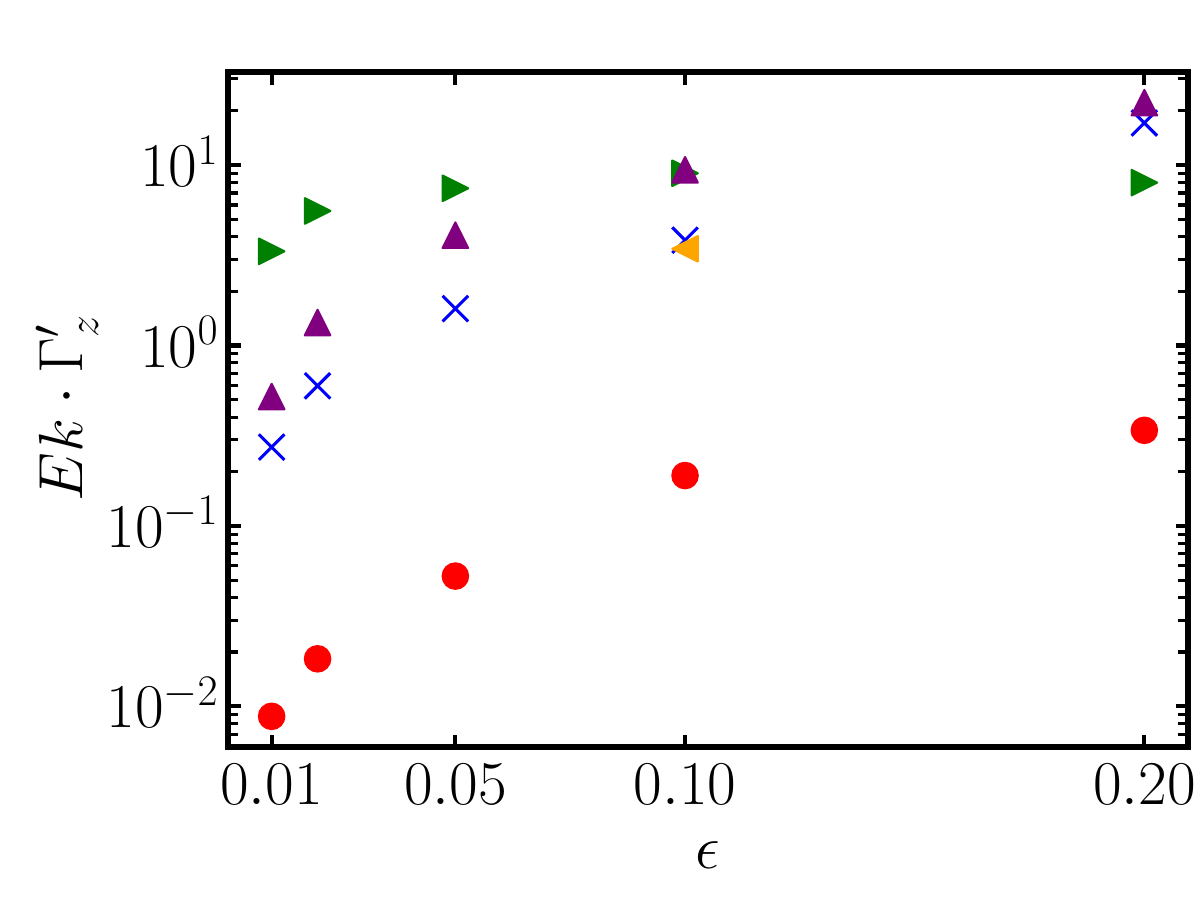}	}
		\subfloat[][]{\includegraphics[width=0.49\textwidth]{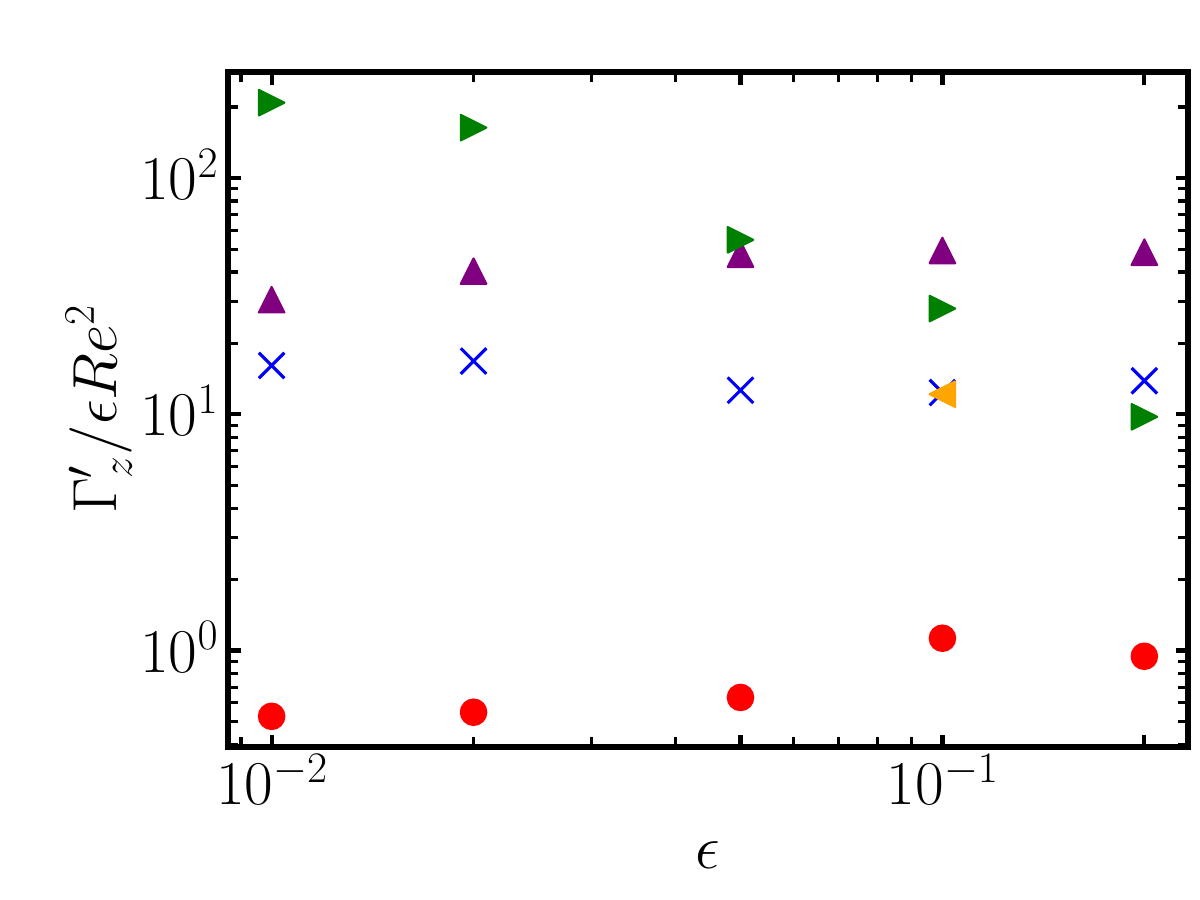}}
	\end{center}
	\caption{(a) Standard deviation of the reduced torque for all cases. (b) Standard deviation of the torque normalized by the dynamical scaling $\epsilon Re^{2}$.
	}
	\label{f:ekt}
\end{figure*}

For large topographic amplitude the torque signal for the $\hlm{32}{16}$ topography becomes less regular.
The (green) time series in Figure \ref{f:torque}(c) is less periodic, although the amplitude of the fluctuations only increases by a factor of about 2.5. The peaks in the frequency spectra are far broader. 
At the same time, the other topographies exhibit much larger torques for $\epsilon  = 0.2$ than $\epsilon  = 0.01,$ and the spectral peaks are of similar width for the two amplitudes.

In order to better understand the effect of topographic amplitude on torque magnitude, we present the standard deviations (time) of the axial torques in Figure \ref{f:ekt}. 
Figure \ref{f:ekt}(a) shows the standard deviation of the reduced torque for all cases at $\Rat = 40.$ In general, larger topographies and larger topographic order yield the largest torques; however, the torques saturate for $\hlm{32}{16}$ at the large amplitudes. The saturation of the torque magnitude is only observed for $\hlm{32}{16}$, so for the present discussion, we will omit the $\hlm{32}{16}$ cases. Afterwards we will explain why we think $\hlm{32}{16}$ may be unique among the studied topographies. 

The increase of the torque with $\epsilon$ can be attributed to two mechanisms. First is the factor of $\epsilon$ in equation \eqref{e:ax_torque}, which suggests a straightforward linear relationship. Second is the change in the flow itself; both the speeds and morphology change with topography. 
In previous work, where a single bump was used, this second effect was negligible. When only the topographic amplitude was changed, flow speeds were almost unaffected and the only morphological difference was the presence of a small cyclonic patch over the bump. The linear scaling with $\epsilon$ was confirmed by \cite{tO25}. 
However, \cite{tO25} were able to test the effect of varied flow speed by testing different values of $\Rat$ and $Ek.$ A dynamic scaling law which accounted for changes in flow speed was proposed, 
\be
\Gstd \sim \epsilon\;Re^{2}Ek^{-1/3}.
\label{e:tq_dynamic}
\ee
The most turbulent simulations presented in that work indicate convergence toward this scaling. 


The $Ek^{-1/3}$ factor arises due to the dominant scales present in the convection. In order for these scales to generate torques, the topography must also contain $Ek^{-1/3}$ scales. The tomographic model and the single bump tested in \cite{tO25} contain these small scales; however, the single harmonic topographies do not.
The single harmonic topographies interact with only one mode, determined by the topographic wavelength rather than the convection. We will demonstrate that the $Ek^{-1/3}$ factor applies to spectrally broad topographies, but we first test the $Re^{2}$ dependence. 

Figure \ref{f:ekt}(b) shows the torques rescaled as in equation \eqref{e:tq_dynamic}, where the $Ek$ dependence is omitted. We find that the linear dependence on $\epsilon$ is roughly consistent with all topographic shapes except $\hlm{32}{16}$ since this case exhibits saturation for $\epsilon > 0.05$. 
Polynomial fits of the form $\Gstd = Re^{2} \;\epsilon^{\mu}$  yield power laws of $ \mu = [1.25\pm0.08,0.92\pm0.05,0.0\pm 0.1,$ and $1.15 \pm 0.04]$ for $\ls\hlm{2}{1},\hlm{8}{4},\hlm{32}{16},\text{ and }h_t\rs$ respectively. The reported uncertainties are determined from the covariance matrix associated with the polynomial fit.
Fitting all of the data (excluding $\hlm{32}{16}$) yields $\mu =0.94\pm 0.45,$ which agrees with the findings of \cite{tO25}, although the uncertainty is large because the magnitude of the torques strongly varies between topographic shapes. 
The dynamical scaling successfully accounts for changes in the flow speed, but it clearly does not collapse the data among the different topographies. 

As has been pointed out, for $\hlm{32}{16}$ only the $ m=16$ components of the pressure field will contribute to the torques (see equation \eqref{e:ax_torque}).
More precisely, only the component of the pressure that is one-half wavelength out of phase with $\hlm{\ell}{m}$ will contribute -- in this case it is the component proportional to $\sin \lb 16 \phi \rb$.
For a topography with sufficiently large amplitude, the largest component of the flow tends to lock to the geostrophic contours. The locked flow does not contribute to the torques since a non-zero phase shift with respect to the topography is necessary. 

\begin{figure}
	\begin{center}
		\subfloat[][]{\includegraphics[width=0.42\textwidth]{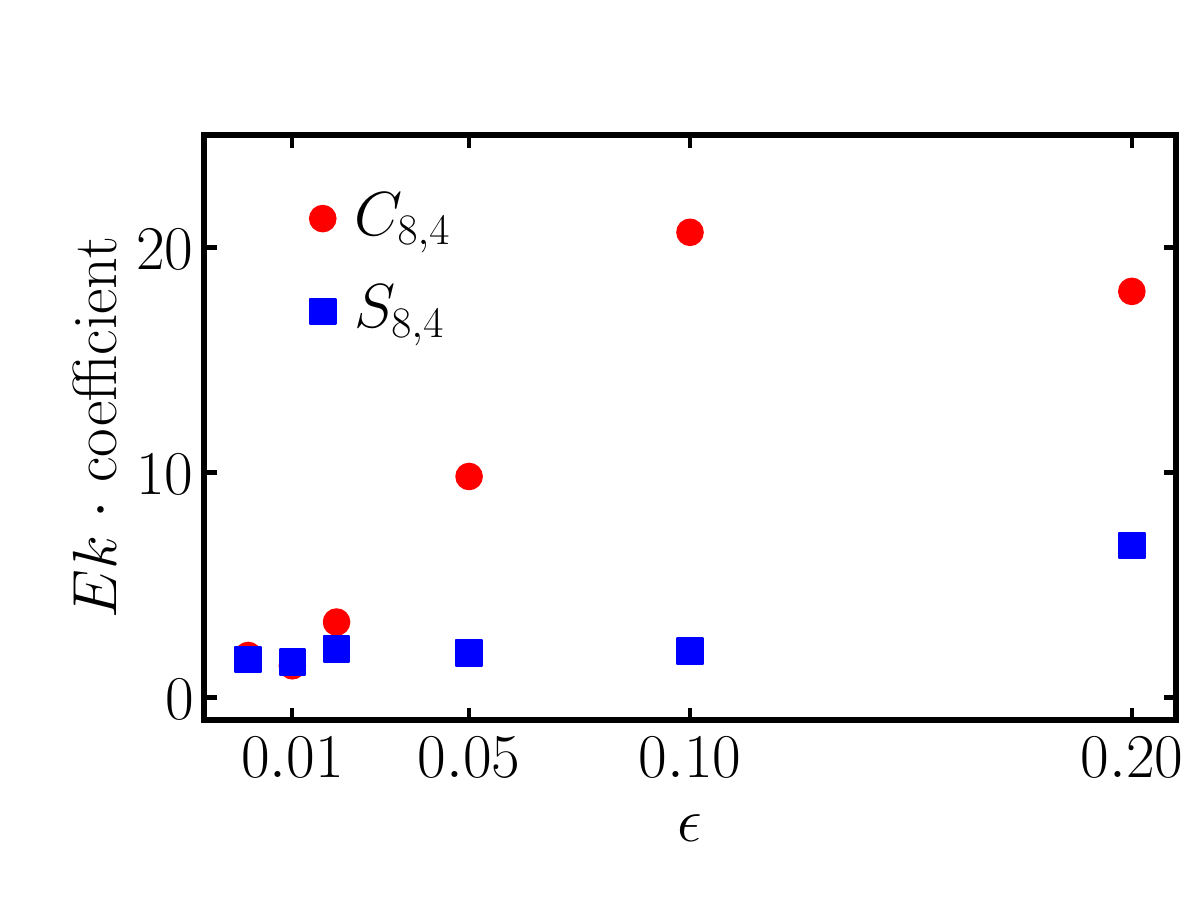}} \hspace{1cm}
		\subfloat[][]{\includegraphics[width=0.42\textwidth]{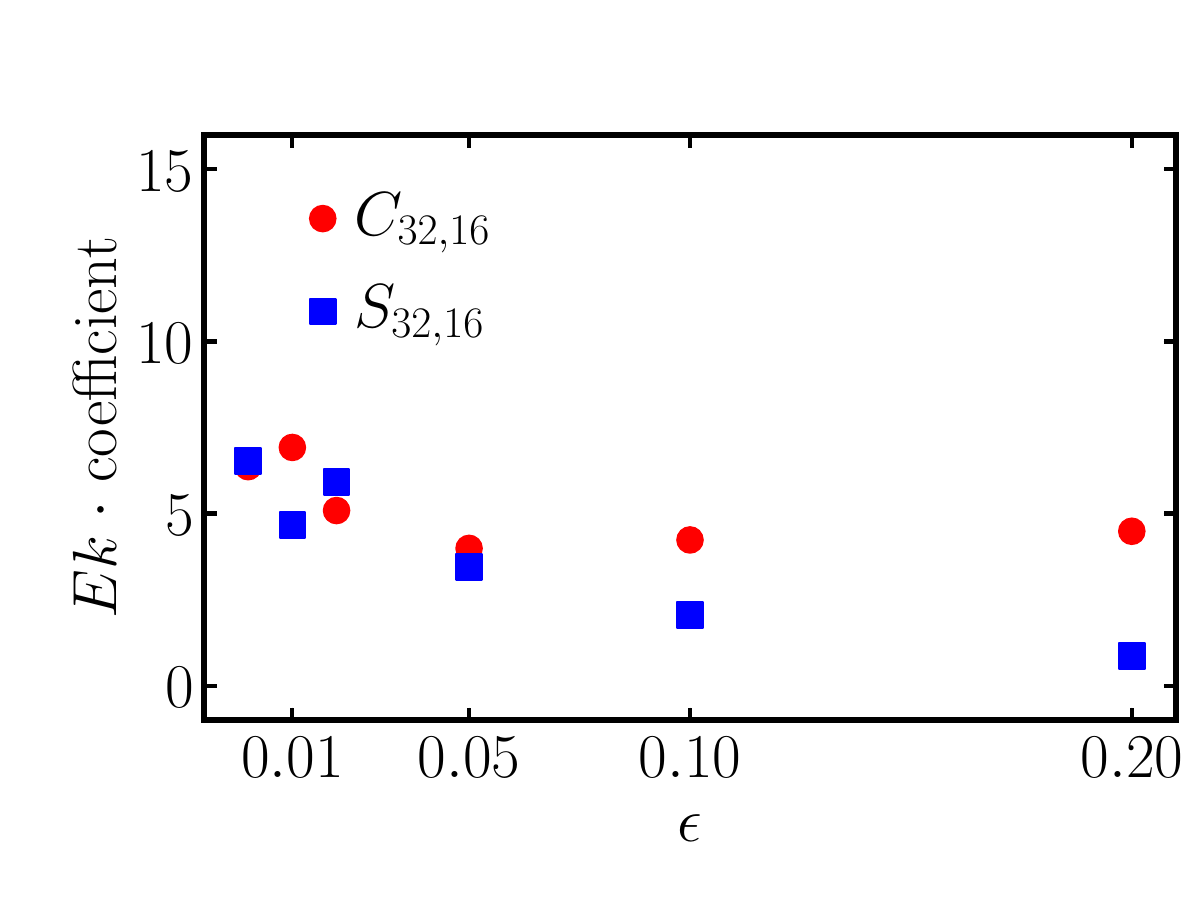}}\\
	\end{center}
	\caption{Rescaled magnitudes of the cosine $(C)$ and sine $(S)$ components for the indicated spherical harmonic of the pressure on the outer boundary for the (a) $\hlm{8}{4}$ and (b) $\hlm{32}{16}$ cases. 
	}
	\label{f:p_disp}
\end{figure}

\begin{figure}
	\begin{center}
		\subfloat[][]{	\includegraphics[width=0.48\textwidth]{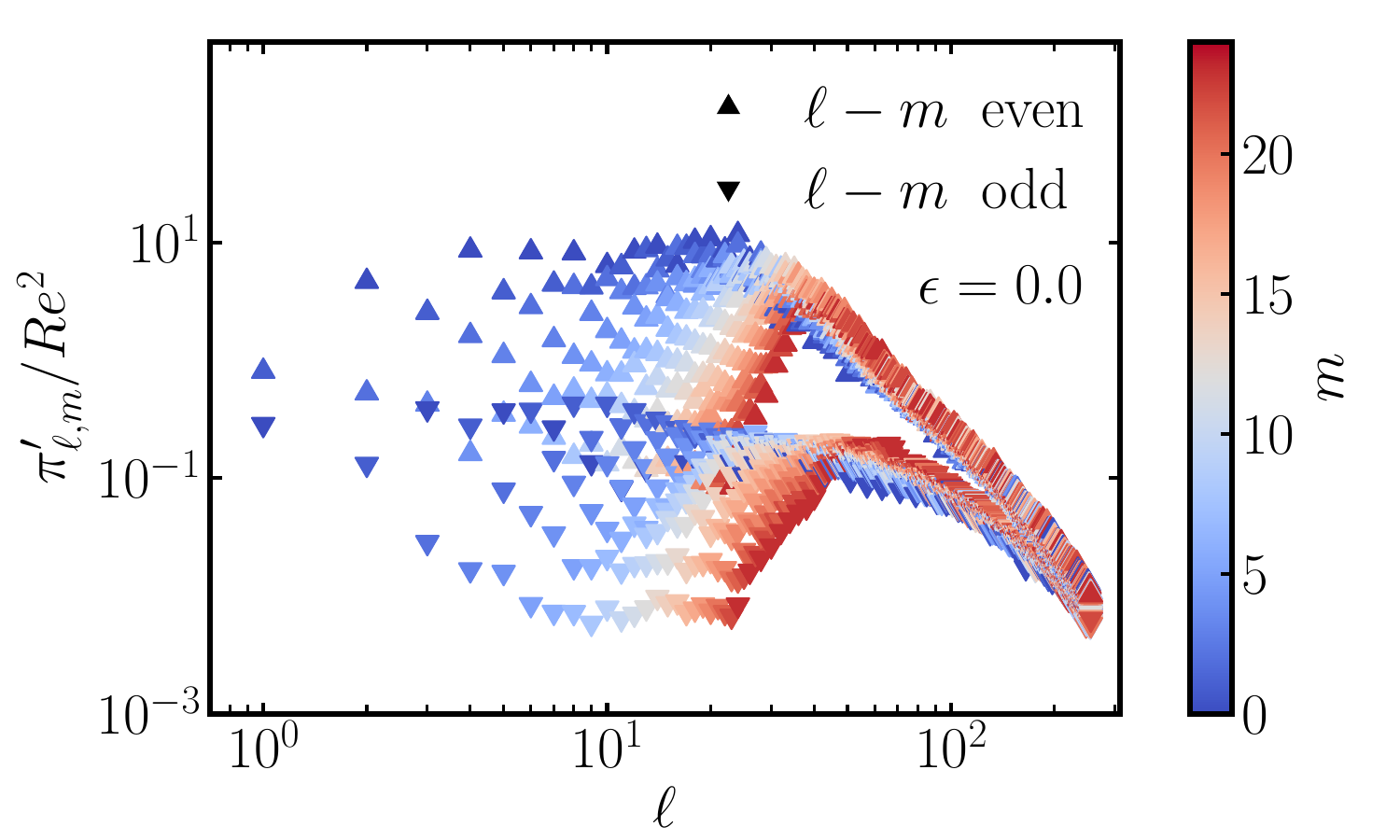}}
		\subfloat[][]{	\includegraphics[width=0.48\textwidth]{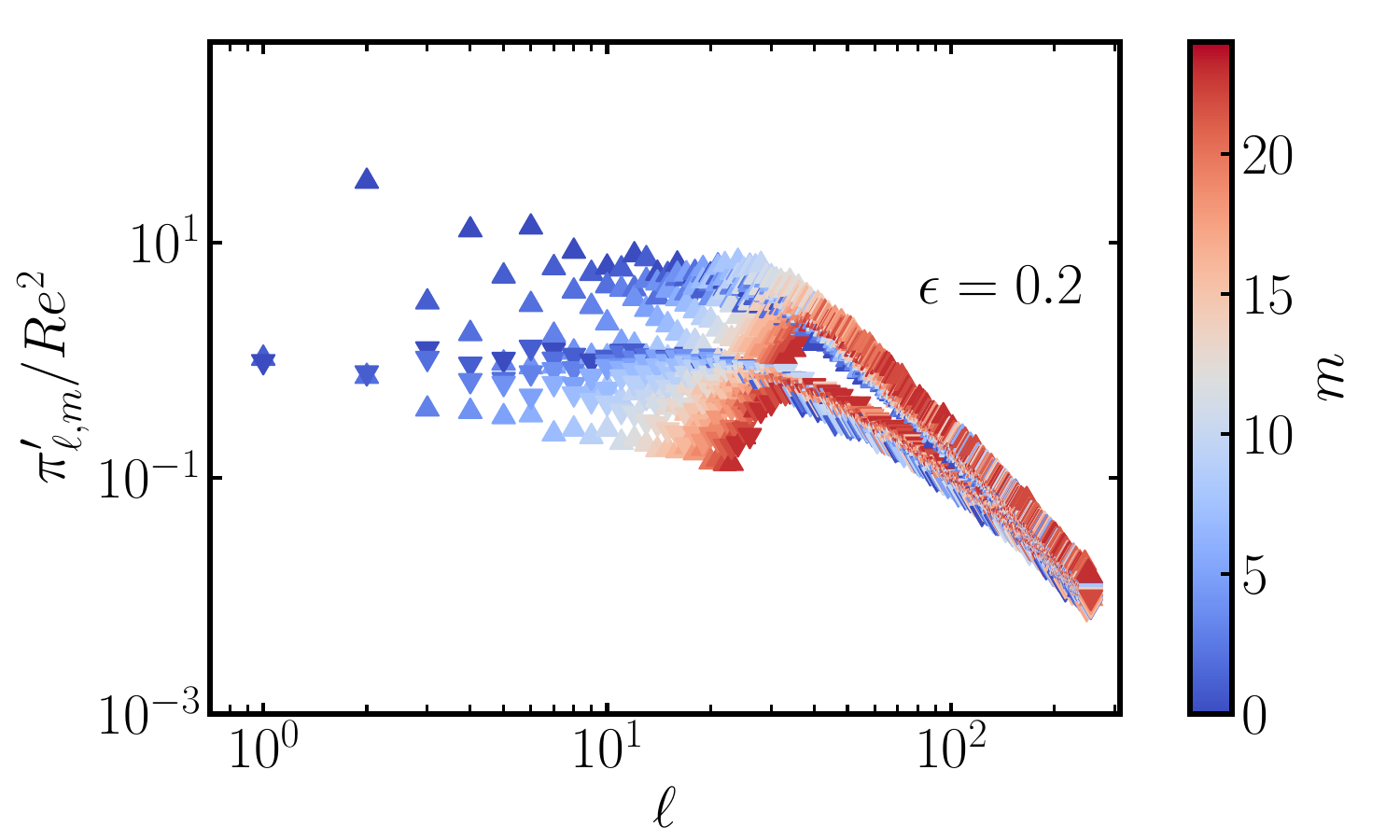}}\\
		\subfloat[][]{	\includegraphics[width=0.48\textwidth]{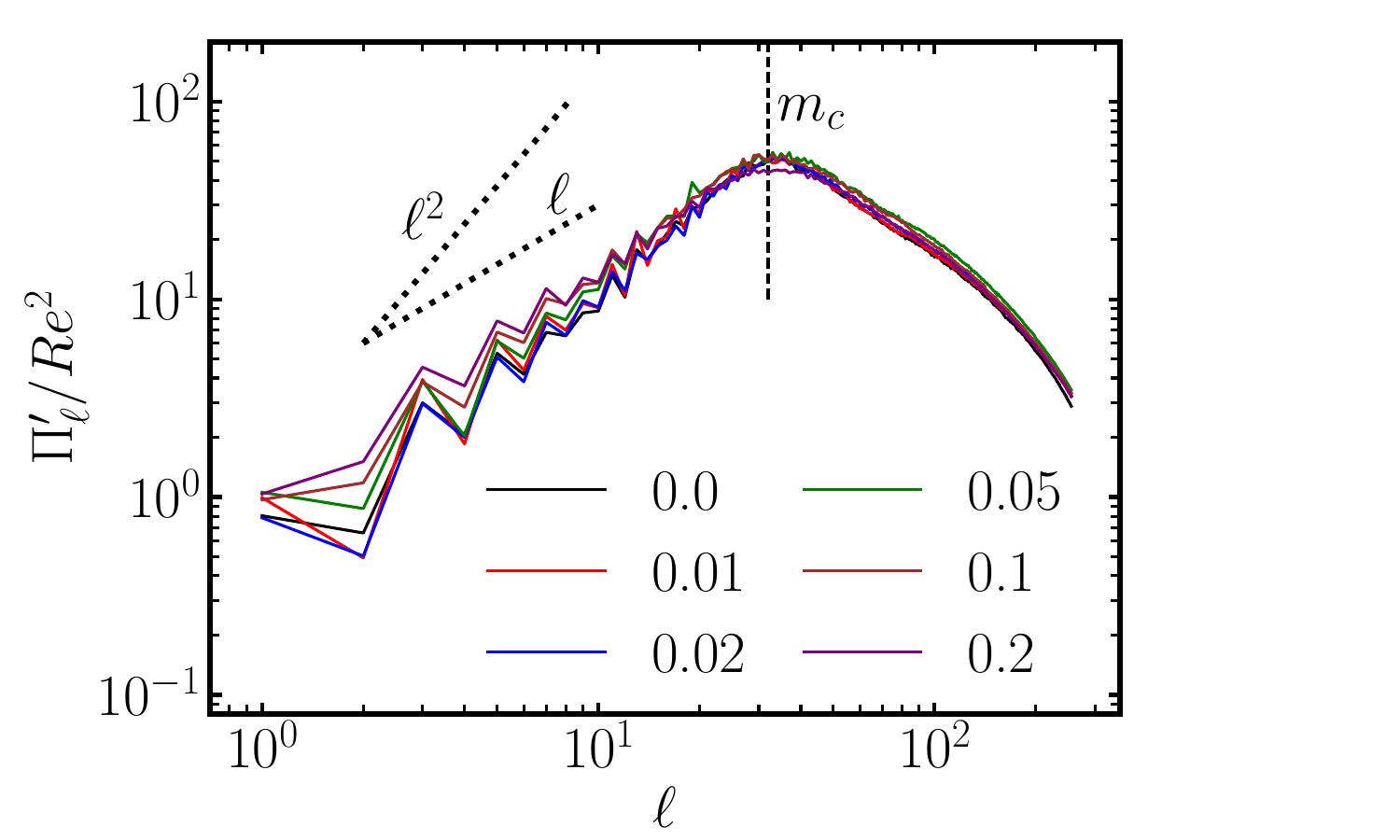}}
		\subfloat[][]{	\includegraphics[width=0.48\textwidth]{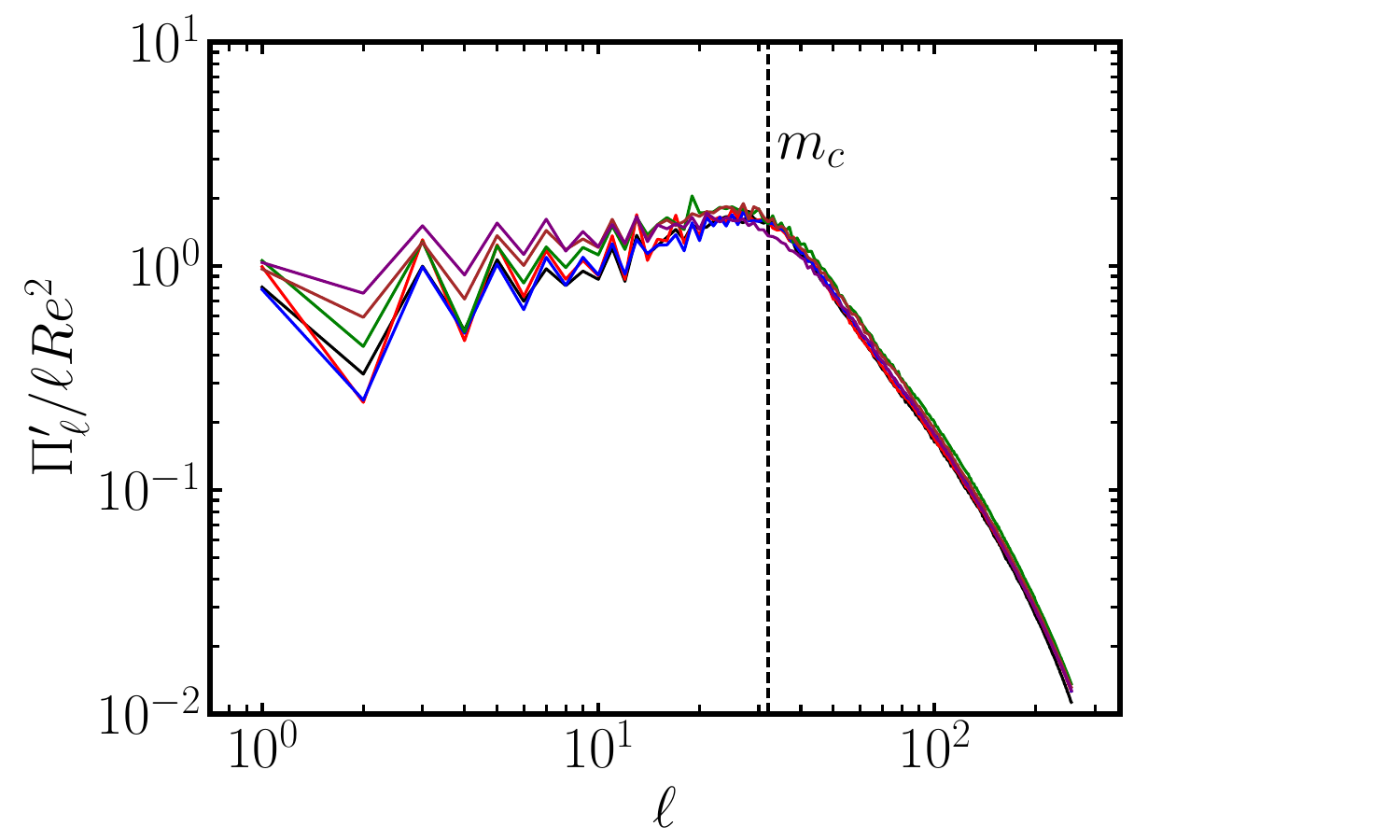}}\\
	\end{center}
	\caption{Power spectra of the fluctuating pressure. (a) Spectra for all $\ell$ up to $m = 24$ for $\epsilon  = 0.0$. We identify two distinct branches corresponding to even and odd values of $\ell-m$. (b) Same as (a) for $\epsilon  = 0.2$ and $h_t$. The branches are less distinguished because the boundary is no longer spherical and the spectra are calculated on $r_{cmb}$. (c) Fluctuating pressure spectra for the tomographic model across all amplitudes. Spectra are summed on $m$ and normalized by $Re^{2}.$ There is a clear distinction between even and odd values of $\ell$ due to the two branches identified in (a) and (b). (d) Same as (c) normalized by $\ell$.  }
	\label{f:tomo_spectra}
\end{figure}

Figures \ref{f:p_disp}(a) and \ref{f:p_disp}(b) show the power in the spherical harmonic component of the pressure that is relevant to the fluctuating torques for the $\hlm{8}{4}$ and $\hlm{32}{16}$, respectively. Within each panel, the red (blue) data corresponds to the power in the cosine (sine) components. 
As the topographic amplitude increases, there is a tendency for more power to be stored in the cosine component for both topographic shapes. This behavior is expected since the geostrophic component of the flow becomes stronger as the amplitude is increased.
However, for the $\hlm{32}{16}$ topography we find that the power in the sine component decreases for sufficiently large $\ep$, which explains why the torque becomes nearly independent of $\ep$.
This trend can be attributed to the tendency for the flow to become preferentially locked to the topography, indicating that there is a resonant interaction between the convection and the topography.
We expect this locking behavior is unique to the $\hlm{32}{16}$ topography because the topographic wavelength is relatively close to the critical wavelength.


Realistic CMB topography is likely characterized by a range of different wavelengths \citep{mP23,rM25,vD25}, and the torques cannot be understood in terms of a single mode in the pressure field. It is therefore important to understand how the fluctuating pressure spectrum within the flow field interacts with such topography. As shown in Figure \ref{f:ht_spec}, the tomographic model, $h_t$, is characterized by a broad power spectrum, so we limit the present discussion to only this topography.
We present the fluctuating pressure spectrum $\phatLM,$ which is calculated from the spherical harmonic decomposition of the fluctuating pressure field on the surface $r = r_{cmb}$ (see equations \eqref{e:phat} and \eqref{e:pfs}).
We note that the surface $r = r_{cmb}$ is not a spherical surface when topography is present. 
The top row of Figure \ref{f:tomo_spectra} shows the fluctuating pressure spectrum $\phatLM$ normalized by $Re^{2}$ for all $\ell$ up to $m = 24.$ 
Figure \ref{f:tomo_spectra}(a) is for $\epsilon  =0 $ and Figure \ref{f:tomo_spectra}(b) is for the tomographic model at  $\epsilon  = 0.2$. The spectra are separated into two distinct branches, which we identify as corresponding to even (upper branch) and odd (lower branch) values of $\ell-m.$ \TOr{There tends to be more power in the even-valued $\ell -m$ spectra because the flows are largely symmetric about the equator (axially invariant).
The branches are less distinguished for $\epsilon  = 0.2$ because the surface $r = r_{cmb}$ over which the spherical harmonic coefficients are calculated is no longer a sphere, although the flow structures themselves retain a similar equatorial symmetry.}
The spectra exhibit a peak, but the location of the peak is dependent on the order $m.$
Furthermore, the scaling of $\phatLM$ with $\ell$ depends on the value of $m$ (at least at low wavenumber).
Our goal is to determine a power law for $\phatLM$ and this is difficult over both the degree $\ell$ and order $m.$
In the bottom row of Figure \ref{f:tomo_spectra} we plot $\phatLM$ summed over $m\geq1$, for which we define
 \be
 \Pi_{\ell}^{\prime} = \sum_{m = 1}^{\ell}\phatLM.
 \label{e:sum_m}
 \ee
 Note that we omit the $m=0$ modes because they do not contribute to the torques.

 Figure \ref{f:tomo_spectra}(c) displays $\PhatLM$ across all topographic amplitudes for the tomographic model. $\PhatLM$ tends to be larger at odd $\ell$ than the nearby even $\ell$. This behavior is due to the tendency of $\phatLM$ to have more power when $\ell - m$ is even (compared to odd) and to the fact that there tends to be more power in smaller $m$. 
We note that if we calculate  $\PhatLM$ by summing over $m\geq2,$ then the values corresponding to even $\ell$ are larger than the nearby odd $\ell$, which illustrates the large contributions the low-$m$ modes make. 
 Regardless, we find that $\PhatLM$ increases up to a peak that is close to $m_{c}=32$. It appears that  $\ell^{2}$ is too strong a scaling for $\PhatLM$ while $\ell$ is slightly too weak. Figure \ref{f:tomo_spectra}(d) shows the same data from Figure \ref{f:tomo_spectra}(c) rescaled by $\ell.$ The data are nearly flattened, although the rescaled quantity still slowly increases up to about $m_{c}$. We performed a power-fit of the data up to $\ell = m_{c}$ to determine $\mu_{\ell}$ for
 \[\PhatLM\sim \ell^{\mu_{\ell}}.\]  
 We found $\mu_\ell={1.4}$ for the control case and $\mu_\ell={1.2}$ for $\epsilon  = 0.2.$
 This variation in $\mu_{\ell}$ may be due to changes in the flow morphology, the increased turbulence, or a combination of the two. Increased turbulence may equipartition the pressure fluctuations at the non-dissipative scales more efficiently. We note that perfect equipartition would set $\mu_{\ell}=1$ \citep{rK67}.
 Previous work has shown that as turbulence is increased, the width of the spectral peak in the kinetic energy spectra tends to broaden as the inverse cascade transfers kinetic energy into larger length scales \citep{cG19,cG25,sM21,tO23}.   
 The extent to which $\mu_{\ell}$ changes with $Re$ is beyond the scope of this paper, although we do stress that $Re>1200$ for all turbulent simulations, and that non-linearities should already be mixing the scales efficiently. The fact that the peak is near $m_{c}$ suggests that the critical wavelength still plays an important role in the underlying convection.

We now demonstrate how we can use $\PhatLM$ to approximate the torques. For the present discussion we will use an undetermined (i.e.~general) value of $\mu_{\ell}$. 
We also need a spectrum for geophysical topography. We will assume Kaula's rule \citep{wK66} holds, that is \[\eta_{\ell,m}\sim \ell^{-2},\]
 where $\eta_{\ell,m}$ are the spherical harmonic coefficients of some topography $h\lb \theta,\phi\rb.$ This approach has been used in previous work on topographic torques \citep{mP23,vD25} and, as we noted previously, is adequate for our tomographic model.
 We can estimate $\Gstd$ as
 \[
		 \Gstd \approx Re^{2}\epsilon \oint h\partial_{\phi}\sum_{\ell}\sum_{m=0}^{\ell} \phatLM P_{\ell}^{m} \lb\cos\theta\rb\lb \cos m\phi + \sin m\phi \rb    \;da
	 \]
 \begin{equation}
		       \approx Re^{2}\epsilon  \sum_{\ell}\sum_{m = 1}^{\ell}\eta_{\ell,m}\; m \;\phatLM .
	\label{e:gst_ap}
 \end{equation}
 This approximation almost certainly over-estimates the torques, because it ignores any cancellation in the integral in equation \eqref{e:ax_torque}; however, we expect that it may be sufficient to determine a scaling. 
 The data in Figure \ref{f:tomo_spectra} indicates that we can truncate the sums at $ \ell = m_{c},$ although it has been suggested that as turbulence is increased the convective length scales of the flow will increase \citep{jA20, cG25}. Therefore we adopt an undetermined cutoff value $\ell_{c}$. The scaling $\PhatLM\sim \ell^{\mu_{\ell}}$ indicates that $\phatLM\sim \ell^{\mu_{\ell}-1}$ since there are $\ell$ $\phatLM$'s which sum together to make $\PhatLM$.  Equation (\ref{e:gst_ap}) scales as
 \[
	 \Gstd \sim  Re^{2}\epsilon \sum_{\ell=0}^{\ell_{c}}\sum_{m=1}^\ell \ell^{\mu_{\ell}-1} \ell^{-2}m
 \]
 and after summing over the $O\lb \ell_{c}^{2}\rb $ modes yields
 \be
 \Gstd\sim Re^{2}\epsilon\;(\ell_{c})^{\mu_{\ell}}.
 \label{e:gst_ap2}
 \ee



  \begin{figure}
  	\begin{center}
  		\includegraphics[width=0.6\textwidth]{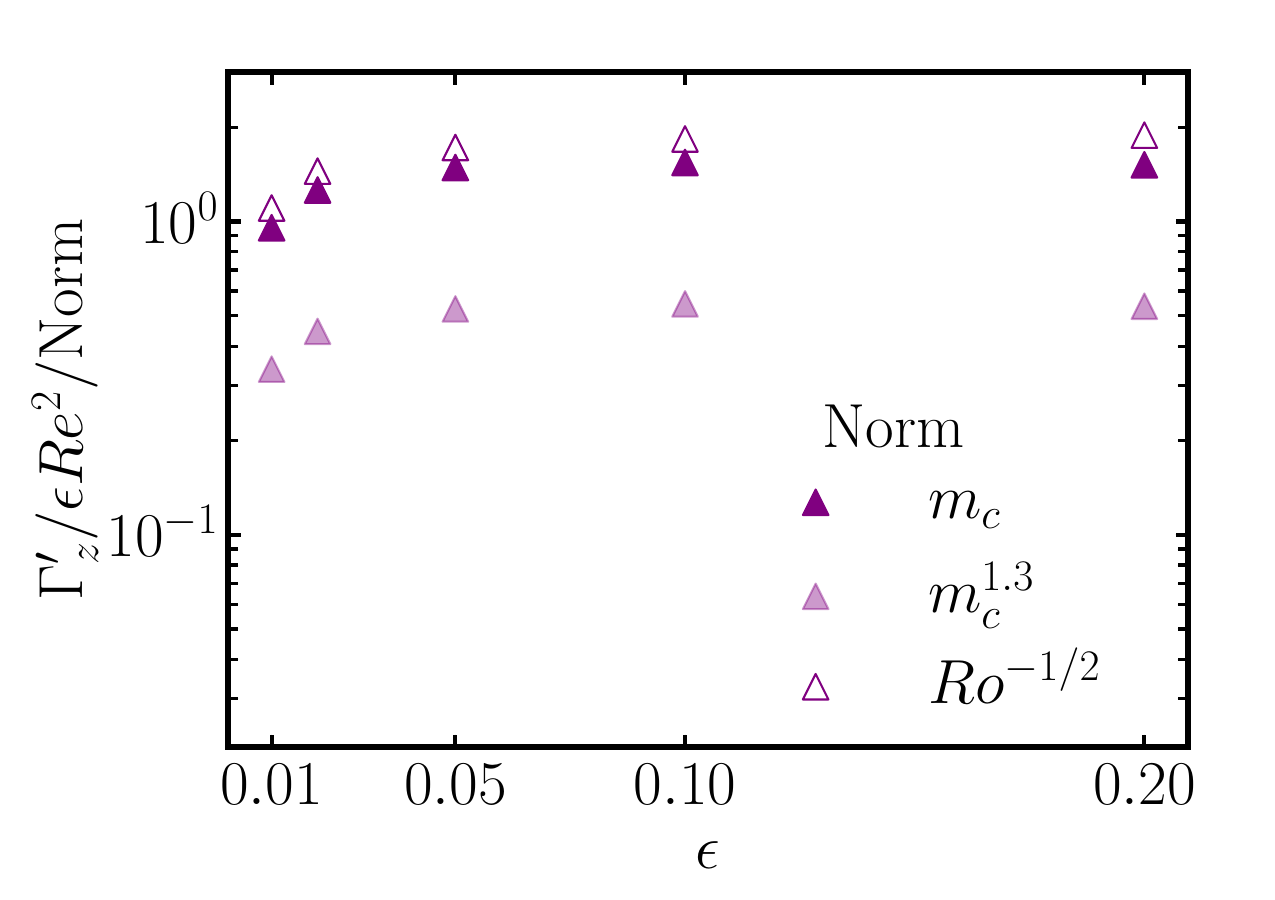}
  	\end{center}
	\caption{Standard deviation of the topographic torque for $h_{t}$, normalized by $\epsilon  Re^{2}$ and the normalization factor associated with the approximation in equation \eqref{e:gst_ap2}. We present normalizations by $m_{c},$ the critical wavelength for convection (closed), $Ro^{-1/2}$ (open), and $m_{c}^{1.3}$ (light-fill). 	}
  	\label{f:tq_bybump}
  \end{figure}

 An appropriate scaling is needed for $\ell_{c}.$ 
 In Figure \ref{f:tq_bybump} we present the torques rescaled as in equation \eqref{e:gst_ap} using two different values for $\ell_c.$ First we use $\ell_{c}=m_{c}$ which agrees with the data but may be inappropriate for more turbulent flows.
 \TOr{
 We also present a scaling based on the Rossby number $Ro=ReEk.$ The CIA theory \citep[e.g.][]{jA20} predicts that the inertial scale is $O(Ro^{-1/2})$, so we test $\ell_{c}\sim Ro^{-1/2}$.}
 Both scalings 
 \TOr{
 ($\ell_c \sim m_c$ and $\ell_c \sim Ro^{-1/2}$)
 } 
 assume $\mu_{\ell }=1,$ which is slightly less than the data in Figure \ref{f:tomo_spectra} suggests.
 Therefore we also rescale by $m_{c}^{1.3},$ where $\mu_\ell = 1.3$ is an average value determined from the power fits for $\PhatLM$. All three scalings yield quantities at or just below $O\lb 1\rb$ and we do not think it is appropriate to throw out any scaling because our approach is heavily simplified. 
  We merely present Figure \ref{f:tq_bybump} to provide justification for the procedure above.
  The $m_{c}$ and $Ro^{-1/2}$ scalings are nearly indistinguishable because $25.7<Ro^{-1/2}<27.9$ in our simulations. However, as turbulence increases the separation between $Ro^{-1/2}$ and $m_c$ will grow.
  We expect that further investigations on the scales of rotating flows, particularly the scales and spectra associated with the pressure fluctuation, will improve the provided scaling procedure.


\section{Discussion}
\label{s:disc}

	\noindent Four results from this study merit particular emphasis: 
	\begin{enumerate}
		\item We confirm the \cite{pB96} instability --- previously derived from a two-dimensional annulus model --- in a fully three-dimensional spherical geometry. 
		
		\item Topography deforms the geostrophic contours which enables buoyancy to perform work on the geostrophic flow and enhances heat and momentum transport. 
		\item Large-amplitude topography overrides the canonical viscous $Ek^{1/3}$ paradigm for the onset of motion and for the time-averaged geostrophic flow, even while the small-scale turbulence retains its viscous fingerprint. 
		\item The dynamical torque scaling $\Gstd \sim \epsilon Re^{2}$ established in our prior single-bump study is preserved across diverse global topographic shapes, supporting its generality. 
			We develop a generalized scaling theory that estimates topographic torques for arbitrary, spectrally broad topography, replacing single-mode approximations and yielding $O\lb 10^{18}\text{ N m}\rb$ torques when extrapolated to core conditions --- consistent with the magnitude required to drive observed $\Delta$LOD variations. 
	\end{enumerate}	
	We discuss each finding in turn below, alongside the secondary results that support them. Lastly, we discuss future directions for this work.


	\subsection{\TOr{Comparison with \cite{pB96}}}

One of the main effects of global scale topography is the ability of the buoyancy force to directly drive flow along geostrophic contours \citep{pB96}. In a purely spherical geometry, the geostrophic contours are circles about the origin, and a radial buoyancy force is unable to perform work on the component of the flow (the geostrophic flow) directed along these contours. 
Instead, geostrophic flow in a sphere is driven entirely by Reynolds stresses.
Non-axisymmetric topography deforms the contours and allows for buoyancy to directly force the geostrophic flow. 
Using the two-dimensional, quasi-geostrophic model for the Busse annulus, BS96 showed that this effect leads to novel instabilities that occur for Rayleigh numbers below the critical value which characterizes the convective Rossby waves that are the primary instability in the absence of topography \citep{sC61,cJ00}.
The simulations presented in this study confirm the presence of this instability in a global geometry and show qualitative agreement with BS96; however, we did not observe the oscillatory mode that BS96 predicted for small values of $\epsilon$. The quantitative discrepancies between our results and BS96 may not be surprising given that their analysis was for a simplified geometry and assumed short topographic wavelengths.
In the turbulent regime we find that the amplitude of the geostrophic flow may be significantly larger than the corresponding flow in the spherical case, depending on the shape and size of CMB topography. 



\subsection{\TOr{Effects on flow morphology and transport}}

The present study shows that globally distributed topography which deforms geostrophic contours leads to significant changes in turbulent mixing, as quantified through global heat ($Nu$) and momentum ($Re$) transport.
Over the investigated range of parameters we find changes in these measurements in excess of 100\% relative to the spherical case with no topography, indicating that the dynamics of this system are strongly influenced by boundary heterogeneities. 
This enhanced transport can be attributed to two separate effects. 
The first is the buoyancy work performed on the geostrophic flow, which is most prominent for topography with high equatorial symmetry and larger topographic amplitudes. 
The flow locks to the geostrophic contours and temperature variations along the contours can significantly increase the buoyancy work performed on the flow. Furthermore, the deformed geostrophic contours yield conveyor-belt-like structures that transport heat and momentum more effectively in comparison to the more chaotic mixing associated with convective Rossby waves.
The second effect is a resonance that we propose arises between the convective instability and the topographic wavelengths.

Equatorially anti-symmetric topography, corresponding to spherical harmonics where $\ell-m$ is odd \TOr{(in this work $\hlm{2}{1}$ and $\hlm{32}{17}$)}, generates smaller amplitude and larger wavenumber deviations \TOr{in the geostrophic contours} than equatorially symmetric topography.
Therefore, the locking phenomenon and the buoyancy interaction are suppressed when the topography is anti-symmetric about the equator.
Nevertheless, the $\hlm{32}{17}$ topography was found to elevate momentum and heat transport, which we attribute to the resonance with the convective instability at low wavenumber.
\TOr{
In the Earth, the LLSVPs are characterized by a large $\ell=2, \text{ }m = 2$ component at the equator \citep[e.g.][]{jR11}, which suggests that the resultant topography may generate larger buoyant forcing than studies without topography may suggest.
The LLSVPs also have power in the odd $\ell-m$ spectra; however, the convective scales in the core are likely much smaller than the scales of topography. Therefore we may not expect the elevated transport due to a resonance with the convective instability to contribute. }

\subsection{\TOr{Topography and the convective-scaling paradigm}}

The kinetic energy spectra show that certain topographic shapes lead to significant differences in the energetically active length scales of the flow. 
The spectra display prominent peaks at the topographic wavenumber and the higher order harmonics. 
The geostrophic contours contain power in the harmonic wavenumbers of the topography and the nonlinearities are also efficient at moving energy between these modes.
The $\hlm{32}{16}$ and $\hlm{32}{17}$ topographies show elevated kinetic energy at short wavelengths, which we attribute to the topographic wavenumber being near \TOr{$m_c$, the critical wavenumber for the onset of convection in a purely} spherical shell. 
Calculations of effective length scales from these spectra show that the critical wavenumber remains important for our investigated range of parameters. 
The $\hlm{8}{4}$ topography is characterized by a significant increase in the dominant length scale as the topographic amplitude is increased, which we attribute to the disparity in length scales between the topography and the underlying convection.


\TOr{
A natural question raised by these results is whether globally distributed topography fundamentally shifts the dominant force balance in rotating convection away from the viscous scales that set the canonical $Ek^{1/3}$ paradigm. The answer that emerges from our simulations is split. At large topographic amplitudes, topography clearly dominates the leading-order balance for the \textit{onset of motion} and for the \textit{time-averaged geostrophic flow}: the \cite{pB96} instability drives flow at Rayleigh numbers below the spherical-shell onset, and the geostrophic component of the time-averaged flow can increase by more than 100\% relative to the spherical case --- both effects driven by buoyancy work along deformed contours rather than by viscous selection. By contrast, the small-scale turbulence retains its viscous fingerprint across all our simulations: the kinetic energy spectra steepen near $m \approx m_c/2$ and the fluctuating pressure spectra peak near $m_c$ regardless of $\epsilon$, indicating that the underlying convective length scale remains set by $Ek^{1/3}$. A regime in which the topographic scale fully replaces the viscous scale --- disrupting the canonical paradigm even for the small-scale turbulence --- would require the topographic and viscous scales to separate asymptotically, which we cannot access at $Ek = 10^{-6}$. Such a regime is plausible at core conditions, where $Ek$ is many orders of magnitude smaller and topographic intrusions of $O(\text{km})$ are large compared to the viscous scale. Identifying when and how this transition occurs --- and whether it modifies the widely reported CIA scaling for $Re$ and $Nu$ \citep[e.g.][]{cG19} --- is a promising direction for future work.
}


\subsection{\TOr{Topographic torque scaling and flow locking}} 

The standard deviation of the topographic torque exhibits a roughly linear dependence on topographic amplitude for most topographic shapes. 
This linear scaling is not as clear in comparison to the previously studied localized bump given the more pronounced effects of topography on the global scale flows. 
For the particular case of the $\hlm{32}{16}$ topography we find that the torque essentially saturates near $\ep \approx 0.05$. 
In the $\hlm{32}{16}$ case the pressure fluctuations in the \TOr{torque-yielding} harmonic decreases with increasing $\epsilon$ for $\epsilon >0.02.$ We attribute this decrease to the convective instability preferentially locking to the mode in phase with the topography. 
The locking of the convection to the topographic pattern leads to no significant increase of the torque with $\ep$ despite the increase in flow speeds. 
Although a preference for the mode in phase with the topography was observed in all topographic shapes, only the $\hlm{32}{16}$ topography exhibited a decrease in the out-of-phase mode (the mode that yields torques).
We find that with the exception of the $\hlm{32}{16}$ topography, the dynamical scaling law $\Gamma_{z}^{\prime}\sim Re^{2}\epsilon$ agrees well with the data. 
Analysis of the frequency content of the time-dependent topographic torque showed that, as for the isolated bump investigated previously, the global scale topography seems to interact with frequencies that scale as $\omega\sim Ek^{-2/3}$, which is indicative of convection dominating the signal.

\subsubsection*{\TOr{A \MakeLowercase{generalized theory for spectrally broad topography}}}
We expect CMB topography to be spectrally broad, and since its shape is unconstrained, we present a procedure for determining the effect of topography based on a scaling of the topographic spectra (i.e., Kaula's rule). We find that the peak of the spherical harmonic spectra for the fluctuating pressure is located at a short wavelength which we associate with the convective scale. 
Our data suggest that the magnitude of the harmonics is $O\lb Re^{2}\rb$ at scales larger than the convective scale. 
Assuming a spectral distribution for the topography, the sum over spectral coefficients can be approximated to determine how the axial torque scales with the convective scale. 
We note that our procedure ignores cancellation in the integral. A more sophisticated approach could account for variations in the sign of the spectra of the topography and pressure fluctuations, but this is beyond of the scope of the present work. 
If we use Kaula's rule \citep{wK66} and approximate the convective scale as $O\lb Ek^{1/3}\rb, $ we determine that the axial torque fluctuations should scale as $\epsilon  Re^{2}Ek^{-\mu_\ell/3},$ where $\mu_\ell$ is a fit parameter chosen to satisfy $\PhatLM\sim\ell^{\mu_\ell}.$ Our data suggest that $\mu_\ell\approx1,$ although further investigation of the fluctuating pressure spectra is required. We note that if we set $\mu_\ell = 1,$ then we find $\Gstd\sim \epsilon Re^{2}Ek^{-1/3}$, which was reported in previous work where a localized topography was used. We think that the argument presented in this previous study, which relied solely on the determination of the magnitude of the pressure gradient and not the convolution with $h\lb \theta,\phi\rb $, may have been relevant to the particular parameter regime tested, but would not hold for arbitrary topography.
Nevertheless, the results presented in this study and the proposed scaling approach agree with the previously reported scaling which, when extrapolated to the parameter regime of the core yields the $O\lb 10^{18}\text{N}\;\text{m}\rb $ torques thought to be necessary to generate $\Delta$LOD \citep{tO25,HdV05}.


\subsection{\TOr{Future work}}

\TOr{Several aspects of this problem warrant further investigation, the first of which is the role of magnetic fields. The phenomena reported here are expected to have significant consequences for any resulting dynamo, and we are currently extending this work to the magnetohydrodynamic regime. For example, the broken equatorial symmetry associated with the $\hlm{32}{17}$ topography may yield a kinetic helicity that is not equatorially antisymmetric. The kinetic helicity of the flow, defined as the dot product of the velocity and vorticity vectors, is known to be a crucial ingredient for the generation of global-scale magnetic field \citep{eP55}; in spheres it is observed to be largely antisymmetric about the equator, so if topography disrupts this behavior the resultant magnetic fields will likely exhibit significant inhomogeneity. We also anticipate that the locking of flow to topography will generate a corresponding locked magnetic field, with ramifications for geomagnetic field observations.}

\TOr{Seismological investigations indicate that CMB topography is not well constrained. We have employed a simple adaptation of the `cluster analysis' or `voting map' procedure of \cite{sC16}, \cite{pK21}, and \cite{vL12} to construct what we have referred to as our tomographic model, enforcing topographic intrusions into the core wherever consensus on the presence of LLSVPs occurs. Our approach used sharp edges for the boundaries of the voting map, which effectively localized the resulting tomographic model. The consequence was a minimal change to the geostrophic contours relative to the sphere. While the topographic torques for this model exhibited similar behavior to those for the single-harmonic topographies (with the exception of $\hlm{32}{16}$), the changes in flow patterns were minimal compared to the $\hlm{8}{4}$ and $\hlm{32}{16}$ topographies at similar topographic amplitude. Future efforts will examine smoother topographic profiles, including models motivated by mantle convection studies (e.g., \citealp{tL07}; \citealp{tL10}). Since a definitive structure for CMB topography is not likely to be known in the near future, further investigation of the fluctuating pressure spectra is required to refine the proposed scaling procedure, which relies only on a power-law spectrum for the topography.}

\TOr{Exploring the effects of topography location is also of interest. Mid-latitude topography alters the geostrophic contours in a similar fashion to the topographies studied here, while equatorial topography splits the geostrophic flows between the two hemispheres, similar to how the inner core separates the contours at the northern and southern poles. Equatorial topography is therefore less likely to generate the buoyantly forced geostrophic flows that we examined in this work.}

\TOr{Finally, the timescale of topographic evolution differs across planetary systems. For the Earth, CMB topography is generated by mantle convection, which evolves on timescales of hundreds of millions of years; the topography is therefore effectively static on the core's convective timescale --- the regime modeled here. By contrast, in thin-shell systems such as icy moons, convection itself can directly generate topography on its own timescale: \cite{tG25} and \cite{jK22} demonstrated this in models of melting in subsurface oceans. The present work focused on a deep spherical shell ($1-\chi = O(1)$), appropriate for the Earth, but many of the dynamics are likely to carry over to thin shells, and the dynamic-topography regime relevant to icy moons constitutes a distinct direction for future work.}

\pagebreak

\section*{Acknowledgments}
This work was funded by the National Science Foundation through CSEDI grant EAR-2201595. The Stampede3 and Anvil supercomputers at the Texas Advanced Computing Center (TACC) and Purdue University, respectively, were made available through allocation PHY180013 from the Advanced Cyberinfrastructure Coordination Ecosystem: Services \& Support (ACCESS) program, which is supported by National Science Foundation grants \#2138259, \#2138286, \#2138307, \#2137603, and \#2138296 \citep{tB23}. Access to the Frontera supercomputer \citep{frontera} at TACC was made available through grant EAR24006. This work also utilized the Alpine high performance computing resource at the University of Colorado Boulder. Alpine is jointly funded by the University of Colorado Boulder, the University of Colorado Anschutz, and Colorado State University.
\section*{Conflict of Interest Disclosure}
\noindent
The authors declare that there are no conflicts of interest for this manuscript.
\section*{Open Research}
\subsection*{Data Availability}
\noindent
The datasets presented in the current study are available by request via the corresponding author.
\\ The data required to generate the visualizations are not available because they are too large to store.

\newpage
\appendix
\renewcommand{\thesection}{Appendix \Alph{section}} 
\counterwithin*{equation}{section}
\renewcommand{\theequation}{\Alph{section}\arabic{equation}}
\counterwithin*{table}{section}
\renewcommand{\thetable}{\Alph{section}\arabic{table}}
\counterwithin*{figure}{section}
\renewcommand{\thefigure}{\Alph{section}\arabic{figure}}
\section{Deformations to the geostrophic contours}
\label{a:oddveven}
We demonstrate that topography that is odd under the transformation \[\theta\rightarrow \pi - \theta\]
generates $O\lb \epsilon^{2}\rb $ variations to the geostrophic contours, while the harmonics that are even generate $O\lb \epsilon\rb $ variations. 
We restrict this analysis to geostrophic contours that form closed loops about the origin.
The following argument is general, but we will restrict the analysis to single harmonics, which is relevant to the present study. 
In this special case, we also obtain the leading order azimuthal dependence.

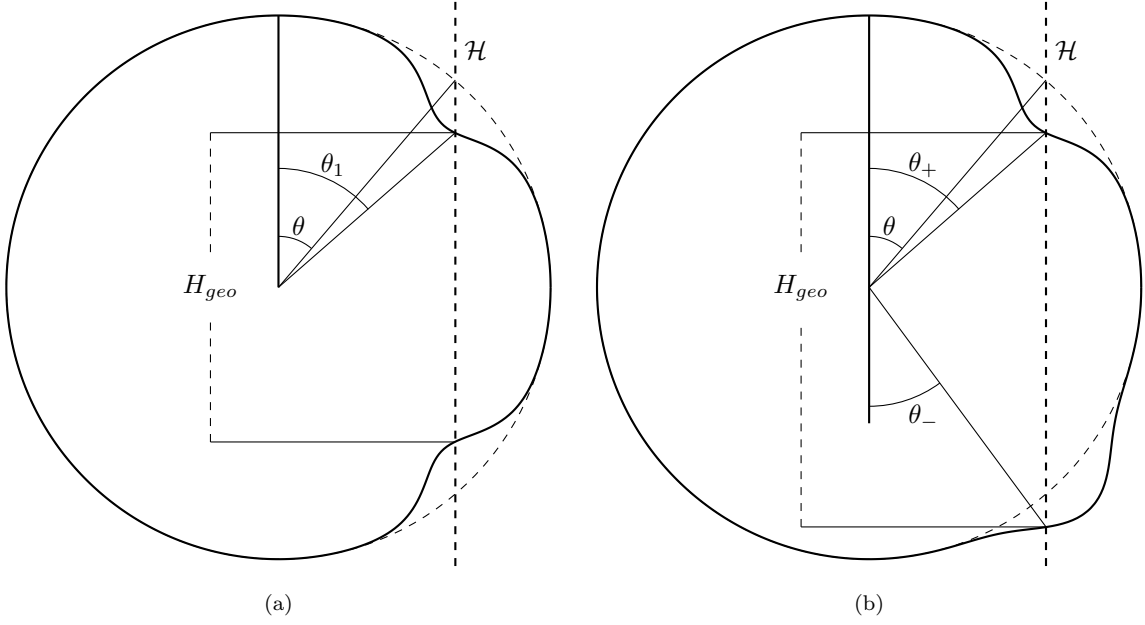
\begin{figure}
	\begin{center}
		\subfloat[][]{
\begin{tikzpicture}[scale = 0.9]

                \draw [color = black, dashed] (0,0) circle (4cm);
                \draw [color = black, thick,dashed] (2.6,4.2) --(2.6,0) ;
		\node at (2.9,3.5) {$\hc$};
		\draw [color = black, thick,dashed] (2.6,0) --(2.6,-4.2) node[left] {};
                \draw [color = black,thick] (0,0) --(0,4) node[left] {};
                \draw [color = black] (0,0) --(2.6,2.275) node[left] {};
                \draw [color = black] (0,0) --(2.6,3.05) node[left] {};
		\draw[color = black] (0,0.75) arc (90:50:0.75);
		\node at (0.3,0.9) {$\theta$};

		\draw[color = black] (0.00,1.75) arc (90:41:1.75);
		\node at (0.8,1.8) {$\theta_{1}$};
                \draw [color = black] (2.6,2.275) --(-1,2.275);
                \draw [color = black] (2.6,-2.275) --(-1,-2.275);
		\draw [color = black,dashed] (-1,-2.275)--(-1,-0.5);
		\draw [color = black,dashed] (-1,2.275)--(-1,0.5);
		\node at (-1,0) {$H_{geo} $};
        \draw [thick, color=black, domain=0:2*pi, samples=200, smooth]
	plot (xy polar cs:angle=\x r, radius={4-0.6*exp(-(\x-7*pi/4)^2/.05)-0.6*exp(-(\x-pi/4)^2/.05)});
\end{tikzpicture}
}
\quad
\subfloat[][]{
	\begin{tikzpicture}[scale = 0.9]

                \draw [color = black, dashed] (0,0) circle (4cm);
                \draw [color = black, thick,dashed] (2.6,4.2) --(2.6,0) ;
		\node at (2.9,3.5) {$\hc$};
		\draw [color = black, thick,dashed] (2.6,0) --(2.6,-4.2) node[left] {};
                \draw [color = black,thick] (0,-2) --(0,4) node[left] {};
                \draw [color = black] (0,0) --(2.6,2.275) node[left] {};
                \draw [color = black] (0,0) --(2.6,3.05) node[left] {};
		\draw[color = black] (0,0.75) arc (90:50:0.75);
		\node at (0.3,0.9) {$\theta$};

		\draw[color = black] (0.00,1.75) arc (90:41:1.75);
		\draw[color = black] (0.00,-1.75) arc (270:306.5:1.75);
		\node at (0.8,1.8) {$\theta_{+}$};
		\node at (0.8,-1.9) {$\theta_{-}$};
                \draw [color = black] (2.6,2.275) --(-1,2.275);
                \draw [color = black] (2.6,-3.525) --(-1,-3.525);
		\draw [color = black] (0,0)--(2.6,-3.525) ;
		\draw [color = black,dashed] (-1,-3.525)--(-1,-0.5);
		\draw [color = black,dashed] (-1,2.275)--(-1,0.5);
		\node at (-1,0) {$H_{geo} $};
        \draw [thick, color=black, domain=0:2*pi, samples=200, smooth]
	plot (xy polar cs:angle=\x r, radius={4+0.6*exp(-(\x-7*pi/4)^2/.05)-0.6*exp(-(\x-pi/4)^2/.05)});
\end{tikzpicture}
}
%
\end{center}
\caption{(a) Topography that is even under the transformation $\theta\rightarrow \pi -\theta.$ (b) Topography that is odd.  $H_{geo}$ is the geostrophic height.}
\label{f:sketch}
\end{figure}

A symmetry property of the spherical harmonics is that
\[\hlm{\ell}{m}\lb \pi,\phi\rb = \lb -1\rb ^{\ell - m}\hlm{\ell}{m}\lb \pi-\theta,\phi\rb.\] 
We separately consider even and odd values of $\ell - m.$ The respective setups are illustrated in Figure \ref{f:sketch}.
\subsection*{$\ell - m$ is even}
Consider the geostrophic height function $H_{geo}$ associated with the boundary defined by equation (\ref{e:rcmb}) with $h\lb \theta,\phi\rb =\hlm{\ell}{m}$. 
Define $\hc\lb \theta,\phi\rb $ to be the extended chord which intersects the spherical surface $r=r_{o}$ at $\lb \theta,\phi\rb $ and $\lb \pi-\theta,\phi\rb.$  
Define $H_{\ell,m}^{\epsilon} = H_{\ell,m}^{\epsilon}\lb \theta,\phi\rb, $ to be the length of the portion of $\hc\lb \theta,\phi\rb $ which terminates at $r = \rcmb.$
The subscripts and superscript on $H_{\ell,m}^{\epsilon}$ indicate the structure and amplitude of the topography respectively. That is $H_{\ell,m}^{\epsilon} = H_{geo}$ for topography $\hLM$ with amplitude $\epsilon .$
We ignore regions near the poles (where the geostrophic cylinders extend from inner core to outer core) and the equator (where $H_{\ell,m}^{\epsilon}\rightarrow0$). 

$\hc$ intersects the surface $r = \rcmb$ at the polar angle $\theta_{1},$ which is slightly different than $\theta.$ There is not a closed expression for $\theta_{1},$ but
\[r_{o}\sin\theta = \ls r_{o} - \epsilon\hLM\lb \theta_{1},\phi\rb \rs\sin\theta_{1} \]
Assume that $\theta_{1}=\theta + \delta_{1},$ where $\delta_{1} \ll1.$
\[
	r_{o}\sin\theta \approx\ls r_{o}-\epsilon \hLM\lb \theta + \delta_{1},\phi\rb \rs \lb \sin \theta+\delta_{1} \cos\theta\rb 
\] 
Taylor expanding $\hLM$ and matching lowest order terms requires
\[\delta_{1} = \frac{\epsilon}{r_{o}}\hLM\lb \theta,\phi\rb \tan \theta \]
By the symmetry, $\hc\lb \theta,\phi\rb $ intersects the bottom hemisphere at the polar angle $\pi -\theta -\delta_{1}.$ Therefore
\[H_{\ell,m}^{\epsilon} = 2\cos \theta_{1}\ls r_{o} - \epsilon \hLM\lb \theta_{1},\phi\rb \rs \approx 2r_{o}\cos\theta -2\epsilon \hLM\lb \theta,\phi\rb \lb \cos\theta + \tan\theta \sin\theta\rb \]
\begin{equation}
	\implies H_{\ell,m}^{\epsilon}\approx 2r_{o}\cos\theta - \epsilon\; \lb \cos\theta + \tan\theta\sin\theta\rb \frac{1}{Q_{\ell}^{m}}P_{\ell}^{m}\lb \theta\rb\;  \cos m\phi.
	\label{e:hg_even}
\end{equation}
More succinctly, the leading order $\phi$ dependence of $H_{\ell,m}^{\epsilon}$ is $\cos m\phi$ at $O\lb \epsilon\rb .$ 

\subsection*{$\ell - m$ is odd}

The setup is the same as the previous section, except now $\hLM\lb \theta,\phi\rb = -\hLM\lb \pi-\theta,\phi\rb.$
We therefore consider the northern $\lb +\rb $ and southern $\lb -\rb $ hemispheres separately. 
In the previous section, the subscript $1$ was used to identify the polar angle $\theta_{1}$ at which $\mathcal{H}$ intersects the spherical boundary. 
The symmetry about the equator meant that we could use $\theta_{1}$ in both hemispheres.
We must now distinguish between the two, so we adopt a $+/-$ notation to refer to the upper ($+)$ and lower $(-)$ hemispheres. 
\[\delta_{\pm}\approx \pm\frac{\epsilon}{r_{o}}\hLM\lb \theta,\phi\rb \tan\theta \]
The geostrophic height is 
  \[H_{\ell,m}^{\epsilon}\lb \theta, \phi\rb = \ls r_{o}-\epsilon \hLM\lb \theta_{+},\phi\rb \rs \cos\theta_{+} +\ls r_{o}-\epsilon \hLM\lb \pi-\theta_{-},\phi\rb \rs \cos\theta_{-} \]
    \[
	    \begin{aligned}
		    \approx &\ls r_{o} - \epsilon \lb \hLM\lb \theta,\phi\rb  + \delta_{+} \frac{\partial \hLM}{\partial\theta}|_{\theta\phantom{\pi -}}\rb \rs \lb \cos \theta - \delta_{+}\sin\theta\rb \\
			    &\ls r_{o} + \epsilon \lb \hLM\lb \theta,\phi\rb  - \delta_{+} \frac{\partial \hLM}{\partial\theta}|_{\pi-\theta}\rb \rs \lb \cos \theta + \delta_{+}\sin \theta\rb
    \end{aligned}
    \]
     Expanding and collecting,
\begin{equation}
        H_{\ell,m}^{\epsilon} = 2 r_{o}\cos \theta + \epsilon \delta_{+}\ls 2\hLM \lb \theta,\phi\rb \sin\theta -  \cos\theta\lb\frac{\partial\hLM }{\partial\theta }|_{\theta\phantom{\pi-}} +\frac{\partial\hLM }{\partial\theta }|_{\pi-\theta}\rb\rs + O\lb \epsilon^{3}\rb
    	    \label{e:hodd}
\end{equation}

    The second term contains the leading order azimuthal dependence. Note that $\delta_{+}$ contains a factor of $\cos m\phi$. We can write the above approximation for $H_{\ell,m}^{\epsilon}$ as \[H_{\ell,m}^{\epsilon}\lb \theta,\phi\rb= 2 r_{o} \cos\theta + \epsilon^{2}\cos 2m\phi \;g\lb \theta\rb, \]
where $g$ is determined from equation \eqref{e:hodd}. 
The azimuthal order is doubled and the deformations are $O\lb \epsilon^{2}\rb. $

\section{Stability curves}
\label{a:bs_stab}
In this appendix we provide the values used to generate the stability curves presented in Figure \ref{f:bs_stab}. The curves are from the analysis of \cite{pB96} (BS96), who use a different variable convention than presented in this study. For clarity, any variable with a star corresponds to the same variable, sans-star, in the BS96 paper.

The analysis adopts the Busse annulus, first proposed by \cite{fB70}, which approximates a mid-latitude annular shell within the sphere. Sinusoidal, single wavelength, topography is imposed on the caps, extending along the rotation direction. 
We consider the mid-latitudes and use $\theta = 45^{\circ}$ (that is the slope parameter describing the annulus endcaps is $\eta^{*}=1$). 
The radial scale is chosen to be on the order of the bump wavelength, assumed to be much smaller than the vertical. That is
\[D^{*}= \frac{1}{m}\ll 2 r_o \cos\lb\theta\rb = 1.62.\]
We note that this requirement is only weakly met for $m = 4.$
The critical parameters in the theory are $\Gamma^{*} $ and $\mathcal{R}^{*},$ which can be related to $\epsilon$ and $Ra$ respectively by
 \[\Gamma^{*} =  Ek^{-1/4} \lb\frac{1}{D^{*}}\rb^{1/2}\epsilon\cos\lb\theta\rb \frac{m}{D^{*}},\]
 \[\mathcal{R}^{*} = Ra Ek \;2r_{0}\cos\lb\theta\rb\frac{\eta^{*}}{2}.\] 
For slow radial variation (that is, $\beta^{*}=O\lb1\rb$ in the language of BS96), stationary convection (dotted lines in Figure \ref{f:bs_stab}) occurs above the curve drawn by 
\[\Gamma_{s}^{*} = \sqrt{\frac{m^{4} + m^{2} \mathcal{R}^{*^2}}{m^{2}\mathcal R^{*}}}.\]
Due to the quadratic dependency on $\mathcal{R}^{*},$ two values of $\mathcal{R}^{*}$ solve the above relation. The critical value is the lesser of the two, although both branches are shown in the Figure.
 At the smallest $\epsilon,$ the onset of convection is oscillatory (solid line in Figure \ref{f:bs_stab}), and occurs above the curve drawn by
 \[\Gamma_{s}^{*} = \sqrt{\frac{2 m^{2}}{\mathcal{R}^{*}}}.\]

 \section{Buoyancy work}
 \label{a:b_work}
 
\begin{figure}
	\begin{center}
		\subfloat[][]{\includegraphics[width=0.45\textwidth]{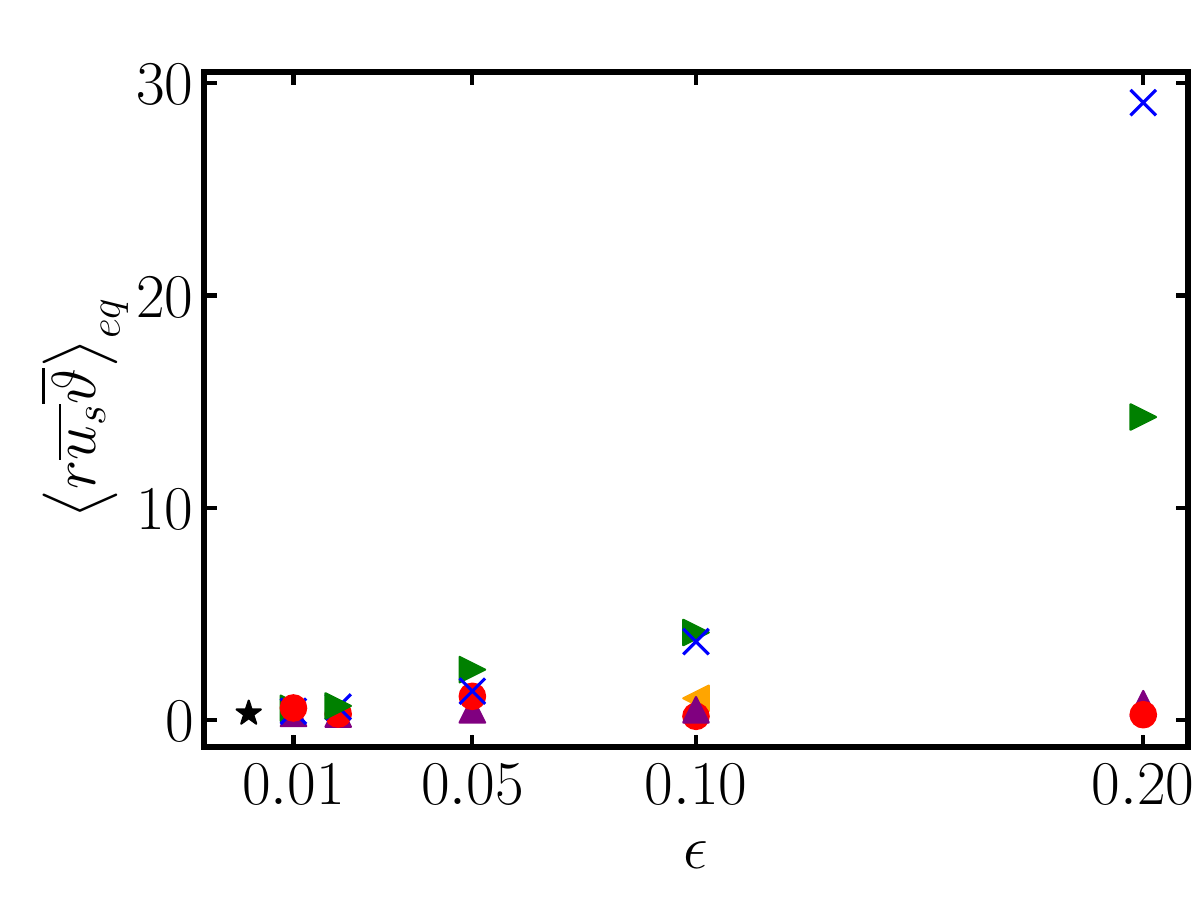}}
		\subfloat[][]{\includegraphics[width=0.45\textwidth]{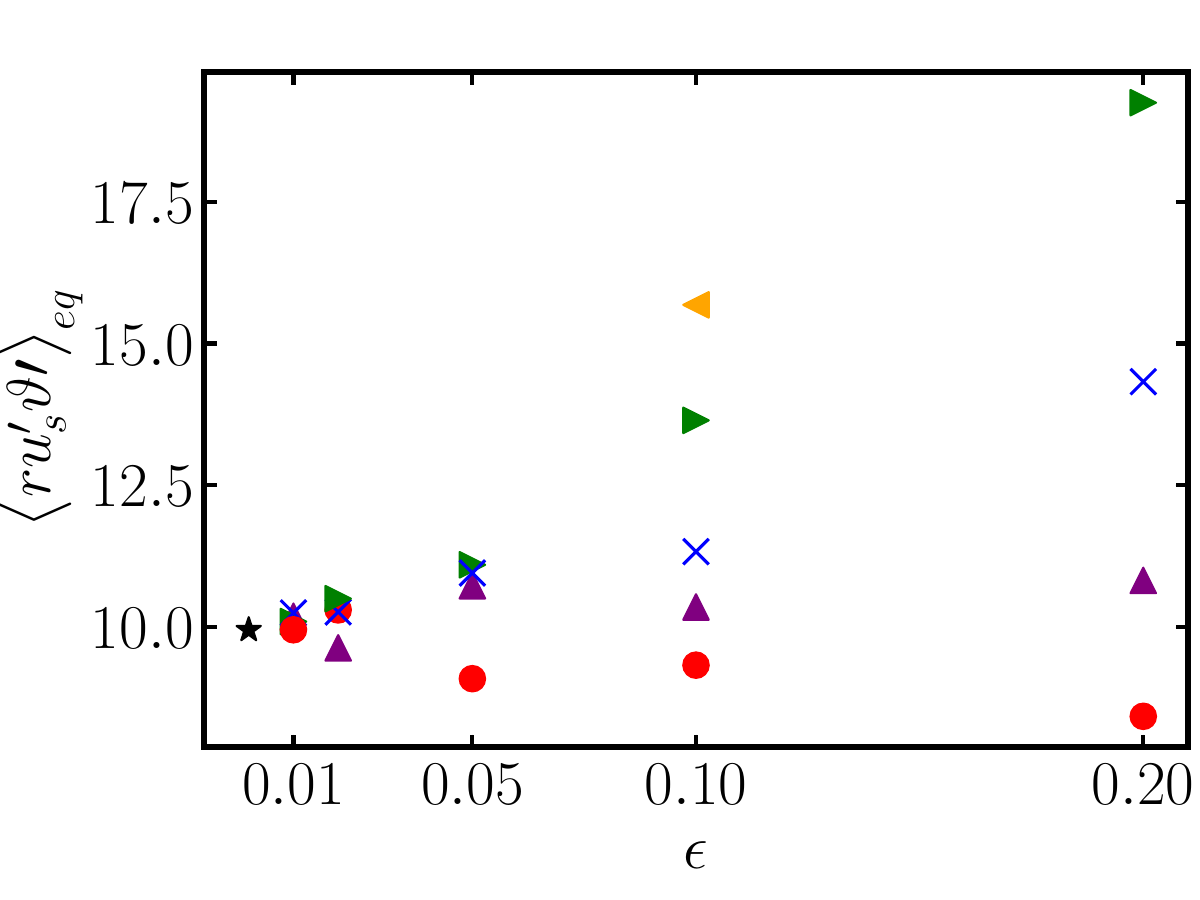}}
	\end{center}
	\caption{Equatorial average of the time averaged buoyancy work: (a) the first and (b) the second terms in equation (\ref{e:buoy_work}). 		
		(a) The mean-mean interaction is near zero for no topography, but tends to increase with $\epsilon.$ The effect is most pronounced in the $\hlm{32}{16}$ and $\hlm{8}{4}$ topographies where the contours are most deformed. (b) The fluctuating interaction, which increases with $\epsilon $ for $\hlm{32}{16}$ and $\hlm{8}{4}$ but tends to decrease for $\hlm{2}{1}$. Note that the vertical scales are different for the two plots, and that the increase due to the mean-mean interaction for $\hlm{32}{16}$ and $\hlm{8}{4}$ is significantly greater than any of the observed increases for the fluctuating interaction. 
	}
	\label{f:b_work}
\end{figure}
We can explain the elevated transport quantities in Figures \ref{f:gs}(a) and \ref{f:gs}(b) by calculating the work done by buoyancy.
The second term on the right-hand side of equation (\ref{e:mom}) is the buoyancy term, which we abbreviate as $\mathbf{F_b}.$
The time-averaged power delivered by buoyancy is $\overline{\mathbf{u}\cdot\mathbf{F_{b}}}$ which can be decomposed into the sum of averaged and fluctuating components (denoted with a prime $\cdot^\prime$). 
\begin{equation}
	\overline{\mathbf{u}\cdot\mathbf{F_b}}
	= \frac{Ra}{Pr}\frac{r}{r_{o}}\lb \overline{u_{r}}\overline{\vartheta} + \overline{u_{r}^{\prime}\vartheta^{\prime}} \rb .
\label{e:buoy_work}
\end{equation}
When the geostrophic contours are circular, the time averaged radial flow is near zero \TOr{(because the geostrophic flow is azimuthal)} and only the second term on the right-hand side of equation (\ref{e:buoy_work}) remains significant. 
As $\epsilon$ is increased, however, we expect the first term, which we refer to as the mean-mean interaction, to contribute.
\TOr{Figure \ref{f:b_work} shows the average values of the first (a) and second (b)} terms in equation (\ref{e:buoy_work}) averaged over the equatorial plane for all simulations performed at $\Rat = 40$.
The mean-mean interaction is significant in the large-amplitude ($>0.05$) topographies when the topographic shape is symmetric about the equatorial plane (i.e., $\hlm{8}{4}$ and $\hlm{32}{16}$). However, the effect is suppressed in the anti-symmetric topographies ($\hlm{2}{1}$ and $\hlm{32}{17}$) and the tomographic model.

With the exception of the $\hlm{2}{1}$ and tomographic models, the fluctuating component of equation \eqref{e:buoy_work} increases with $\epsilon$ for all topographies, although the effect is smaller than the mean-mean interaction. 
\TOr{
	It is difficult to assign a causal relationship between the global transport quantities and the fluctuating buoyancy work.
	As speeds increase we expect the buoyancy work (via the factor of $\mathbf{u}$) to increase, although one could conversely argue that excess buoyancy work yields additional kinetic energy and therefore greater flow speeds. 
	There is a clear mechanism to explain the increase of the mean-mean interaction with increasingly deformed geostrophic contours (see equation \eqref{e:m_avg2}),
	but the interpretation for the fluctuating component is less clear.
In light of the kinetic energy spectra (Figure \ref{f:spectra}) we primarily attribute the increased fluctuating buoyancy work (which is most pronounced in the $\hlm{32}{16}$ and $\hlm{32}{17}$ topographies) to a resonance between the convective wavelengths and the topography, which we believe generates the ``bellies'' that we observe in the kinetic energy spectra for $\hlm{32}{16}$ and $\hlm{32}{17}$, although we stress that increased mean flows can generate increased fluctuating buoyancy work via non-linearities in the flow.
}

\TOr{ 
	Lastly, we point out that in the particular case of $\hlm{32}{17}$ we can attribute the elevated values of $Re$ and $Nu$ (Figure \ref{f:gs}) solely to the fluctuating buoyancy work, since the mean-mean interaction is near 0 (and unchanged from the control).
This result corroborates the data from Figure \ref{f:re_geo_a}(a) and Figure \ref{f:re_geo_a}(b), which displays an unchanged geostrophic Reynolds number and an elevated ageostrophic Reynolds number. 
}


 \section{Viscous torques}
 \label{a:visc_t}
\begin{figure}
	\begin{center}
		\includegraphics[width=0.5\textwidth]{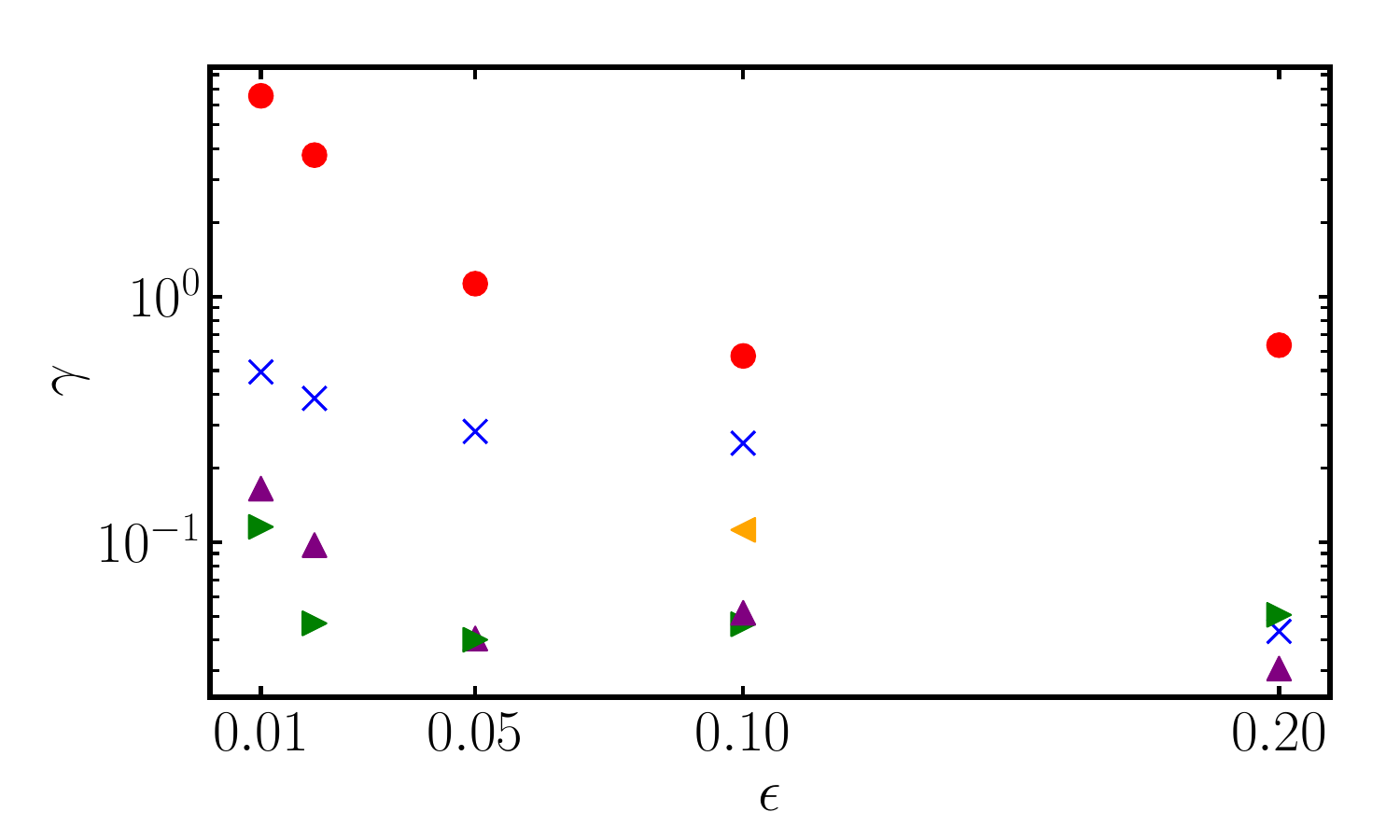}
	\end{center}
	\caption{Ratio of the standard deviations of the viscous to topographic torques ($\gamma$) calculated at the CMB. We plot against topographic amplitude $\epsilon .$ Colors and markers correspond to the legend in Figure \ref{f:gs}. Values are less than unity for all parameters except for the $\hlm{2}{1}$ topography at small $\epsilon$, which is associated with the smallest topographic torques. We expect $\gamma$ to decrease with $\Rat$ \citep{tO25}.}
	\label{f:viscrat}
\end{figure}
Previous work \citep[e.g.][]{bB07} has suggested that viscous torques do not play a large role in variations in LOD. In our previous study, we found that the ratio of viscous to topographic torques decreased as $\Rat$ increased. Therefore, viscous torques were not a major focus of the present work.
In Figure \ref{f:viscrat} we show $\gamma,$ the ratio of the standard deviations of the viscous to topographic torques. 
Except for the $\epsilon<0.1$, $\hlm{2}{1}$ topographies, where the topographic torques are very small, we find that $\gamma<1,$ indicating that topographic torques dominate over viscous torques. 
\cite{pR12} indicate that viscous torques should scale as $Re Ek^{-1/2}$ which is $O\lb 10^{6}\rb$ in our simulations and about the same size as our topographic torques. However, in the core $Re Ek^{-1/2}$ is likely much smaller than $\Gstd \sim Re^{2}m_{c}\;\epsilon$.
\def\newblock{\hskip .11em plus .33em minus .07em} 
\bibliography{References,journal_full}

\renewcommand{\thetable}{\arabic{table}}

\begin{table}
\begin{center}
\begin{tabular}{  c c c c c c  }
\hline
$\epsilon$ &$0.01$ &$0.02$ & $0.05$ &$0.10$ & $0.2$ \\
\hline
\hline
$\hlm{2}{1}$& 29.74 &29.74 &29.74 &29.74 &29.75 \\
$\hlm{8}{4}$& 29.74 &29.74 &29.74 &29.74 &29.75 \\
$\hlm{32}{16}$&29.74 &29.74 &29.74 &29.75 &29.76 \\
$\hlm{32}{17}$&29.74 &29.74 &29.74 &29.74 &29.75 \\
$h_{t}$& 30.01 &30.33 &31.31 &32.99 &36.47 \\
\hline
\end{tabular}
\end{center}
\caption{The surface areas of the outer surfaces for all geometries}
\label{t:sa}
\end{table}
\begin{table}
\begin{center}
\begin{tabular}{c c c c c c c c c c}
\hline
$\epsilon$ & Topography & $\Rat$ & $Ek$ & $\vartheta(r_{cmb})$ & $Re$ & $Re_w$ & $Re_{\gchar}$ & $Re_{\achar}$ & $Nu$\\
\hline
\hline
$0.0$ & -- & $40$ & $10^{-6}$ & 0 &$1278$ & $532.3$ & $936.1$ & $721.9$ & $6.1$ \\
 $0.01$ & $\hlm{2}{1}$ & $40$ & $10^{-6}$ & 0 &$1293$ & $540.9$ & $950.2$ & $729.0$ & $5.8$ \\
 $0.02$ & $\hlm{2}{1}$ & $40$ & $10^{-6}$ & 0 &$1293$ & $545.3$ & $946.9$ & $738.8$ & $5.8$ \\
 $0.05$ & $\hlm{2}{1}$ & $40$ & $10^{-6}$ & 0 &$1290$ & $545.2$ & $938.9$ & $728.0$ & $5.9$ \\
 $0.1$ & $\hlm{2}{1}$ & $40$ & $10^{-6}$ & 0 &$1298$ & $566.0$ & $910.9$ & $720.7$ & $6.1$ \\
 $0.2$ & $\hlm{2}{1}$ & $40$ & $10^{-6}$ & 0 &$1338$ & $602.6$ & $913.1$ & $710.5$ & $6.4$ \\
 $0.01$ & $\hlm{8}{4}$ & $40$ & $10^{-6}$ & 0 &$1303$ & $539.3$ & $964.7$ & $733.5$ & $5.8$ \\
 $0.02$ & $\hlm{8}{4}$ & $40$ & $10^{-6}$ & 0 &$1333$ & $545.4$ & $1006$ & $735.8$ & $5.9$ \\
 $0.05$ & $\hlm{8}{4}$ & $40$ & $10^{-6}$ & 0 &$1590$ & $567.2$ & $1257$ & $747.3$ & $6.2$ \\
 $0.1$ & $\hlm{8}{4}$ & $40$ & $10^{-6}$ & 0 &$1760$ & $558.8$ & $1503$ & $730.9$ & $6.4$ \\
 $0.2$ & $\hlm{8}{4}$ & $40$ & $10^{-6}$ & 0 &$2484$ & $726.9$ & $2156$ & $836.4$ & $10.0$ \\
 $0.01$ & $\hlm{32}{16}$ & $40$ & $10^{-6}$ & 0 &$1262$ & $547.6$ & $870.5$ & $725.6$ & $5.9$ \\
 $0.02$ & $\hlm{32}{16}$ & $40$ & $10^{-6}$ & 0 &$1304$ & $561.5$ & $879.1$ & $734.2$ & $6.1$ \\
 $0.05$ & $\hlm{32}{16}$ & $40$ & $10^{-6}$ & 0 &$1647$ & $616.5$ & $1339$ & $709.2$ & $7.0$ \\
 $0.1$ & $\hlm{32}{16}$ & $40$ & $10^{-6}$ & 0 &$1792$ & $739.8$ & $1448$ & $885.7$ & $9.0$ \\
 $0.2$ & $\hlm{32}{16}$ & $40$ & $10^{-6}$ & 0 &$2023$ & $811.8$ & $1657$ & $1091$ & $13.1$ \\
 $0.1$ & $\hlm{32}{17}$ & $40$ & $10^{-6}$ & 0 &$1679$ & $920.8$ & $957.9$ & $955.2$ & $10.3$ \\
 $0.01$ & $h_{t}$ & $40$ & $10^{-6}$ & 0 &$1305$ & $538.2$ & $942.4$ & $738.5$ & $6.1$ \\
 $0.02$ & $h_{t}$ & $40$ & $10^{-6}$ & 0 &$1288$ & $525.9$ & $934.0$ & $706.3$ & $6.4$ \\
 $0.05$ & $h_{t}$ & $40$ & $10^{-6}$ & 0 &$1309$ & $562.4$ & $929.5$ & $759.1$ & $6.4$ \\
 $0.1$ & $h_{t}$ & $40$ & $10^{-6}$ & 0 &$1379$ & $577.8$ & $958.1$ & $756.5$ & $6.8$ \\
 $0.2$ & $h_{t}$ & $40$ & $10^{-6}$ & 0 &$1510$ & $631.7$ & $1073$ & $781.2$ & $7.1$ \\
 \hline
 \end{tabular}
\end{center}
\caption{Run parameters for turbulent simulations. The aspect ratio is $\chi = 0.35$ and the Prandtl number is $Pr = 1$. Resolution was $230400$ elements each with $8^3$ grid points ($1.18\times10^{8}$ total points).}
\label{t:p_space}
\end{table}
\begin{table}
\begin{center}
\begin{tabular}{c c c c c c c c c c}
\hline
$\epsilon$ & Topography & $\Rat$ & $Ek$ & $\vartheta(r_{cmb})$ & $Re$ & $Re_w$ & $Re_{\gchar}$ & $Re_{\achar}$ & $Nu$\\
\hline
\hline
$0.01$ & $\hlm{8}{4}$ & $0.01$ & $10^{-6}$ & 0&$*$ & $*$ & $*$ & $*$ & $*$ \\
 $0.01$ & $\hlm{8}{4}$ & $0.1$ & $10^{-6}$ & 0&$*$ & $*$ & $*$ & $*$ & $*$ \\
 $0.01$ & $\hlm{8}{4}$ & $1.0$ & $10^{-6}$ & 0&$13.0$ & $6.89\times 10^{-1}$ & $20.2$ & $1.53$ & $1.00$ \\
 $0.02$ & $\hlm{8}{4}$ & $0.01$ & $10^{-6}$ & 0&$*$ & $*$ & $*$ & $*$ & $*$ \\
 $0.02$ & $\hlm{8}{4}$ & $0.1$ & $10^{-6}$ & 0&$*$ & $*$ & $*$ & $*$ & $*$ \\
 $0.02$ & $\hlm{8}{4}$ & $1.0$ & $10^{-6}$ & 0&$37.6$ & $2.33$ & $34.1$ & $3.21$ & $1.03$ \\
 $0.05$ & $\hlm{8}{4}$ & $0.01$ & $10^{-6}$ & 0&$*$ & $*$ & $*$ & $*$ & $*$ \\
 $0.05$ & $\hlm{8}{4}$ & $0.1$ & $10^{-6}$ & 0&$10.2$ & $2.35\times 10^{-1}$ & $13.8$ & $4.38\times 10^{-1}$ & $1.02$ \\
 $0.05$ & $\hlm{8}{4}$ & $1.0$ & $10^{-6}$ & 0&$55.6$ & $4.06$ & $50.7$ & $5.70$ & $1.15$ \\
 $0.1$ & $\hlm{8}{4}$ & $0.01$ & $10^{-6}$ & 0&$4.52$ & $7.26\times 10^{-2}$ & $4.23$ & $7.97\times 10^{-2}$ & $1.04$ \\
 $0.1$ & $\hlm{8}{4}$ & $0.1$ & $10^{-6}$ & 0&$25.0$ & $7.56\times 10^{-1}$ & $23.2$ & $9.57\times 10^{-1}$ & $1.18$ \\
 $0.1$ & $\hlm{8}{4}$ & $1.0$ & $10^{-6}$ & 0&$87.9$ & $7.49$ & $82.1$ & $10.0$ & $1.43$ \\
 $0.2$ & $\hlm{8}{4}$ & $0.01$ & $10^{-6}$ & 0&$8.11$ & $4.02\times 10^{-1}$ & $7.58$ & $5.36\times 10^{-1}$ & $1.19$ \\
 $0.2$ & $\hlm{8}{4}$ & $0.1$ & $10^{-6}$ & 0&$37.6$ & $2.64$ & $34.4$ & $3.80$ & $1.48$ \\
 $0.2$ & $\hlm{8}{4}$ & $1.0$ & $10^{-6}$ & 0&$135$ & $18.7$ & $120$ & $29.9$ & $2.23$ \\
\hline
 \end{tabular}
\end{center}
\caption{Run parameters for subcritical simulations.  Values denoted with a $*$ correspond to runs that were prohibitively expensive to saturate. The aspect ratio is $\chi = 0.35$ and the Prandtl number is $Pr = 1$. Resolution was $33600$ elements each with $8^3$ grid points ($1.72\times10^{7}$ total points).}
\label{t:p_space2}
\end{table}\begin{table}
\begin{center}
\begin{tabular}{c c c c c c c c c c}
\hline
$\epsilon$ & Topography & $\Rat$ & $Ek$ & $\vartheta(r_{cmb})$ & $Re$ & $Re_w$ & $Re_{\gchar}$ & $Re_{\achar}$ & $Nu$\\
\hline
\hline
$0.01$ & $\hlm{8}{4}$ & $0.01$ & $10^{-6}$ & $\vartheta_c$&$**$ & $**$ & $**$ & $**$ & $**$ \\
 $0.01$ & $\hlm{8}{4}$ & $0.464$ & $10^{-4}$ & $\vartheta_c$&$**$ & $**$ & $**$ & $**$ & $**$ \\
 $0.01$ & $\hlm{8}{4}$ & $1.0$ & $10^{-6}$ & $\vartheta_c$&$**$ & $**$ & $**$ & $**$ & $**$ \\
 $0.02$ & $\hlm{8}{4}$ & $0.01$ & $10^{-6}$ & $\vartheta_c$&$**$ & $**$ & $**$ & $**$ & $**$ \\
 $0.02$ & $\hlm{8}{4}$ & $0.464$ & $10^{-4}$ & $\vartheta_c$&$**$ & $**$ & $**$ & $**$ & $**$ \\
 $0.02$ & $\hlm{8}{4}$ & $1.0$ & $10^{-6}$ & $\vartheta_c$&$**$ & $**$ & $**$ & $**$ & $**$ \\
 $0.05$ & $\hlm{8}{4}$ & $0.01$ & $10^{-6}$ & $\vartheta_c$&$**$ & $**$ & $**$ & $**$ & $**$ \\
 $0.05$ & $\hlm{8}{4}$ & $0.464$ & $10^{-4}$ & $\vartheta_c$&$**$ & $**$ & $**$ & $**$ & $**$ \\
 $0.05$ & $\hlm{8}{4}$ & $1.0$ & $10^{-6}$ & $\vartheta_c$&$**$ & $**$ & $**$ & $**$ & $**$ \\
 $0.1$ & $\hlm{8}{4}$ & $0.01$ & $10^{-6}$ & $\vartheta_c$&$**$ & $**$ & $**$ & $**$ & $**$ \\
 $0.1$ & $\hlm{8}{4}$ & $0.464$ & $10^{-4}$ & $\vartheta_c$&$3.39$ & $2.90\times 10{-1}$ & $3.16$ & $3.68\times 10^{-1}$ & $1.03$ \\
 $0.1$ & $\hlm{8}{4}$ & $1.0$ & $10^{-6}$ & $\vartheta_c$&$**$ & $**$ & $**$ & $**$ & $**$ \\
 $0.2$ & $\hlm{8}{4}$ & $0.01$ & $10^{-6}$ & $\vartheta_c$&$**$ & $**$ & $**$ & $**$ & $**$ \\
 $0.2$ & $\hlm{8}{4}$ & $0.464$ & $10^{-4}$ & $\vartheta_c$&$6.43$ & $7.65\times10{-1}$ & $5.67$ & $1.67$ & $1.17$ \\
 $0.2$ & $\hlm{8}{4}$ & $1.0$ & $10^{-6}$ & $\vartheta_c$&$**$ & $**$ & $**$ & $**$ & $**$ \\
\hline
 \end{tabular}
\end{center}
\caption{Same as Table \ref{t:p_space2} with heterogeneous temperature boundary condition.  Values denoted with a $**$ correspond to stable parameters that had negative kinetic energy growth rates.} 
\label{t:p_space3}
\end{table}


\end{document}